\documentclass[letterpaper]{JHEP3} %\input epsf.tex

\usepackage{epsfig}

\newcommand{\beq}{\begin{equation}}
\newcommand{\eeq}{\end{equation}}
\newcommand{\beqa}{\begin{eqnarray}}
\newcommand{\eeqa}{\end{eqnarray}}
\newcommand{\bea}{\begin{eqnarray}}
\newcommand{\eea}{\end{eqnarray}}

\newcommand{\ZZ}{\ensuremath{\mathcal{\delta{\cal{Z}}}}}
\newcommand{\ZZb}{\ensuremath{\mathcal{\delta \overline{{\cal{Z}}}}}}
\newcommand{\vb}{\right|}
\newcommand{\LL}{L}
\newcommand{\GG}{\ensuremath \mathbf{G}}

\renewcommand{\thepage}{\arabic{page}}

\title{\LARGE \bf Radiative Effects on the Chiral Square}
\author{Eduardo Pont\'{o}n and Lin Wang\\
Department of Physics, Columbia University, \\ 
538 W. 120th St, New York, NY 10027, USA \\
\email{eponton@phys.columbia.edu, lwang@phys.columbia.edu}} 
\abstract{ We consider general field theories in six dimensions, with
two of the dimensions compactified on a $T_{2}/Z_{4}$ orbifold.
Six-dimensional Weyl fermions propagating on this background give rise
to a chiral zero-mode, which makes them interesting for
phenomenological applications.  The compact two-dimensional space is
flat and has three conical singularities.  We consider the one-loop
structure of these theories, and show that the presence of logarithmic
divergences requires the introduction of counterterms precisely at
these three singular points.  We also show that the corresponding
localized operators are rotationally symmetric in the plane of the two
extra dimensions, as expected from the geometry about the
singularities.  We derive the propagators for spin-0, spin-1/2 and
spin-1 fields in momentum space, in such a way that the appropriate
boundary conditions are satisfied.  This allows us to efficiently
calculate loop diagrams in any given model.  We give general
expressions for the mass splittings among Kaluza-Klein modes within a
given level.  Our results can also be used to obtain interesting
KK-parity preserving interactions among Kaluza-Klein modes.  We pay
special attention to the components of six-dimensional gauge fields
that transform as scalars under the four-dimensional Lorentz group.
These states provide a characteristic signature for these scenarios.
In particular, we find that they can easily be the lightest particles
in the Kaluza-Klein spectrum.  }

\preprint{\hepph{} \\ CU-TP-1142 \\ 
} 

\begin{document}

%%%%%%%%%%%%%%%%%%%%%%%%%%%%%%%%%%%%%
\section{Introduction}
\label{sec:intro}

Recent years have seen a surge of interest in theories with extra
dimensions.  This is due, in part, to their potential to explain
various unanswered questions in the standard model of particle
physics.  A second, no less important, reason is that many of these
theories are amenable to verification or falsification in the next
decade.  Our ability to see the new physics and identify it as coming
from a higher dimensional structure depends sensitively on which
fields can probe the extra dimensions.

One of the most straightforward extensions of the standard model is
the assumption that {\textit{all}} standard model particles propagate
in more than four dimensions, also called Universal Extra Dimensions
or UED's.  The additional dimensions are compact and would manifest
themselves in Kaluza-Klein (KK) towers associated with each and every
standard model field.  These scenarios naturally contain a dark matter
candidate, which can account for the observed dark matter energy
density provided the compactification scale is around the electroweak
(EW) scale~\cite{Servant:2002aq}.  The six-dimensional case has a
number of additional attractive properties.  The requirements of
anomaly cancellation and fermion mass generation lead to the
prediction that the number of fermion generations is a multiple of
three \cite{Dobrescu:2001ae}.  Also, an exact discrete symmetry of the
compactified theory provides a natural explanation for the stability
of matter, even if baryon number is violated near the EW scale
\cite{Appelquist:2001mj}.  In addition, neutrinos are forced to be
Dirac fermions~\cite{Appelquist:2002ft}.  Six-dimensional theories
have also been considered in \cite{Green:1984bx}.

The phenomenology of the UED scenarios is rather interesting, and the
five-dimensional case has received considerable attention
\cite{Cheng:2002iz}.  The interactions arising from bulk operators
preserve KK-number, which is closely related to momentum conservation
in the extra dimensions.  An important consequence is that the heavy
modes can only be pair produced by such interactions and the effective
low-energy theory is simply the standard model, up to loop effects.
As a result, the bounds on the compactification scale are of order a
few hundred GeV \cite{Appelquist:2000nn}, and the KK states should be
accessible in high-energy collider experiments.  It is essential to
notice that a successful phenomenology can only be obtained when the
compact space contains singularities, that allow for a chiral
low-energy theory.  Such singularities can support operators that
induce couplings among KK modes not induced by the bulk interactions
\cite{Georgi:2000ks}.  Equally important is the fact that they give
the leading contribution to the mass splittings among the states
within a given KK level.  For these reasons, the localized operators
are essential in determining the phenomenology of these scenarios.
Other studies emphasizing the role of localized terms in extra
dimensional scenarios have appeared in \cite{Ponton:2001hq}.

In this paper we consider field theories in six dimensions.  Our aim
is to understand in detail the quantum structure of these theories, in
particular with regard to operators localized at the singular points.
We assume a flat spacetime background and that two of the six
dimensions are compactified on a ``chiral square'', as described in
\cite{Dobrescu:2004zi, Burdman:2005sr}.  Related studies have appeared
in \cite{Csaki:2002ur,Hashimoto:2004xz}.  The chiral square
compactification has the following simple description: starting from a
two dimensional square region, {\textit{adjacent}} sides of the square
are identified in pairs.  This can be contrasted with the torus
construction where {\textit{opposite}} sides are identified.  The
``chiral'' square has the topology of a two-dimensional sphere, but
the ``curvature'' is localized at three conical singularities.  We
will assign them coordinates $(x^{4},x^{5}) = (0,0)$, $(L,L)$ and
$(0,L) \sim (L,0)$.  The first two singularities have a deficit angle
of $3\pi/2$, while the latter has a deficit angle of $\pi$.  These
conical singularities play an important role in determining the
physics of these scenarios.

Fields propagating on the chiral square background can belong to four
different classes, that may be characterized by the boundary
conditions imposed on the sides of the fundamental square region:
\beq
\Phi(x^\mu, y, 0)  =  e^{i n \pi/2} \Phi(x^\mu, 0, y) ~, 
\hspace{1cm} n = 0,1,2~{\rm or}~3~,
\label{scalar-bc}
\eeq
where $x^{\mu}$ are coordinates for the non-compact dimensions and $0
\leq y \leq L$ parametrizes one pair of identified sides of the
square.  A similar condition holds for the second pair (see
Ref.~\cite{Dobrescu:2004zi} for further details).  In addition, the
derivatives normal to the ``edges'' of the square satisfy the
``smoothness'' condition
\beqa
\left.\partial_5\Phi \vb_{(x^4, x^5)=(y,0)}  = 
- e^{i  n \pi /2} 
\left.\partial_4\Phi \vb_{(x^4, x^5)=(0,y)} ~.
\label{scalar-bc2}
\eeqa
We will label the four classes of fields by the integer $n$, appearing
in Eq.~(\ref{scalar-bc}), that characterizes the boundary
conditions.\footnote{There is a second category of fields satisfying
``twisted'' boundary conditions that never give rise to a zero-mode,
but we do not consider them here.  In the orbifold construction, this
corresponds to starting with a ``torus'' with anti-periodic
identifications, before moding out by the discrete $Z_{4}$.  See
\cite{Dobrescu:2004zi} for further details.} It is understood that $n$
is defined modulo 4.  Notice that only those fields that satisfy
boundary conditions corresponding to $n=0$ admit a zero-mode, with an
associated flat profile.  Furthermore, when considering 6D Weyl
fermions, $\Psi_{\pm}$ [we use $+$ and $-$ to label the 6D chiralities
and reserve left and right to refer to the 4D chiralities], one finds
that their 4D left- and right-handed chiralities obey boundary
conditions corresponding to integers that differ by one: $n^{\pm}_{L}
- n^{\pm}_{R} = \pm 1$, where the sign depends on the 6D chirality of
the fermion in question.  Hence, fermions propagating on this space
naturally lead to a chiral low-energy theory: at most one of the left-
or right-handed chiralities has a zero mode.  This compactification is
equivalent to a $T^{2}/Z_{4}$ orbifold \cite{Dobrescu:2004zi}.

Gauge fields propagating in six dimensions give rise to 4-dimensional
spin-1 fields plus two scalar states per KK level.  One linear
combinations of these scalar states becomes the longitudinal
polarization of the massive spin-1 fields, while the orthogonal
combination remains as an additional scalar degree of freedom.  This
last phenomenon only occurs in six or higher dimensions, and the
observation of such scalar states in the adjoint representation of the
gauge group may be taken as a signature of the present class of
scenarios.  Following Ref.~\cite{Burdman:2005sr}, we refer to them as
``spinless adjoints''.

Another important property of these theories is that one can impose a
Kaluza-Klein or KK-parity, defined on KK modes by
\beq
\label{KKparity}
\Phi^{(j,k)} (x^\mu) \mapsto (-1)^{j+k} \, \Phi^{(j,k)} (x^\mu) ~,
\label{KKParity}
\eeq
where $\Phi$ stands for a field of any spin, and $j,k$ are integers
labeling the KK level.  The KK-parity has a geometrical interpretation
as a rotation by $\pi$ about the center of the chiral square.

It is important to keep in mind that higher dimensional field theories
should be regarded as effective theories with a cutoff $\Lambda$,
above which a more fundamental UV completion is required.  Integrating
out the (unknown) physics at the scale $\Lambda$ determines, in
principle, the coefficients of various operators through a matching
calculation.  Of course, even if we knew the UV completion, such a
calculation could be in practice very difficult to perform.
Therefore, in the spirit of effective theories, we simply allow for
all operators consistent with the ``low-energy'' symmetries, and
regard their coefficients as free parameters to be determined, if
possible, by experiment.

The operators one can write fall in two distinct classes.  Bulk
operators, such as the kinetic terms for the various bulk fields or
the associated gauge interactions, and operators localized at the
three singular points mentioned above.  The renormalization program
for these scenarios requires localized counterterms to absorb
divergences in the quantized theory.  In fact, we will see by an
explicit computation that the necessary localized counterterms reside
precisely at the conical singularities, and have the structure
\beq
\left[ \delta(x^{4}) \delta(x^{5}) 
+ \delta(L-x^{4}) \delta(L-x^{5}) \right] {\cal{O}}_{1}
+ \delta(x^{4}) \delta(L-x^{5})  {\cal{O}}_{2}~. 
\label{loc_ops}
\eeq
The fact that the operators at $(0,0)$ and $(L,L)$ have identical
coefficients is a consequence of KK-parity.  Localized operators at
$(0,L)$ have coefficients that are, in general, unrelated to those on
the previous two conical singularities.  Therefore, in the 6D theory,
each type of localized operator is characterized by two parameters.
This should be contrasted with the 5D case, where a single parameter
per operator is sufficient.

The operators appearing in ${\cal{O}}_{i}$ have dimensionful
coefficients, suppressed by the scale $\Lambda$.  The most important
ones are those with the lowest dimensionality.  These are kinetic
terms such as
\beq
{\cal{O}}_{i} = - \frac{1}{4} \, \hat{r}^{i}_{A}
F_{\mu\nu} F^{\mu\nu} +  
\frac{\hat{r}^{i}_{\Psi}}{\Lambda^{2}} \, 
i \overline{\Psi} \Gamma^{\mu} D_{\mu} \Psi + \cdots~,
\label{localized_kinetic}
\eeq
where $F_{\mu\nu}$ is the field strength of a generic gauge field, and
$\Psi$ stands for a generic 6D fermion.  We are assuming that the
gauge bulk kinetic term operator has a coefficient $-1/(4 g^{2}_{6})$,
where $g_{6}$ is the 6D gauge coupling constant with mass dimension
-1, so that the gauge field has mass dimension 1, as in four
dimensions.  We also defined dimensionless coefficients
$\hat{r}^{i}_{A}$ and $\hat{r}^{i}_{\Psi}$, with $i=1,2$.

The operators in Eq.~(\ref{localized_kinetic}) are very important in
determining the physics of these scenarios.  They give the leading
contributions to the mass splittings within states in a given
KK-level.  They also induce interactions among KK modes that do not
arise from bulk operators.  In fact, the interactions among KK-modes
induced by bulk operators satisfy well defined rules that follow from
the integrals over the extra dimensional space of the KK-mode
wavefunction profiles.  These ``tree-level'' selection rules are
closely related to momentum conservation in the extra dimension,
except that a reversing of momentum is allowed and the momenta along
the two compact dimensions can be interchanged.  We refer to these
type of interactions as ``KK-number'' preserving.  Interactions
arising from localized operators, on the other hand, lead to KK-number
violating transitions, which are of great phenomenological interest
\cite{Cheng:2002iz,PhenoPaper}.  The only constraint is that they
should satisfy the KK-parity symmetry of Eq~(\ref{KKParity}).

It is clearly very important to have an idea of how large the
corresponding mass splittings and KK-number violating couplings are.
As mentioned before, the values of the dimensionless coefficients in
Eq.~(\ref{localized_kinetic}) at the scale $\Lambda$ should be taken
as free parameters.  However, the values that are relevant to answer
the previous question are those at the scale of the corresponding KK
state, which is in general lower than $\Lambda$.  Those values can be
found by renormalization group (RG) evolution, with the ``bare''
coefficients at the scale $\Lambda$ providing the initial conditions.
The RG running is determined by the physics below $\Lambda$.
Furthermore, it leads to a logarithmic enhancement, so that one can
expect the ``low-energy'' contribution to dominate over the ``bare''
one.\footnote{Of course, in many instances the separation between the
cutoff and KK scales may be of order 10, so that the log may be of
order just a few.} To the extent that the logarithm is sufficiently
large, the size of the dimensionless coefficients is set by the
physics below $\Lambda$.

In this paper, we will calculate the contribution associated with the
KK-modes below $\Lambda$ at one-loop order.  It is natural to ask to
what extent one can trust the results of a one-loop calculation.  To
answer this question, it is necessary to be more specific about how to
choose the cutoff scale of the theory.  A conservative approach is to
identify the cutoff $\Lambda$ with the lowest scale where
perturbativity is lost in some sector of the theory.  For example, if
the field content is that of the standard model, and the 4D effective
low-energy theory is identified with the standard model, $\Lambda$ is
the scale at which the $SU(3)_{C}$ gauge interactions get strong.  To
be more precise, we define strong coupling to correspond to the case
where the loop expansion breaks down.  That is, all loop orders are
equally important and there is no small expansion parameter.  This
criterion provides a way of estimating the dimensionless coefficients
of any operator in the theory, when they are expressed in terms of the
cutoff scale $\Lambda$, following the rules of Naive Dimensional
Analysis or NDA \cite{Manohar:1983md}.  In extra dimensional theories,
the possibility of having operators localized on subspaces of reduced
dimensionality requires an extension of the NDA rules as first studied
in \cite{Chacko:1999hg}.

However, in practical situations there are additional interactions
that are weak at the scale $\Lambda$, e.g. gravitational or the
electroweak and Yukawa interactions (other than those associated with
the top quark).  Loops involving such interactions have a natural
expansion parameter in terms of the corresponding coupling.  It is
also natural to assume that bare operators involving, say, only
particles interacting through electroweak interactions, have a
corresponding suppression.  This comments also hold for flavor
violating transitions that are suppressed in the standard model.  At
any rate, based on phenomenological constraints, such an assumption
about the size of the coefficients of certain operators induced by the
UV completion seems necessary.  As long as the assumption is
technically natural, in the sense that said size is of the order of
loop effects, we are willing to take it as part of the definition of
the scenarios we are interested in.

In the absence of a known UV completion for the kind of theories we
study here, we assume that the order of magnitude of the coefficients
of bulk and localized operators are no larger than the loop induced
effects.  For the strongly interacting sector, this agrees with the
NDA rules.  For weakly interacting particles our assumption amounts to
the statement that the size of the bare coefficients, induced by the
physics that was integrated out at the scale $\Lambda$, is no larger
than the effects of the physics in the theory below $\Lambda$, which
is well approximated by the lowest order term in the loop expansion.

Thus, there are two qualitatively different cases: for particles that
interact strongly, the best one can hope for is to {\textit{estimate}}
the size of the coefficients of local operators in the higher
dimensional theory.  The one-loop contribution to localized operators
such as those in Eqs.~(\ref{loc_ops}) and (\ref{localized_kinetic})
for the case of quarks and gluons can only be taken as indicative of
the order of magnitude of the effect.  Higher orders in the loop
expansion give equally important contributions.  On the other hand,
for weakly interacting particles one can hope that the one-loop
computation is a good approximation and the corresponding effects are
under control.

There are also finite one-loop effects that contribute to the mass
splittings as well as to the KK-number violating interactions.  Some
of the finite contributions to the mass splittings can be calculated
in the context of a simple torus compactification, as done in
\cite{Cheng:2002iz}.  These effects are subdominant, not being
logarithmically enhanced.  However, the finite contributions to
certain KK-number violating interactions can be of phenomenological
interest.  We mention here two important cases: the couplings of
KK-parity even states to a pair of zero-mode gauge bosons, and the
coupling of KK-parity even spinless adjoints to a pair of zero-mode
fermions.

In the first instance, we notice that the couplings of zero-mode gauge
bosons are rather constrained by the unbroken 4D gauge invariance
associated with these massless spin-1 fields.  To be specific,
consider the coupling of a $(1,1)$ KK-gluon to two gluons.  The
effective 4D operator must take the form of a product of three field
strengths, one associated with each of the spin-1
fields.\footnote{Note that after KK decomposition, the cubic terms in
the non-abelian gauge kinetic term of Eq.~(\ref{localized_kinetic})
naively lead to a vertex between two gluons and a $(1,1)$ state.
However, they also lead to mixing between the zero-mode and the heavy
KK modes.  The unbroken 4D gauge invariance insures that this system
contains a massless state.  Furthermore, there exists a basis where
both kinetic and mass mixings between this state and the massive ones
are absent.  Therefore, dimension four operators that would induce a
$(0,0)$--$(0,0)$--$(1,1)$ vertex do not exist.} This effective
four-dimensional, KK-number violating operators can arise from
localized operators as in Eq.~(\ref{loc_ops}) with, e.g.,
\beqa
{\cal{O}}_{1,2} \sim \frac{\hat{r}}{2^{3} \Lambda^{2}} 
\, \GG \! \cdot \! \GG\!  \cdot \! \GG~,
\label{LocalizedGGG}
\eeqa
%
%%
%\beqa
%\frac{\hat{r}}{2^{3} \Lambda^{2}} f^{abc} 
%G_{\mu\nu}^{a} G^{\nu\lambda \, b} G^{\mu \, c}_{\lambda}~.
%\label{LocalizedGGG}
%\eeqa
%%
where $\GG$ stands for the gluon field strength and the dots denote
appropriate contractions both for Lorentz and gauge indices.  Using
NDA to estimate the dimensionless coefficient, we find $\hat{r} \sim
N_{c}/l_{4}$, where $l_{4} = 16\pi^{2}$, and $N_{c}$ is the number of
colors.  Integrating over the extra dimensions, we find an effective
4D operator
\beqa
\frac{1}{2^{3} M_{c}^{2}}
\left( \frac{M_{c}}{\Lambda} \right)^{2} \left( \frac{N_{c} }{l_{4}} \right)  
 \GG^{(0,0)} \! \cdot \! \GG^{(0,0)} \! \cdot \! \GG^{(1,1)}~,
\label{effectiveGGG}
\eeqa
where we chose the KK scale $M_{c} = 1/R$ as the mass scale
suppressing the operator.  Although the fields are not canonically
normalized, the coefficients of the various kinetic terms are of order
$1/g_{4}^{2}$, where $g_{4}$ is the observed $SU(3)_{C}$ coupling,
which is of order one at the KK scale.  Therefore, the size of the
effect can be read directly from Eq.~(\ref{effectiveGGG}).

The ratio $M_{c}/\Lambda$ can be determined by matching the 6D and 4D
gauge coupling constants.  Neglecting the contribution to the kinetic
term from the localized gauge operator in (\ref{localized_kinetic}),
this is simply $g_{6}^{2}/(\pi R)^{2} \sim g_{4}^{2} = {\cal{O}}(1)$.
Since NDA gives $g_{6}^{2} \Lambda^{2} \sim l_{6}/N_{c}$, where $l_{6}
= 128 \pi^{3}$ is a 6D loop factor, we find $(M_{c}/\Lambda)^{2} \sim
N_{c} \pi^{2}/l_{6} \sim N_{c}/(2 l_{4})$.  This suggests that
operators such as (\ref{LocalizedGGG}) are generated at two-loop
order.  In fact, this is easy to see from the fact that at one loop
only a finite number of states contribute to KK-number violating
operators, and the finiteness follows from the corresponding statement
in 4D QCD. At higher loop order, one encounters infinite KK sums that
require the localized counterterms in Eq.~(\ref{LocalizedGGG}).

Nevertheless, we expect a non-vanishing, finite one-loop induced
vertex between a $(1,1)$ gluon and two zero-mode gluons.  This
corresponds to an effective 4D operator as in (\ref{effectiveGGG}),
but without the factor $(M_{c}/\Lambda)^{2}$.  This finite effect
clearly dominates over the expected contributions from the physics
integrated out at the scale $\Lambda$.

The second example where finite effects may play an important role is
the coupling of the KK-parity even spinless adjoints to zero-mode
fermions or gluons.  The coupling to gluons is similar to the
couplings of heavy gluons to gluons just discussed.  The coupling to
fermions arising from localized operators in the 6D theory proceeds
through operators like\footnote{Note that the operator $i
\overline{\Psi}_{1} \Gamma^{M} \Gamma^{N} \Psi_{2} \, G_{MN}$ has
lower dimensionality.  However, this operator flips chirality and is
forbidden by gauge invariance for the standard model field content,
unless $\Psi_{1} = \Psi_{2}$, in which case it does not contain two
fermion zero-modes.}
\beqa
{\cal{O}}_{1,2} \sim \frac{\hat{r}^{\prime}}{\Lambda^{2}} 
\overline{\Psi} \, \Gamma^{M} \Gamma^{N} \Gamma^{L} 
\Psi \partial_{L} G_{MN}~,
\label{LocalizedFFA}
\eeqa
where $M,N = 0,1,\ldots,5$ run over the 6-dimensional Lorentz indices.
Again, the NDA estimate corresponds to a two-loop effect, and the
physical coupling is dominated by a finite one-loop contribution.
Although quite interesting for phenomenological applications, the
calculation of such finite effects is beyond the scope of this work.

In this paper we compute the one-loop logarithmically divergent
contributions to the localized kinetic terms of scalar, fermion and
gauge fields.  This allows us to calculate the leading contributions
to the mass splittings and to certain KK-number violating interactions
such as those between KK-parity even gauge bosons and zero-mode
fermions, which are of phenomenological interest \cite{PhenoPaper}.
Recall that the power-law divergences renormalize bulk operators and
do not induce mass splittings.  Our main results for the localized
operators are given in Eqs.~(\ref{Arc_scalar}), (\ref{Arc_fermion}),
(\ref{Arc_gauge}), (\ref{Frc_gauge}), (\ref{Frc_Yukawa}),
(\ref{rcn0}), (\ref{rcn0_Yukawa}), (\ref{rcSpinless_gauge}),
(\ref{rcSpinless_fermion}) and (\ref{rcSpinless_scalar}) in the body
of the paper, and are summarized in
Eqs.~(\ref{localOps1_summary})--(\ref{r2}).  We give here the mass
shifts for fields of various spins, that can be read from those
localized operators [see also Tables~\ref{tableABmatter},
\ref{tableABYukawa} and \ref{tableBBpspinless2}].

In the following expressions, $g_{4}$ and $\lambda_{4,i}$ are the
4-dimensional gauge and Yukawa couplings, respectively.  $C_{2}(F)$ is
the eigenvalue of the Casimir operator in the representation of the
fields $F = A_{\mu}$, $\Psi$ or $\Phi$, while ${\rm{Tr}}(T^{a} T^{b})
= T(F) \delta_{ab}$, where $T^{a}$ are the generators in the
representation of the field $F$.  We consider 6D gauge fields which
comprise 4D spin-1 and spin-0 components, 6D Weyl fermions that give
rise to a zero-mode of any 4D chirality (we do not consider fermions
satisfying $n=2$ boundary conditions), and complex 6D scalar fields
satisfying any of the four boundary conditions $n=0,1,2$ or $3$.  The
6D fermions can have any of the two 6D chiralities, $\Psi_{\pm}$.
Notice that the Yukawa couplings require the presence of fermions with
opposite 6D chiralities.

The mass-shifts are different for KK-parity even and KK-parity odd
states, as a result of the localized operators at the conical
singularity with coordinates $(0,L)$.  We obtain:

\vspace{3mm}
\noindent
\textbf{For KK-parity odd states, $\mathbf (-1)^{j+k} = -1$: }
\begin{itemize}
\item Spin-1 fields:
\beqa
\frac{\delta M^{A}_{j,k}}{M_{j,k}}  &=& \frac{g_4^2}{16 \pi^2} 
\ln \frac{\Lambda^2}{\mu^2} \left[ \frac{14}{3} C_{2}(A) - 
\frac{2}{3} \sum_\Psi T(\Psi) + \sum_\Phi T(\Phi) 
\times \left\{%
\begin{array}{r}
     -5/12 \\
     1/12 \\
     3/12\\
     1/12 \\
\end{array}%
\right. \hspace{3mm}
\right]~.
\label{OddGaugeMass}
\eeqa
\item Spinless adjoints:
\beqa
\frac{\delta M^{S\!A}_{j,k}}{M_{j,k}}  &=& \frac{g_4^2}{16 \pi^2} 
\ln \frac{\Lambda^2}{\mu^2} \left[ 8 \, C_{2}(A) - 
4 \sum_\Psi T(\Psi) + \sum_\Phi T(\Phi) 
\times \left\{%
\begin{array}{r}
     13/4 \\
     -1/4 \\
     -11/4\\
     -1/4 \\
\end{array}%
\right. \hspace{3mm}
\right]~.
\label{OddSpinlessMass}
\eeqa
\item Spin-1/2 fields:
\beqa
\frac{\delta M^{\Psi_{+}}_{j,k}}{M_{j,k}} &=& \frac{1}{16 \pi^2} 
\ln \frac{\Lambda^2}{\mu^2} \left[ 4 \sum_{{\rm gauge}} g^{2}_{4} 
C_{2}(\Psi) + \sum_{i} |\lambda_{4,i}|^{2}
\times \left\{%
\begin{array}{r}
     5/8 \\
     7/8 \\
     -3/8\\
     -9/8 \\
\end{array}%
\right. \hspace{3mm}
\right]~.
\label{OddFermionMass}
\eeqa
For chirality $-$ fermions, $\Psi_{-}$, the second and fourth lines
are exchanged.
\item Spin-0 fields:
\beqa
\frac{\delta \left( M^{\Phi}_{j,k} \right)^{2}}{M^{2}_{j,k}} &=& \frac{1}{16 \pi^2} 
\ln \frac{\Lambda^2}{\mu^2} \left[ \sum_{{\rm gauge}} g^{2}_{4} 
C_{2}(\Phi) 
\times \left\{%
\begin{array}{c}
     15/4 \\
     15/4 \\
     7/4 \\
     15/4 \\
\end{array}%
\right. 
+ \sum_{i} |\lambda_{4,i}|^{2}
\times \left\{%
\begin{array}{r}
     2 \\
     0 \\
     0\\
     0 \\
\end{array}%
\right. \hspace{3mm}
\right]~.
\label{OddScalarMass}
\eeqa
\end{itemize}
In the equations for the spin-1 fields and spinless adjoints the sums
run over \textit{6D Weyl} fermions, $\Psi_{\pm}$, that give rise to a
zero-mode.  The four lines in the sums over the \textit{complex} 6D
scalars, $\Phi$, list the results for scalars obeying $n=0,1,2$ and
$3$ boundary conditions, in that order.  The terms proportional to
$C_{2}(A)$ include the contributions of the complete 6D gauge
multiplet, i.e. both the 4D spin-1 components, as well as the spinless
adjoints.  The sums in the equation for the fermions are over its
gauge interactions, and Yukawa interactions with scalars obeying any
of the four types of boundary conditions.  Similar comments apply to
the mass-shift of scalars.  In this latter case, the four lines refer
to the boundary conditions obeyed by the corresponding scalar, ordered
as just mentioned.  When the scalar satisfies $n=0$ boundary
conditions [first line in Eq.~(\ref{OddScalarMass})], there is also a
localized bare mass contribution [see Eq.~(\ref{Phi_mass-shifts})].
In these formulae, $\mu$ is the renormalization scale, and should be
taken of the order of the scale of the corresponding KK state, e.g.
the tree-level mass $M_{j,k} = \sqrt{j^{2}+k^{2}}/R$.

\vspace{3mm}
\noindent
\textbf{For KK-parity even states, $\mathbf (-1)^{j+k} = +1$:}

\begin{itemize}
\item Spin-1 fields:
\beqa
\frac{\delta M^{A}_{j,k}}{M_{j,k}}  &=& \frac{g_4^2}{16 \pi^2} 
\ln \frac{\Lambda^2}{\mu^2} \left[ \frac{17}{3} C_{2}(A) - 
\frac{2}{3} \sum_\Psi T(\Psi) + \sum_\Phi T(\Phi) 
\times \left\{%
\begin{array}{r}
     -1/2 \\
     1/6 \\
     1/6 \\
     1/6\\
\end{array}%
\right. \hspace{3mm}
\right]~.
\label{EvenGaugeMass}
\eeqa
\item Spinless adjoints:
\beqa
\frac{\delta M^{S\!A}_{j,k}}{M_{j,k}}  &=& \frac{g_4^2}{16 \pi^2} 
\ln \frac{\Lambda^2}{\mu^2} \left[ 9 \, C_{2}(A) - 
4 \sum_\Psi T(\Psi) + \sum_\Phi T(\Phi) 
\times \left\{%
\begin{array}{r}
     7/2 \\
     -1/2 \\
     -5/2\\
     -1/2 \\
\end{array}%
\right. \hspace{3mm}
\right]~.
\label{EvenSpinlessMass}
\eeqa
\item Spin-1/2 fields:
\beqa
\frac{\delta M^{\Psi_{+}}_{j,k}}{M_{j,k}} &=& \frac{1}{16 \pi^2} 
\ln \frac{\Lambda^2}{\mu^2} \left[ 5 \sum_{{\rm gauge}} g^{2}_{4} 
C_{2}(\Psi) + \sum_{i} |\lambda_{4,i}|^{2} 
\times \left\{%
\begin{array}{r}
     3/4 \\
     3/4 \\
     -1/4\\
     -5/4\\
\end{array}%
\right. \hspace{3mm}
\right]~.
\label{EvenFermionMass}
\eeqa
For chirality $-$ fermions, $\Psi_{-}$, the second and fourth lines
are exchanged.
\item Spin-0 fields:
\beqa
\frac{\delta \left( M^{\Phi}_{j,k} \right)^{2}}{M^{2}_{j,k}} &=& \frac{1}{16 \pi^2} 
\ln \frac{\Lambda^2}{\mu^2} \left[ \sum_{{\rm gauge}} g^{2}_{4} 
C_{2}(\Phi) 
\times \left\{%
\begin{array}{c}
     11/2 \\
     15/4 \\
     0\\
     15/4 \\
\end{array}%
\right. 
+ \sum_{i} |\lambda_{4,i}|^{2} 
\times \left\{%
\begin{array}{r}
     2 \\
     1 \\
     0 \\
     1 \\
\end{array}%
\right. \hspace{3mm}
\right]~.
\label{EvenScalarMass}
\eeqa
For scalars satisfying $n=1$ or $n=3$ boundary conditions (second and
fourth lines), the Yukawa contribution given here applies when
\textit{both} 6D fermions in the loop give rise to a zero-mode.  If
\textit{only one} of the fermions contains a zero-mode, the Yukawa
contribution has the opposite sign [see
subsection~\ref{sec:ScalarYukawa} for more details].

\end{itemize}

A nontrivial check of the previous mass-shift formulae can be obtained
by considering a supersymmetric theory.  As a first example, consider
supersymmetric (SUSY) QCD in six dimensions.  This theory contains a
6D gauge field and a 6D Weyl fermion.  For concreteness assume that
the fermion has 6D $+$ chirality, $\Lambda_{+}$, and that it gives
rise to a left handed zero-mode.  In 4D, $N=1$ language these fields
arrange themselves into a vector multiplet, $V = (A_{\mu},
\lambda_{+L})$, and a chiral multiplet, $H = (A_{-},
\lambda^{c}_{+R})$, where $A_{\pm} = A_{4} \pm i A_{5}$ are the
spinless adjoints, and $\Lambda_{+} = \lambda_{+L} + \lambda_{+R}$.
The boundary conditions break the higher dimensional supersymmetry
down to 4D, $N=1$ SUSY: $V$ satisfies $n=0$ boundary conditions, while
$H$ satisfies $n=1$ boundary conditions.  Therefore, the above
formulae should predict that $A_{\mu}$ and $\lambda_{+L}$ present the
same mass-shift, as well as $A_{+}$ and $\lambda_{+R}$.  Actually,
since at each KK level $\lambda_{+L}$ and $\lambda_{+R}$ combine into
a Dirac fermion, the complete 4D, $N=2$ supermultiplet should present
the same mass-shift, a fact that is easy to check from
Eqs.~(\ref{OddGaugeMass})--(\ref{OddFermionMass}) [and independently
from Eqs.~(\ref{EvenGaugeMass})--(\ref{EvenFermionMass})].  One can also
check some of the terms coming from the Yukawa interactions by adding
a hypermultiplet, i.e. a 6D Weyl spinor, $\Psi_{-}$, assumed to have a
left-handed zero-mode, and two complex 6D scalars, $\Phi$ and
$\Phi^{c}$.  In 4D, $N=1$ language these decompose into two chiral
multiplets, $Q = (\Phi, \psi_{-L})$ and $Q^{c} = (\Phi^{c},
\psi^{c}_{-R})$, where $\Psi_{-} = \psi_{-L} + \psi_{-R}$.  Now $Q$
satisfies $n=0$ boundary conditions, while $Q^{c}$ satisfies $n=3$
boundary conditions.  In the SUSY limit, the gauginos interact with
the scalars and fermions in the hypermultiplet with Yukawa couplings
of strength $\lambda_{4} = \sqrt{2} g_{4}$.  Taking this into account,
the mass-shifts for the KK parity odd gauge bosons, gauginos and
spinless adjoints are all proportional to $4\,C_{2}(A) - T(Q)$.
Similarly, the mass-shifts for the KK-parity even states are
proportional to $5\,C_{2}(A) - T(Q)$.  To check that the fermion and
scalar fields in the hypermultiplet present a common mass-shift
requires inclusion of the effects from the trilinear and quartic
scalar self-interactions in the scalar mass formulae, which we have
not computed.  Nevertheless, the mass-shift for the hypermultiplet can
be obtained from the fermion mass-shift formulae given above.

Notice that there are a couple of qualitative differences compared to
the mass shifts one obtains in a 5-dimensional theory.  First, in 6D
the fermions give a negative contribution to the masses of the gauge
bosons, whereas in 5D the fermion contribution vanishes as a result of
a cancellation between the left- and right-handed components of the 5D
Dirac fermion \cite{Cheng:2002iz}.  In the 6-dimensional case, one can
trace the surviving contribution to the existence of additional
states.  For example, the mass shift of the $(2,0)$ states, which play
a role akin to the second level states in 5D, receives no contribution
from $(1,0)$ states due to a cancellation similar to the 5D case, but
it receives a contribution from the $(1,1)$ states, which have no
analog in 5D. A second difference is related to the Yukawa
contributions to the fermion masses, due to couplings to scalars with
a zero-mode: in 5D such a contribution is negative, but in 6D it is
found to be positive.  The positive sign is special to six dimensions,
and originates in the existence of two 6D chiralities.  The Yukawa
coupling necessarily involves two fermions of opposite 6D chiralities,
and this translates into a relative physical phase that accounts for
the previous result.  Finally, one can also see that the spinless
adjoints receive a negative contribution from their gauge interactions
with the fermions, as do the gauge bosons.  However, the coefficient
in the spinless adjoint formula is larger than for the gauge bosons.
As a result, the spinless adjoints are lighter than their spin-1
counterparts.  This may have interesting consequences for dark matter,
since the lightest KK particle, in the 6D standard model context, is
the hypercharge spinless adjoint.  It can also affect the collider
phenomenology in an interesting way \cite{PhenoPaper}.

The general formulae Eqs.~(\ref{OddGaugeMass})--(\ref{EvenScalarMass})
can be easily applied to various models of interest.  As explained
before, when applied to strongly interacting particles they should be
taken only as indicative of the order of magnitude of the effect.
However, when applied to weakly interacting particles, such as the
electroweak gauge bosons or leptons, they should reliably give the
leading contribution to the corresponding mass shifts.

This paper is organized as follows: in Section~\ref{scalarcase} we
develop the technical ingredients that are necessary to perform the
one-loop computation.  This requires finding propagators that encode
correctly the boundary conditions implied by Eqs.~(\ref{scalar-bc})
and (\ref{scalar-bc2}).  We do so in the context of a scalar field
theory.  We then give the propagators for fermion fields
(Section~\ref{fermioncase}) and gauge fields
(Section~\ref{gaugecase}).  The latter include both the spin-1 and
spin-0 components (under the 4D Lorentz group), as well as the ghost
fields.  Section~\ref{radiative} contains our main results.  We
compute the one-loop corrections to the gauge boson two-point function
in Subsection~\ref{sec:Gauge_self}, to the fermion two-point function
in Subsection~\ref{sec:Fermion_self}, to the scalar two-point function
in Subsection~\ref{sec:scalar_self}, and to the two-point functions of
the ``spinless adjoints'' in Subsection \ref{sec:Spinless_self}.  We
summarize and conclude in Section~\ref{sec:conclusions}.  We also
include two appendices.  In Appendix~\ref{App:MomSpace} we give
details on how to relate the KK-number and momentum space
representations of the propagators.  In Appendix~\ref{App:Propagators} 
we give details of
the derivation of the 6D propagators associated with fermion and 6D
gauge fields, propagating on the chiral square background.

%%%%%%%%%%%%%%%%%%%%%%%%%%%%%%%%%%%%%
\section{The Scalar Case: Generalities}
\label{scalarcase}

Our first goal is to develop the tools necessary to perform loop
calculations in an efficient manner.  One approach would be to
decompose the bulk fields into a set of Kaluza-Klein modes satisfying
the appropriate boundary conditions, and derive the effective
four-dimensional theory to read the relevant vertices (given by
integrals over KK wavefunctions).  These can then be used to calculate
any quantity of interest.  An alternative approach is to work in
momentum space in the extra dimensions, and include the correlations
arising from the boundary conditions in the form of the propagators.
This latter approach has the advantage that the effects of the
boundary conditions appear only in the propagators, and are therefore
universal.  The vertices conserve momentum in the standard sense and
can be read as in the $T^{2}$ compactification.  We adopt the second
approach since it is simpler to generalize to various types of
interactions.  The first step is then to understand how to encode the
boundary conditions in the form of the propagators.

We start by deriving some useful general relations in the context of a
scalar field theory.  We first write down the general expression for
the scalar propagator in ``Kaluza-Klein space'', where the boundary
conditions are manifest.  We can then use this representation as a
starting point for deriving the momentum space expression of the
propagator, that correctly includes the effects of the boundary
conditions.  We also discuss the generalizations needed in the
presence of Kaluza-Klein mixing.  This will allow us to understand the
structure of the propagator when radiative effects are included.

%%%%%%%%%%%%%%%%%%%%%%%%%%%%%%%%%%%%%
\subsection{Propagators and Boundary Conditions}

We start from the KK expansion of a 6-dimensional scalar field
\beq
\Phi_{n}(x^{\mu}; z) = \frac{1}{L} \, {\sum_{j,k}}' \,
\phi^{(j,k)}(x^{\mu}) f_{n}^{(j,k)}(z)~,
\label{scalarKK}
\eeq
where we use the shorthand notation $z = (x^{4}, x^{5})$.  These
coordinates range over the fundamental square $0 \leq x^{4},x^{5} \leq
L$.  The KK wavefunctions, $f_{n}^{(j,k)}(z)$, satisfy the boundary
conditions appropriate for the ``chiral square'', as derived in
Ref.~\cite{Dobrescu:2004zi}.  They may be written as
\beq
f_{n}^{(j,k)}(z) = \frac{1}{2(1+\delta_{j,0}\delta_{k,0})}
\left[ h^{(j,k)}(z) +
e^{i\theta} h^{(k,-j)}(z) + e^{2i\theta} h^{(-j,-k)}(z) +
e^{3i\theta} h^{(-k,j)}(z) \right]~,
\label{KKf}
\eeq
in terms of the momentum space wavefunctions
\beq
h^{(j,k)}(z) = e^{i(j x^{4} + k x^{5})/R}~,
\label{planewaves}
\eeq
where $R = L/\pi$, and $\theta = n \pi/2$ with $n = 0,1,2,3$.  The
integer $n$ labels the possible consistent boundary conditions that
result after imposing the folding identifications described in
\cite{Dobrescu:2004zi}.  Notice that the KK towers contain a zero-mode
only for $n=0$.  The $'$ superscript in the summation in
Eq.~(\ref{scalarKK}) indicates that the KK sums run over the
restricted range $j > 0$, $k \geq 0$ and $j=k=0$.

Given the KK wavefunctions and spectrum, we can immediately write down
the expression for the propagator.  It is convenient to work in
configuration space in $x^{4}$, $x^{5}$, since this will allow us to
easily project onto KK-number or momentum space as needed.  However,
we do work in momentum space for the four non-compact dimensions from
the beginning, i.e. we use a mixed position and momentum space
representation.  The general representation of the scalar propagator
in the compactified theory is
\beqa
G_{n}(p; z; z^{\prime}) &=&
\int d^{4}x \, e^{ipx} 
\langle \Phi_{n}(x; z) \Phi^{\dagger}_{n}(0;z^{\prime}) \rangle
\nonumber \\
&=& \frac{1}{L^{2}} \, {\sum_{j,k}}' \, g_{S}^{j,k} f_{n}^{(j,k)}(z)
\left[ f_{n}^{(j,k)}(z^{\prime}) \right]^{*}~,
\label{DiagGenProp}
\eeqa
where $g_{S}^{j,k}$ is the 4-dimensional scalar propagator (in the
KK-number representation)
\beq
g_{S}^{j,k}  = \frac{i}{p^{2} - M_{j,k}^{2}}~,
\label{GKKScalar}
\eeq
and in the present case the spectrum is given by
\beq
M_{j,k}^{2} = \frac{j^{2} + k^{2}}{R^{2}}~.
\eeq
In fact, this propagator satisfies
\beq
(p^{2} + \partial_{4}^{2} + \partial_{5}^{2})
G_{n}(p; z; z^{\prime}) = i \, \delta^{(2)}(z - z^{\prime})~,
\label{GScalarEqn}
\eeq
since
\beq
(\partial_{4}^{2} + \partial_{5}^{2} + M_{j,k}^{2})
f_{n}^{(j,k)}(z) = 0~,
\eeq
and the $f_{n}^{(j,k)}(z)$ form a complete set for functions
satisfying the appropriate boundary conditions:\footnote{If one wants
to interpret Eq.~(\ref{completeness}) outside the fundamental region
$0 \leq x^{4},x^{5} \leq L$, the $\delta$-function on its r.h.s.
should be extended in a manner consistent with the relevant boundary
conditions, e.g. $\delta^{(2)}({\cal{R}}(z) - z^{\prime}) = e^{-i n
\pi/2} \delta^{(2)}(z - z^{\prime})$, $\delta^{(2)}(z -
{\cal{R}}(z^{\prime})) = e^{i n \pi/2} \delta^{(2)}(z - z^{\prime})$
and $\delta^{(2)}({\cal{R}}(z) - {\cal{R}}(z^{\prime})) =
\delta^{(2)}(z - z^{\prime})$, where $\cal{R}$ stands for a
counterclockwise rotation by $\pi/2$ in the $z$-plane.  It should also
be extended periodically, with period $2L$ along both $x^{4}$ and
$x^{5}$, outside $-L \leq x^{4},x^{5} \leq L$.}
\beq
\frac{1}{L^2} \,
{\sum_{j,k}}' \, f_n^{(j,k)}(z) \left[ f_n^{(j,k)}(z^{\prime}) \right]^*
= \delta^{(2)}(z - z^{\prime})~.
\label{completeness}
\eeq
It is also worth keeping in mind the orthonormality relations
\beq
\frac{1}{L^{2}} \int_{0}^{L}  d^{2}z \, f_{n}^{(j,k)}(z)
\left[ f_{n}^{(j',k')}(z) \right]^{*} =
\delta_{j,j'} \delta_{k,k'}~,
\label{orthof}
\eeq
which ensure the canonical normalization of the KK fields,
$\phi^{(j,k)}(x^{\mu})$.

%%%%%%%%%%%%%%%%%%%%%%%%%%%%%%%%%%%%%
\subsection{Propagators in Momentum Space}
\label{MomSpace}

Now suppose we want to work in momentum space, as opposed to KK-number
space, in the compactified dimensions: $(p^{4}, p^{5}) = (m/R,l/R)$,
where $m$ and $l$ are {\textit{arbitrary}} integers, i.e. we define
\beqa
G_{p, n}^{(m,l; m',l')} &\equiv& \left( \frac{1}{2 L} \right)^{2}
\int_{-L}^{L} \, d^{2}z \, d^{2}z^{\prime}
e^{i (p_{4} x^{4} + p_{5} x^{5})} e^{-i (p^{\prime}_{4} x^{\prime 4}
+ p^{\prime}_{5} x^{\prime 5})} G_{n}(p; z; z^{\prime})~.
\label{Gp4p5}
\eeqa
Note that we are letting the integration run over the extended range
$-L \leq x^{4}, x^{5} \leq L$; it is understood that $G_{n}(p; z;
z^{\prime})$ has been analytically continued outside the fundamental
region $0 \leq x^{4}, x^{5} \leq L$.  The factor of $(1/2 L)^{2}$ was
introduced so that $G_{p, n}^{(m,l; m',l')}$ has mass dimension $-2$,
as in four dimensions. 

By using the representation of the propagator given in
Eq.~(\ref{DiagGenProp}) we automatically obtain from Eq.~(\ref{Gp4p5})
the momentum space representation satisfying the appropriate boundary
conditions.  In switching from KK-number to momentum space one
encounters the integrals
\beqa
\frac{1}{4L^{2}} \int_{-L}^{L} d^{2}z \left[h^{(m,l)}(z) \right]^{*}
f_{n}^{(j,k)}(z) &=& \frac{1}{2 \left[1+\delta_{j,0}\delta_{k,0} \right]}
\, \hat{\delta}(m,l;j,k;n)~,
\label{rotationmatrix}
\eeqa
where we used the explicit form of the KK wavefunctions given in
Eq.~(\ref{KKf}), as well as the orthonormality relations
\beq
\frac{1}{4L^{2}} \int_{-L}^{L}  d^{2}z \, h^{(m,l)}(z)
\left[ h^{(m',l')}(z) \right]^{*} = \delta_{m,m'} \delta_{l,l'}~,
\label{orthoh}
\eeq
that hold for the standard plane waves given in Eq.~(\ref{planewaves}).
We also defined a ``generalized'' Kronecker delta
\beq
\hat{\delta}(m,l;m',l';n) =
\delta_{m,m'} \delta_{l,l'} +
e^{i\theta} \delta_{m,l'} \delta_{l,-m'} +
e^{2i\theta} \delta_{m,-m'} \delta_{l,-l'} +
e^{3i\theta} \delta_{m,-l'} \delta_{l,m'}~,
\label{delta}
\eeq
where $\theta = n \pi/2$.  Note that when the quantum numbers
$m,l,m',l'$ are all taken positive, Eq.~(\ref{delta}) coincides with
the standard two-dimensional Kronecker delta.  The additional terms
take into account the boundary conditions and depend on the integer
$n$.  However, note also that for the case of the zero-mode (which
only arises for $n=0$) one has $\hat{\delta}(m,l;0,0;n = 0) = 4
\delta_{m,0} \delta_{l,0}$, with an extra factor of 4.

%%%%%%%%%%%%%%%%%%%%%%%%%%%%%%%%%%%%%
\subsubsection{Diagonal Propagators}
\label{Diagonal}

For a propagator with the general representation in KK-number space,
\beq
G_{n}(p; z; z^{\prime}) = \frac{1}{L^{2}} \, {\sum_{j,k}}' \, g_{j,k} \,
f_{n}^{(j,k)}(z) \left[f_{n}^{(j,k)}(z^{\prime})\right]^{*}~,
\label{GenProp}
\eeq
we can also write
\beqa
G_{n}(p; z; z^{\prime}) &=&
\frac{1}{4 L^{2}} \sum_{m,l} \sum_{m',l'} G_{p, n}^{(m,l; m',l')}
h^{(m,l)}(z) \left[ h^{(m',l')}(z^{\prime}) \right]^{*}~,
\label{GMomRep}
\eeqa
with $G_{p, n}^{(m,l; m',l')}$ as defined in Eq.~(\ref{Gp4p5}).
Notice that the sums in Eq.~(\ref{GMomRep}) run over {\textit{all}}
integers.  We will reserve the labels $m$ and $l$ to denote the
momentum along the compact dimensions in units of $1/R$, hence running
unrestricted over integer values, while leaving the labels $j$ and $k$
to denote the KK numbers, which can take integer values on the
restricted range $j > 0$, $k \geq 0$ and $j=k=0$.  For a scalar field
one finds the simple result, derived in Appendix~\ref{App:MomSpace},
\beqa
G_{p, n}^{(m,l; m',l')} &=& \frac{i}{p^{2} - M_{m,l}^{2}} \, 
\hat{\delta}(m,l;m',l';n)~.
\label{Gml}
\eeqa
Therefore, the information about the boundary conditions is contained
in the ``generalized'' $\delta$-functions defined in
Eq.~(\ref{delta}).  Note that for the zero mode, $G_{p, n=0}^{(m,l;
0,0)} = 4 (i/p^{2}) \delta_{m,0} \delta_{l,0}$, one finds an
additional factor of 4 in the momentum representation, compared to the
zero-mode propagator $g^{0,0}_{S} = i/p^{2}$ in the KK-number
representation.

%%%%%%%%%%%%%%%%%%%%%%%%%%%%%%%%%%%%%
\subsubsection{Kaluza-Klein Mixing}
\label{KKmixing}

It is also useful to derive the relation between the KK-number and
momentum space representations of the propagator when transitions
among different KK states are allowed.  This situation arises when
interactions are included and the unperturbed KK states defined in
Eqs.~(\ref{scalarKK}) and (\ref{KKf}) are not exact mass eigenstates.
These KK-number violating transitions are of great phenomenological
interest and we will consider how they arise in greater detail in
later sections.

Here we simply note that quite generally we can represent the full
propagator either in the KK-number basis or the momentum basis, along
the lines discussed previously:
\beqa
G(p; z; z^{\prime}) &=&
\frac{1}{L^{2}} \, {\sum_{j,k}}' {\sum_{j',k'}}' \, g_{(j,k);(j',k')} \,
f_{n}^{(j,k)}(z) \left[f_{n}^{(j',k')}(z^{\prime})\right]^{*}
\nonumber \\
&=&
\frac{1}{4 L^{2}} \sum_{m,l} \sum_{m',l'} G_{n}^{(m,l; m',l')}
h^{(m,l)}(z)
\left[ h^{(m',l')}(z^{\prime}) \right]^{*}~,
\label{connection}
\eeqa
the only new ingredient being the possibility of non-diagonal KK
transitions.  In Appendix~\ref{App:MomSpace} we derive the general
relation between the expansion coefficients $G_{n}^{(m,l; m',l')}$ and
$g_{(j,k);(j',k')}$.  The specific form of the momentum space
coefficients $G_{n}^{(m,l; m',l')}$ encodes the appropriate boundary
conditions.

A special case of interest arises when the momentum quantum numbers
$m$ and $l$ take on positive values:
\beq
g_{(j,k);(j',k')} = G_{n}^{(j,k; j',k')}
\hspace{1cm} {\textrm{for}}~j,j' > 0~\textrm{and}~k,k' \geq 0~.
\label{momKK1}
\eeq
Also, when a zero mode is involved we get:
\beqa
g_{(j,k);(0,0)} &=& \frac{1}{2} G_{n}^{(j,k; 0,0)}
\hspace{1.2cm} \textrm{for}~ j > 0, k \geq 0 ~,
\nonumber \\
g_{(0,0);(j',k')} &=& \frac{1}{2} G_{n}^{(0,0; j',k')}
\hspace{1cm} \textrm{for}~ j' > 0, k' \geq 0 ~,
\label{momKK2} \\
g_{(0,0);(0,0)} &=& \frac{1}{4} G_{n}^{(0,0; 0,0)} ~.
\nonumber
\eeqa
Note that the factor of 4 relating $g_{(0,0);(0,0)}$ and $G_{n}^{(0,0;
0,0)}$ is similar to the one found in the diagonal case studied in
subsection \ref{Diagonal}.

Relations~(\ref{momKK1}) and (\ref{momKK2}) are useful to obtain the
KK-number expansion coefficients, which contain all physical
information, from the momentum space expansion coefficients, which are
more easily calculated in certain situations.

%%%%%%%%%%%%%%%%%%%%%%%%%%%%%%%%%%%%%
\subsection{KK-Number violating Structure due to Localized Operators}
\label{sec:localized}

As mentioned in the Introduction, the chiral square compactification
has conical singularities at the corners of the fundamental square
region, with coordinates $(x^{4},x^{5}) = (0,0)$, $(L,L)$ and $(0,L)$.
In this section we consider the effect of operators localized at these
special points.  In fact, the calculation of loops involving bulk
interactions reveals logarithmic divergences that require counterterms
localized precisely at these points.  In later sections we shall show
by explicit computation that the divergences at one-loop order have
precisely this property.  It will be useful to define the shorthand
notation
\beq
\delta_{c}(z) \equiv 
\delta(x^{4}) \delta(x^{5}) 
+ \delta(L-x^{4}) \delta(L-x^{5}) 
+ c \, \delta(x^{4}) \delta(L-x^{5})~, 
\label{loc_kkp}
\eeq
where $c$ is a dimensionless coupling that parametrizes the strength
of the operators at $(0,L)$ relative to those at $(0,0)$ and $(L,L)$.
We assume that operators localized at $(0,0)$ and $(L,L)$ appear with
identical coefficients, as required by KK-parity.

Let us consider the most general set of localized kinetic terms for a
complex scalar field $\Phi$
\beqa
& &
\frac{1}{4} \, \delta_{c_{1}}\!(z) \times r_{1} \, \partial_{\mu} \Phi^{\dagger} 
\partial^{\mu} \Phi + 
\left[ \frac{1}{4} \, \delta_{c_{2}}\!(z) \times 
r_{2} \, \Phi^{\dagger}(\partial_{+} 
\partial_{-} \Phi) + {\rm{h.c.}} \right]
\nonumber \\ [0.4em]
& & \mbox{} -
\frac{1}{4} \, \delta_{c_{3}}\!(z) \times r_{3}(\partial_{+} \Phi^{\dagger}) 
(\partial_{-} \Phi) -
\frac{1}{4} \, \delta_{c'_{3}}\!(z) \times r'_{3} (\partial_{-} \Phi^{\dagger}) 
(\partial_{+} \Phi)~,
\label{localOp}
\eeqa
where
\beq
\partial_{\pm} = \partial_{4} \pm i \partial_{5} ~.
\label{partialpm}
\eeq
The constants $r_{i}$, $c_{i}$ for $i=1,2,3$ and $r'_{3}$, $c'_{3}$
are arbitrary.  For convenience, we extracted a factor of $\left( 1/2
\right)^{2}$ to account for an enhancement due to the KK wavefunctions
in Eq.~(\ref{KKf}), evaluated at the singular points.  Note that the
$c_{i}$ are dimensionless, but the $r_{i}$ have dimensions of length
squared.  Notice also that the coefficients of the 4D-like kinetic term
need not be the same as for the kinetic operators with derivatives in
the compact directions, $\partial_{\pm}$.  However, we assumed a
rotational symmetry in the transverse $x^{4}$--$x^{5}$ plane that
forces the $\partial_{4}$ and $\partial_{5}$ derivatives to appear on
an equal footing.  This is natural given the rotational symmetry of
the conical singularities, and will be explicitly checked by the
one-loop computation in the following sections.

Using the scalar propagator representation given in
Eq.~(\ref{DiagGenProp}), the contribution of the first term in
Eq.~(\ref{localOp}) to the two-point function is
\beqa
\put(-25,-11.5){
\resizebox{3cm}{!}{\includegraphics{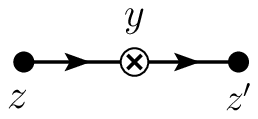}}
}
\hspace*{2cm}
&=& \int_{0}^{L} d^{2}y \, G(p; z', y) \left[ \frac{1}{4} \, \delta_{c_{1}}(y) \,
i r_{1} p^{2} \right] G(p; y, z)
\nonumber \\
&=& \left( \frac{1}{L^{2}} \right)^{2} {\sum_{j,k}}' {\sum_{j',k'}}' 
g_{S}^{j,k} f_{n}^{(j,k)}(z') 
\left[ i r_{1} p^{2} K_{c_{1}}^{(j, k)( j', k' )} \right]
g_{S}^{j',k'}
\left[ f_{n}^{(j',k')}(z) \right]^{*}~,
\label{InsertionMixing}
\eeqa
where the cross represents an insertion of the localized 4D-like
kinetic term in Eq.~(\ref{localOp}), and
\beqa
\label{localcouplings}
K_{c}^{(j, k)( j', k' )} &=& \frac{1}{4} \left\{
\left[ f_{n}^{(j,k)}(0,0) \right]^{*} f_{n}^{(j',k')}(0,0) +
\left[ f_{n}^{(j,k)}(L,L) \right]^{*} f_{n}^{(j',k')}(L,L) \right.
\nonumber \\ [0.4em]
& & \left. \mbox{} +
c \left[ f_{n}^{(j,k)}(0,L) \right]^{*} f_{n}^{(j',k')}(0,L)
\right\}~,
\eeqa
with the KK wavefunctions $f_{n}^{(j,k)}$ defined in Eq.~(\ref{KKf}).
Insertions of the kinetic terms involving derivatives in the extra
dimensions can be treated along the same lines with the help of the
relations
\beq
\partial_{\pm} f_{n}^{(j,k)}(z)
= i \, r_{j,\pm k} M_{j,k} f_{n \mp 1}^{(j,k)}(z) ~,
\label{delf}
\eeq
where the $r_{j,k}$ are complex phases defined by
\beq
r_{j,k} \equiv \frac{j + i k}{\sqrt{j^2+k^2}} ~.
\label{rjk}
\eeq

We see that Eq.~(\ref{InsertionMixing}) has the general KK-number
mixing structure of Eq.~(\ref{connection}).  In order to interpret the
result of the one-loop computations we present in the following
sections, it is useful to consider the momentum space representation
of the previous process.  Using Eq.~(\ref{momKK1}) we write the
contribution to the two-point function in momentum space,
$(p^{4},p^{5})=(m/R,l/R)$, when $m,l>0$, for the four types of
boundary conditions $n=0,1,2$ or $3$.

Consider first scalar fields satisfying $n=0$ boundary conditions,
i.e. having a zero-mode in its KK spectrum.  Assuming for simplicity
that $r_{2}$, $c_{2}$ are real, we get 
\beqa G_{n}^{(m,l;m',l')} &=&
\put(-10,-3){ \resizebox{3cm}{!}{\includegraphics{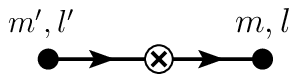}} }
\nonumber \\ [0.4em] 
&=& g_{S}^{m,l} \, \left\{ i \frac{r_{1}}{L^{2}}
p^{2} K_{c_{1}}^{(m,l)( m', l')} - i \frac{r_{2}}{L^{2}} \left(
M^{2}_{m,l} + M^{2}_{m',l'} \right) K_{c_{2}}^{(m,l)( m', l')}
\right\} \, g_{S}^{m',l'}~,
\label{KKviolatingStructure}
\eeqa
where, for these boundary conditions, Eqs.~(\ref{KKf}) and
(\ref{localcouplings}) give
\beqa
\label{Knzero}
K_{c}^{(m, l)( m', l' )} &=& 
\frac{
4 \left[ 1+ (-1)^{m + l + m' + l'} \right] + 
c [(-1)^{m} + (-1)^{l}] [(-1)^{m'} + (-1)^{l'}] }
{4 \left(1+ \delta_{m, 0} \delta_{l, 0} \right) 
\left(1+ \delta_{m', 0} \delta_{l', 0} \right)}~.
\eeqa
It is also easy to see that the operators proportional to $r_{3}$ and
$r'_{3}$ in Eq.~(\ref{localOp}) vanish in this case.

Note that, as a result of the operators at $(0,0)$ and $(L,L)$ being
equal, the induced KK transitions are non-vanishing only when $m+l$
and $m'+l'$ are both even or both odd, thus preserving precisely the
KK parity defined by $(-1)^{m + l}$.  Furthermore, it is useful to
notice that Eq.~(\ref{Knzero}) gives rise to precisely three different
non-vanishing cases, which we find convenient to order as follows
\beqa
\begin{tabular}{lcl}
  Case 1a: & $(-1)^{m+l} = (-1)^{m'+l'} = +1$, & $m-m'$ even, \\ [0.3em]
  Case 1b: & $(-1)^{m+l} = (-1)^{m'+l'} = +1$, & $m-m'$ odd, \\ [0.3em]
  Case 2: & $(-1)^{m+l} = (-1)^{m'+l'} = -1$. \\
\end{tabular}
\label{cases}
\eeqa
That is, localized operators distinguish between KK-number transitions
among KK-parity even states and KK-parity odd states.  Notice that the
distinction arises from operators localized at $(0,L)$.  Furthermore,
for KK-parity even transitions, there is a difference depending on
whether $m-m'$ is even or odd.

For $n=2$ boundary conditions, the momentum space two-point function
with insertions of the localized operators in Eq.~(\ref{localOp}) has
the structure of Eq.~(\ref{KKviolatingStructure}) with \beqa
K_{c}^{(m, l)( m', l' )} &=& \frac{ c [(-1)^{m} - (-1)^{l}] [(-1)^{m'}
- (-1)^{l'}] } {4 \left(1+ \delta_{m, 0} \delta_{l, 0} \right) \left(1+
\delta_{m', 0} \delta_{l', 0} \right)}~.
\label{Kntwo}
\eeqa
In this situation, we obtain three cases that we label as follows
\beqa
\begin{tabular}{lcl}
  Case 1: & $(-1)^{m+l} = (-1)^{m'+l'} = +1$, \\ [0.3em]
  Case 2a: & $(-1)^{m+l} = (-1)^{m'+l'} = -1$, & $m-m'$ even, \\ [0.3em]
  Case 2b: & $(-1)^{m+l} = (-1)^{m'+l'} = -1$, & $m-m'$ odd. \\ [0.3em]
\end{tabular}
\label{cases2}
\eeqa
Note that for $n=2$, it is the transitions between KK-parity odd states
that contain two subcases, 2a and 2b, while for $n=0$, the subcases
appeared when KK-parity even transitions are considered, 1a and 1b.

Finally, for $n=1$ or $n=3$ boundary conditions, the localized 4D-like
kinetic term does not contribute to the two-point function as a result
of the vanishing of the KK wavefunctions at the three conical
singularities.  Only the operators proportional to $r_{3}$ and
$r'_{3}$ in Eq.~(\ref{localOp}) give a nonvanishing contribution:
\beqa G_{n=1,3}^{(m,l;m',l')} &=& g_{S}^{m,l} \, \left\{ - i
\frac{r_{3}}{L^{2}} \, r_{m,l} r^{*}_{m',l'} M_{m,l} M_{m',l'}
K_{c_{3}}^{(m,l)( m', l')} \right.  \nonumber \\
& & \hspace{2cm} \left. \mbox{} -
i \frac{r'_{3}}{L^{2}} \, r^{*}_{m,l} r_{m',l'} M_{m,l} M_{m',l'}
K_{c'_{3}}^{(m,l)( m', l')}
\right\} \, g_{S}^{m',l'}~,
\label{KKviolatingStructure13}
\eeqa
To obtain this, we used Eq.~(\ref{delf}) as well as the fact that
$\left[f^{(j,k)}_{1}(z)\right]^{*} = - f^{(j,k)}_{3}(z)$, from
Eq.~(\ref{KKf}).  When $n=3$, $K_{c_{3}}^{(m,l)( m', l')}$ is as given
in Eq.~(\ref{Knzero}) and $K_{c'_{3}}^{(m,l)( m', l')}$ as in
Eq.~(\ref{Kntwo}), while for $n=1$, it is the other way around.
Notice that for $n=1$ or $3$ one generically gets four different
non-vanishing results, corresponding to cases 1a, 1b, 2a and 2b in
Eqs.~(\ref{cases}) and (\ref{cases2}).

By checking that loop contributions to the two point function have the
structure of Eqs.~(\ref{KKviolatingStructure}) and
(\ref{KKviolatingStructure13}) we shall be able to confirm in later
sections that they correspond to localized operators as in
Eq.~(\ref{localOp}).

%%%%%%%%%%%%%%%%%%%%%%%%%%%%%%%%%%%%%
\section{Chiral Fermions}
\label{fermioncase}

Having discussed various general properties in the scalar case, and
before we can tackle the calculation of one-loop radiative corrections
in this class of theories, we need to consider the form of the
propagator in momentum space for fields transforming non-trivially
under the 6D Lorentz group.  In this section, we treat the case of 6D
chiral fermions, $\Psi_{\pm}$, where $+$ or $-$ label the 6D
chirality, according to the projection operators $P_{\pm} =
\frac{1}{2} (1 \pm \overline{\Gamma})$, with $\overline{\Gamma}$ the
6D chirality operator.  In the following section we consider the
propagators associated with 6D gauge fields, which contain both spin-1
and spin-0 components under the unbroken 4D Lorentz group.

As shown in \cite{Dobrescu:2004zi}, fermions propagating in six
dimensions with the two extra dimensions compactified on the chiral
square have a chiral zero-mode.  Starting from the free fermion action
($\Gamma^{M}$ are $8 \times 8$ Dirac $\Gamma$-matrices, and
$M=0,1,\ldots,5$ runs over the 6D Lorentz indices)
\beq
S_\Psi = \int d^4 x \int_0^{\LL} dx^4 \int_0^{\LL} dx^5 \,
\frac{i}{2}
\left[\overline{\Psi} \, \Gamma^M \partial_M P_{\pm} \Psi 
- \left(\partial_M \overline{\Psi} \right) 
\Gamma^M P_{\pm }\Psi \right] ~,
\label{fermion-action}
\eeq 
and imposing the identification of adjacent sides of the square region
$0 < x^{4}, x^{5} < L$, one can show that the two 4-dimensional
chiralities, $\Psi_{\pm L}$ and $\Psi_{\pm R}$, contained in
$\Psi_{\pm} \equiv P_{\pm} \Psi$, satisfy boundary conditions
determined by integers $n^\pm_L$ and $n^\pm_R$ such that
\beq
n^\pm_R = n^\pm_L \mp 1 \; {\rm mod} \, 4~.
\label{nLnR}
\eeq
This shows that only a chiral zero-mode is allowed.  The propagator
associated with this system is obtained by inverting the operator
appearing in Eq.~(\ref{fermion-action}), taking care of imposing the
appropriate boundary conditions.  We leave the details to
Appendix~\ref{App:Propagators}.  The resulting propagator in the
momentum space representation takes the form
\beq
G_{p}^{\pm, (m,l; m',l')} = P_\pm \Gamma^{M} p_{M} \,
g_{S}^{m,l}
\left[ P_{R} \, \hat{\delta}(m,l; m',l';n^{\pm}_{L}) +
P_{L} \, \hat{\delta}(m,l; m',l';n^{\pm}_{R})\right]~,
\label{FullFermionPropagator}
\eeq
where $p_{M} = (p_{\mu}, p_{4}, p_{5})$ with $p_{4} = -m/R$ and $p_{5}
= -l/R$ [the minus signs arising from the Minkowski metric], and
$n^{\pm}_{L}$, $n^{\pm}_{R}$ are related by Eq.~(\ref{nLnR}).
$g_{S}^{m,l}$ is the scalar propagator given in Eq.~(\ref{GKKScalar}).
When using Eq.~(\ref{FullFermionPropagator}), one should be careful to
remember that the 4D chirality projectors, $P_{L,R}$, distinguish
between the $\Gamma^{\mu}$ and $\Gamma^{4,5}$ terms in $\Gamma^{M}
p_{M}$.  In particular, $P_{L,R}$ commute with $\Gamma^{4}$ and
$\Gamma^{5}$.

%%%%%%%%%%%%%%%%%%%%%%%%%%%%%%%%%%%%%
\section{Gauge fields}
\label{gaugecase}

We turn now our attention to the propagators associated with the 6D
gauge system.  We start from the action
\beq
S = \int\,d^4x \int_{0}^{L}dx^4 \int_{0}^{L} dx^5\,
\left(-\frac{1}{4}F_{MN}F^{MN} + {\cal L}_{GF}\right)~,
\label{s1}
\eeq
where the indices $M$, $N$ run over the six spacetime coordinates.
After compactification, the components of the 6D gauge field naturally
separate into $A_{\mu}$ ($\mu = 0,1,2,3$) and $A_{4}$, $A_{5}$.  While
the former are part of a spin-1 field, the latter constitute two
scalar degrees of freedom from a four-dimensional point of view.

A convenient choice of gauge arises by requiring that the mixings of
$A_\mu$ with $A_4$ and $A_5$ vanish:
\beq
{\cal L}_{GF}=-\frac{1}{2\xi}
\left[\frac{}{} \partial_\mu A^\mu - \xi\left(\partial_4A_4 +
\partial_5A_5 \right)\right]^2~,
\label{gf}
\eeq
where $\xi$ is a gauge fixing parameter.  This gauge clarifies how the
physical degrees of freedom are encoded in the 6D field, $A_{M}$: at
each massive KK level, one linear combination of the two scalars,
$A_{4}$ and $A_{5}$ is eaten by the massive $A_{\mu}$ fields, while
the orthogonal combination remains as an additional scalar degree of
freedom.  This system was studied in detail in \cite{Burdman:2005sr},
where the appropriate boundary conditions and resulting interactions
were worked out.  We assume that $A_{\mu}$ satisfies boundary
conditions corresponding to $n=0$, so that the spin-1 KK towers have a
zero-mode.  In other words, we do not consider the case where the
gauge symmetry is broken by the boundary conditions.  One also finds
that the linear combinations
\beq
A_{\pm} = A^{4} \pm i A^{5}
\label{pmdefns}
\eeq
satisfy boundary conditions give by $n=3$ for $A_{+}$ and $n=1$ for
$A_{-}$.  In particular, there are no zero-modes in the scalar sector.
We refer to these scalars as ``spinless adjoints''.

In the following subsections we derive the propagators for the spin-1
and spin-0 components, as well as the ghost fields associated with the
gauge fixing term Eq.~(\ref{gf}).

%%%%%%%%%%%%%%%%%%%%%%%%%%%%%%%%%%%%%
\subsection{The Spin-1 Components}

Apart from the boundary conditions, which are treated as for the
scalar case in section~\ref{scalarcase}, the derivation of the spin-1
propagator in momentum space is identical to the 4D derivation.  We
obtain
\beq
G_{\mu\nu, p}^{(m,l; m',l')} = g_{\mu\nu}^{m,l} \, \hat{\delta}(m,l;m',l';0)~.
\label{GGaugemom}
\eeq
where
\beq
g_{\mu\nu}^{m,l}  = - \left[ \eta_{\mu\nu} -
(1 - \xi) \frac{p_{\mu} p_{\nu}}{p^{2} - \xi M_{m,l}^{2}} \right]
g_{S}^{m,l}~,
\label{GKKGauge}
\eeq
and $g_{S}^{m,l}$ is given in Eq.~(\ref{GKKScalar}).
We recognize the 4-dimensional propagators appropriate to the gauge
fixing term (\ref{gf}) as those for a (massive) gauge field in an
$R_{\xi}$ gauge.

%%%%%%%%%%%%%%%%%%%%%%%%%%%%%%%%%%%%%
\subsection{The Spin-0 Components}

As shown in Appendix~\ref{App:Propagators}, the momentum space
propagator, defined as the inverse of the quadratic operator
associated with the $A_{4}$--$A_{5}$ system in the free Lagrangian,
can be more easily derived in the $A_{\pm}$ basis, defined by
Eq.~(\ref{pmdefns}), where the boundary conditions are well defined.
The result is a $2\times 2$ matrix with components
\beqa
\hspace{-5mm}
\left(  
\begin{array}{cc}
G_{p,++}^{(m,l; m',l')} & G_{p,+-}^{(m,l; m',l')} \\ [0.4em]
G_{p,-+}^{(m,l; m',l')} & G_{p,--}^{(m,l; m',l')}
\end{array}
\right) &=&
\nonumber \\ [0.4em]
& & \hspace{-4.5cm}
\left(
\begin{array}{cc}  
\left( g^{m,l}_{h} + g^{m,l}_{\phi} \right)
\hat{\delta}(m,l; m',l';3) &
- r_{m,l}^{2} \, \left( g^{m,l}_{h} - g^{m,l}_{\phi}
\right)
\, \hat{\delta}(m,l; m',l';1)
\\ [0.4em]
- r_{m,l}^{*2} \, \left( g^{m,l}_{h} - g^{m,l}_{\phi}
\right)  \,
\hat{\delta}(m,l; m',l';3)
&
\left( g^{m,l}_{h} + g^{m,l}_{\phi} \right)
\hat{\delta}(m,l; m',l';1)
\end{array}
\right)~,
\label{GA4A5mom}
\eeqa
where
\beq
g^{m,l}_{h} = \frac{i}{p^{2} - M^{2}_{m,l}}~, \hspace{1cm}
g^{m,l}_{\phi} = \frac{i}{p^{2} - \xi M^{2}_{m,l}}~,
\label{ghphi}
\eeq
and the complex phases $r_{m,l}$ were defined in Eq.~(\ref{rjk}). 
The pole structure shown in Eq.~(\ref{ghphi}) reveals that the
$A_{4}$--$A_{5}$ system has two degrees of freedom at each KK level,
one with mass $M_{m,l}$ and the second with mass $\sqrt{\xi}M_{m,l}$.
The scalar mode with the $\xi$-dependent mass corresponds to the
longitudinal degree of freedom of the massive spin-1 gauge fields, as
required by the higher dimensional Higgs mechanism.

Taking into account the fact that $A_{-}^{\dagger} = A_{+}$, one can
show that the relation between the various components $G_{++}$,
$G_{+-}$, $G_{-+}$ and $G_{--}$ in Eq.~(\ref{GA4A5mom}), and the
tree-level two-point functions that enter the Feynman rules is given
by
\beqa
\langle A_{+}^{m,l} A_{+}^{m',l' \dagger} \rangle
&=& \frac{1}{2} \left[ G_{p,++}^{(m,l; m',l')} + G_{p,--}^{(-m',-l'; -m,-l)} 
\right]
\,\, = \,\, G_{p,++}^{(m,l; m',l')}~,
\nonumber \\
\langle A_{+}^{m,l} A_{+}^{m',l'} \rangle
&=& \frac{1}{2} \left[ G_{p,+-}^{(m,l; m',l')} + G_{p,+-}^{(-m',-l'; -m,-l)} 
\right]
\,\, = \,\, G_{p,+-}^{(m,l; m',l')}~,
\label{ApAm}
\eeqa
together with their complex conjugates, were we chose to express all
correlators in terms of $A_{+}$ and $A_{+}^{\dagger}$.  The second
equalities follow from the explicit solution for $G^{(m,l;m',l')}$
given in Eq.~(\ref{GA4A5mom}).

%%%%%%%%%%%%%%%%%%%%%%%%%%%%%%%%%%%%%
\subsection{Faddeev-Popov Ghosts}

The ghost Lagrangian associated with the gauge fixing term, Eq.~(\ref{gf}),
is
\beq
-\bar{c}^a \left[ \partial_{\mu} D^{\mu} - \xi \left(\partial_{4} D_4 + 
\partial_{5} D_5 \right) \right] c^a ~,
\label{ghostaction}
\eeq
where the ghost fields, $c^{a}$, satisfy the same boundary conditions
as $A_{\mu}$, i.e. given by $n=0$.  By comparing to the derivation of
the scalar propagator, Eq.~(\ref{Gml}), it is easy to see that the
ghost propagator in momentum space is given by
\beq
G_{\xi, p}^{(m,l; m',l')} = \frac{i}{p^{2} - \xi M_{m,l}^{2}}
\, \hat{\delta}(m,l;m',l';0)~,
\label{GGhostmom}
\eeq
i.e., it has a $\xi$-dependent mass given by $\sqrt{\xi} M_{m,l}$.

%%%%%%%%%%%%%%%%%%%%%%%%%%%%%%%%%%%%%
\section{Radiative Corrections}
\label{radiative}

Having at our disposal the propagators for 6D scalars, Weyl fermions
and gauge fields that correctly encode the boundary conditions
appropriate in the chiral square background, we are in a position to
consider the one-loop structure of the theory.  In this section we
compute the quantum corrections to the two-point functions (sometimes
we will refer to these as self-energies, even though they also include
mixing among KK states) for gauge, fermion and scalar fields and show
that, besides a renormalization of the bulk kinetic terms, the two
point correlation functions contain logarithmic divergences
corresponding to counterterms localized at the conical singularities
with coordinates $(x^{4},x^{5}) = (0,0)$, $(L,L)$ and $(0,L)$.

%%%%%%%%%%%%%%%%%%%%%%%%%%%%%%%%%%%%%
\subsection{Gauge Boson Two-Point Function}
\label{sec:Gauge_self}

We first compute the one loop contributions to the gauge boson
two-point function.  We consider in turn scalar matter, fermionic
matter and the gauge self-interactions themselves.  We present the
scalar case in considerable detail to show how the KK-number violating
structure corresponding to localized operators arises.  The lessons
thus learned carry straightforwardly to the remaining types of
one-loop graphs.

%%%%%%%%%%%%%%%%%%%%%%%%%%%%%%%%%%%%%
\subsubsection{Scalar Matter}
\label{sec:scalarloop}

We start by evaluating the one-loop diagrams arising from scalars
minimally coupled to gauge bosons.  We consider 6D scalar fields
satisfying any of the possible boundary conditions, labeled by
$n=0,1,2$ or $3$.  We evaluate these diagrams in detail to show how
the operators localized at the three conical singularities, with
coordinates $(x^{4},x^{5}) = (0,0)$, $(0,L)$ and $(L,L)$, arise.  We
need only consider the divergent pieces.  For fixed KK numbers of the
external lines, $(j,k)$ and $(j',k')$, these split into two different
categories:
\begin{itemize}
\item
Terms with a KK-number structure identical to the tree-level
propagator, Eq.~(\ref{GGaugemom}), i.e. proportional to
$\hat{\delta}(j,k;j',k';0)$.  After KK summation, the result grows as
a power of the number of KK modes.  This power-law divergence
renormalizes 6D bulk operators, and is of no interest to us here.
\item 
Terms with a KK-number structure different from the tree-level
propagator (i.e. KK-number violating).  At one loop, only a finite
number of KK states contribute, and the divergences are only
logarithmic, as in a 4D computation.  We would like to see that the
KK-number violating structure that arises here corresponds precisely
to the one induced by operators localized at $(0,0)$, $(0,L)$ and
$(L,L)$.
\end{itemize}

\FIGURE[t]{
\vspace*{-5mm}
\centerline{
   \resizebox{13cm}{!}{\includegraphics{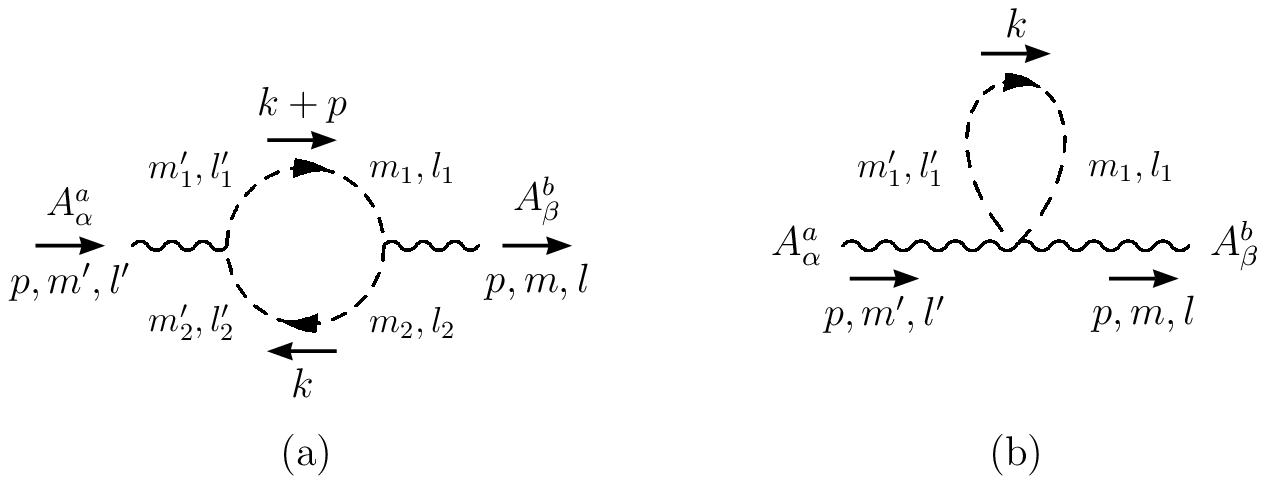}}
}
\caption{One-loop contributions to the gauge boson self energy due to
a minimally coupled scalar.  The external four-dimensional momentum is
denoted by $p$.  The momenta along the extra dimensions are simply
denoted by the corresponding integer quantum number according to
$p_{4} = m/R$, $p_{5} = l/R$, etc.  Since the momentum in the compact
dimensions is not conserved, each line is labeled by two sets of
momenta.}
\label{fig:momenta}
}
\FIGURE[b]{
\vspace*{-5mm}
\centerline{
   \resizebox{14.5cm}{!}{\includegraphics{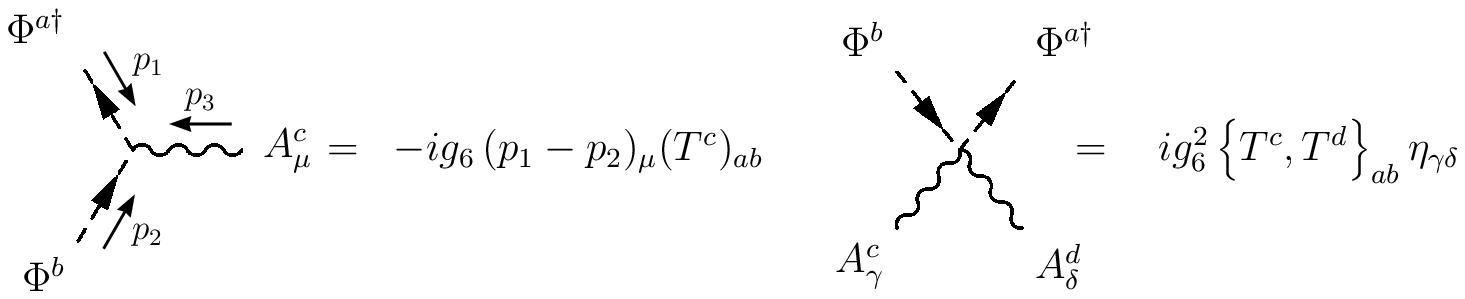}}
}
\caption{Momentum space Feynman rules for scalars minimally coupled to
gauge bosons in the representation $T$.}
\label{fig:Feynman3}
}
In order to see the above features it is useful to concentrate on the
two-point function, as opposed to the amputated diagrams.  Let us
start with the diagram involving the interaction between two scalars
and two gauge bosons, as in Figure~\ref{fig:momenta} $(b)$.  In
momentum space, we have
\beq
\langle A^{(j,k)}_{\mu} A^{(j',k')}_{\nu} \rangle^{(b)} = 
\left(\frac{1}{4L^{2}}\right)^{2}
\left(\frac{1}{4} \right)
\sum_{m,l} \sum_{m',l'} G_{\mu,p}^{\alpha(j,k;m,l)}
\left[ \langle A^{(m,l)}_{\alpha} A^{(m',l')}_{\beta} 
\rangle_{\rm{amp}}^{(b)} \right]
G^{\beta(m',l';j',k')}_{\nu, p}~,
\label{Two-point-quartic}
\eeq
where each propagator carries a factor of $1/(4L^{2})$, as in
Eq.~(\ref{connection}), and the additional factor of $1/4$ arises from
the integration over interaction points in spacetime, $\int_{0}^{L}
d^{2}y = (1/4) \int_{-L}^{L} d^{2}y$.  Using the Feynman rules shown
in Figure~\ref{fig:Feynman3}, the amputated function in
Figure~\ref{fig:momenta} $(b)$ is
\beqa
\langle A^{(m,l)}_{\alpha} A^{(m',l')}_{\beta} \rangle_{\rm{amp}}^{(b)} 
&=& 2 i g_{6}^{2} \, {\rm{Tr}}(T^{a} T^{b}) \, \eta_{\alpha\beta}
\sum_{m_{1},l_{1}}
\int \frac{d^{D}k}{(2\pi)^{D}} G_{k,n}^{(m_{1},l_{1}; m'_{1},l'_{1})}~,
\label{scalarmatterquartic}
\eeqa
where the scalar propagator, $G_{k,n}^{(m_{1},l_{1}; m'_{1},l'_{1})}$,
was defined in Eq.~(\ref{Gml}).  It is understood that, due to
momentum conservation, $m'_{1} = m_{1} + m' - m$ and $l'_{1} = l_{1} +
l' - l$.  Notice that the factor of $1/(4L^{2})$ associated with the
internal propagator disappears after integrating $\int_{-L}^{L}
d^{2}y$, according to the orthonormality condition,
Eq.~(\ref{orthoh}).

Consider the KK sums in the two point function
(\ref{Two-point-quartic}).  The first term in
$\hat{\delta}(m_{1},l_{1};m'_{1},l'_{1};n)$ [see Eq.~(\ref{delta})]
leads to
\beq
\sum_{m,l} \sum_{m',l'} \hat{\delta}(j,k;m,l;0) 
\left[ \delta_{m_{1},m'_{1}} \delta_{m_{2},m'_{2}} \right] 
\hat{\delta}(m',l';j',k';0) = 4 \, \hat{\delta}(j,k;j',k';0)~,
\label{KKnumberconserving}
\eeq
which has the KK-number structure of the tree-level propagator,
arising from the bulk kinetic terms.  Since $G_{k,n}^{(m_{1},l_{1};
m'_{1},l'_{1})} = i(k^{2} - M_{m_{1},l_{1}}^{2})^{-1} \,
\hat{\delta}(m_{1},l_{1};m'_{1},l'_{1};n)$, the 4D momentum integral
in Eq.~(\ref{scalarmatterquartic}) is quadratically divergent.  In
dimensional regularization, the divergent part is proportional to
$M_{m_{1},l_{1}}^{2} \sim m^{2}_{1} + l^{2}_{1}$, and it is clear that
the contribution from Eq.~(\ref{KKnumberconserving}) diverges like
$\sum_{m_{1},l_{1}} (m^{2}_{1} + l^{2}_{1})$.  This power-law
divergence corresponds to a renormalization of the bulk gauge kinetic
term.

The contribution due to the three remaining terms in
$\hat{\delta}(m_{1},l_{1};m'_{1},l'_{1};n)$ lead to a KK-number
violating structure corresponding to localized operators at the three
corners of the chiral square, as follows.  These terms fix the loop
momenta $m_{1}$, $l_{1}$ in terms of $m-m'$ and $l-l'$ according to
\beq
\begin{array}{lcll}
    \delta_{m_{1},l'_{1}} \delta_{l_{1},-m'_{1}} & \rightarrow & 
    m_{1} = \frac{1}{2} \left[ (m-m') - (l-l') \right], & 
    l_{1} = \frac{1}{2} \left[ (m-m') + (l-l') \right], \\ [0.5em]
    \delta_{m_{1},-m'_{1}} \delta_{l_{1},-l'_{1}} & \rightarrow & 
    m_{1} = \frac{1}{2} (m-m'), & 
    l_{1} = \frac{1}{2} (l-l'), \\ [0.5em]
    \delta_{m_{1},-l'_{1}} \delta_{l_{1},m'_{1}} & \rightarrow & 
    m_{1} = \frac{1}{2} \left[ (m-m') + (l-l') \right], & 
    l_{1} = \frac{1}{2} \left[ (m-m') - (l-l') \right]~. 
\end{array}
\label{allowedKKnumbers}
\eeq
Clearly, for given $(m,l)$ and $(m',l')$ only a finite number of
states contribute to the KK sum.  Also, since $m_{1}$ and $l_{1}$ must
be integers, only when $(-1)^{m+l+m'+l'} = +1$ can the result be
nonvanishing, in accordance with the KK-parity assignment $(-1)^{m+l}$
for states with quantum numbers $(m,l)$.  Like for the KK-number
conserving terms, the (logarithmically) divergent contribution for
each nonvanishing term in the sum is proportional to
$m_{1}^{2}+l_{1}^{2}$ (in dimensional regularization).  To obtain the
correct KK-number violating structure, it is necessary to contract
with the external propagators, as in Eq.~(\ref{Two-point-quartic}).
In this process, all the crossed terms, such as $mm'$ or $ll'$ cancel
out and only terms proportional to $M_{m,l}^{2}+M_{m',l'}^{2}$
survive.

%%%%%%%%%%%%%%%%%%%%%%%%%%%%%%%%%%%%%%
%%% Table for scalar functions A and B in the matter sector
%%%%%%%%%%%%%%%%%%%%%%%%%%%%%%%%%%%%%%
\TABLE[t]{
%\begin{center}
\begin{tabular}{|c||rccc|rccc|}
  \hline
   &  &  & $A_{S}$ & & & & $B_{S}$ &
\rule{0mm}{5mm} \\ [0.4em]
\hline
\hline
& & $n=0$ & $n=1,3$ & $n=2$ & & $n=0$ & $n=1,3$ & $n=2$
\rule{0mm}{5mm} \\ [0.4em]
\hline
  (a) & $\frac{1}{3} \times \left\{%
\begin{array}{ll}
     \\
     \\
     \\
\end{array}%
\right.$ \hspace*{-8mm}
&
$\begin{array}{c}
    3 \\
    2 \\
    5/2 \\
\end{array}$%
&
$\begin{array}{c}
    -1 \\
    0 \\
    -1/2 \\
\end{array}$%
&
$\begin{array}{c}
    -1\\
    -2 \\
    -3/2 \\
\end{array}$%
& $(M_{m,l}^2+M_{m',l'}^2) \times \left\{%
\begin{array}{c}
    \\
    \\
    \\
\end{array}%
\right.$ \hspace*{-8mm}
&
$\begin{array}{c}
    5/2 \\
    2 \\
    9/4 \\
\end{array}$%
&
$\begin{array}{c}
    -1/2 \\
    0 \\
    -1/4 \\
\end{array}$%
&
$\begin{array}{c}
    -3/2 \\
    -1 \\
    -7/4 \\
\end{array}$%
\rule{0mm}{5mm} \\ [0.4em]
\hline
  (b) & & 0 & 0 & 0 & $ - (M_{m,l}^2+M_{m',l'}^2) \times \left\{%
\begin{array}{c}
    \\
    \\
    \\
\end{array}%
\right.$ \hspace*{-8mm}
&
$\begin{array}{c}
    5/2 \\
    2 \\
    9/4 \\
\end{array}$%
&
$\begin{array}{c}
    -1/2 \\
    0 \\
    -1/4 \\
\end{array}$%
&
$\begin{array}{c}
    -3/2 \\
    -1 \\
    -7/4 \\
\end{array}$%
\rule{0mm}{5mm} \\ [0.4em]
\hline
\end{tabular}
%\end{center}
\caption{Scalar functions $A_{S}$ and $B_{S}$ for scalar matter loops,
as defined via Eq.~(\ref{self-matter}), corresponding to the diagrams
$(a)$ and $(b)$ of Fig.~\ref{fig:momenta}.  These are computed in
dimensional regularization.  We give the results for scalars
satisfying the four types of boundary conditions, labeled by
$n=0,1,2,3$, allowed on the chiral square compactification of
Ref.~\cite{Dobrescu:2004zi}.  For a given $n$ and for each diagram,
there are three possible cases depending on KK-parity and whether
$m-m'$ is even or odd, as listed in Eq.~(\ref{cases}): the first two
cases correspond to even-even mixings with $m-m'$ even and $m-m'$ odd,
and the third case to odd-odd mixings.  All KK-parity violating
transitions vanish.  In all cases the gauge violating coefficient
$B_{S}$ vanishes when both diagrams are added.}
\label{tableABmatter}
}%

A straightforward computation allows us to write the KK-number violating
contribution to the two-point function in momentum space as
\beq
\left( \frac{1}{4L^{2}} \right) G_{\mu,p}^{\alpha(j,k)}
\left\{
- i \frac{g_{4}^2}{16\pi^2} T(\Phi) \delta_{ab} 
\,\Gamma\!\left(\frac{\epsilon}{2}\right) \left[A_{S} (p^2 \eta_{\alpha\beta} - 
p_{\alpha}p_{\beta}) - B_{S} \eta_{\alpha\beta} \right]
\right\}
G^{\beta(j',k')}_{\nu, p}~,
\label{self-matter}
\eeq
where $\epsilon = 4-D$ and ${\rm{Tr}}(T^{a} T^{b}) = T(\Phi)
\delta_{ab}$, with $T(\Phi) = 1/2$ for matter in the fundamental
representation of $SU(N)$.  The scalar functions $A_{S}$, $B_{S}$ for
the four types of boundary conditions are given in
Table~\ref{tableABmatter}, and the explicit factor of $1/(4L^{2})$ in
Eq.~(\ref{self-matter}) should be identified with the one appearing in
Eq.~(\ref{connection}), while the remaining factor of $1/L^{2}$ was
absorbed in the 4D gauge coupling, $g_{4}^{2} = g_{6}^{2}/L^{2}$.

We see that for each type of boundary condition ($n=0,1,2$ or $3$)
obeyed by the scalar running in the loop, there are three different
cases, that we ordered as in Eq.~(\ref{cases}).  This is precisely the
KK-number violating structure that arises from quadratic operators
involving fields satisfying $n=0$ boundary conditions (the gauge
field), localized at the points $(0,0)$, $(L,L)$ and $(0,L)$, as
discussed in Section~\ref{sec:localized}.  Notice that the above three
cases are distinct precisely as a result of the operator localized at
$(0,L)$, proportional to $c$ in Eq.~(\ref{loc_kkp}).

For the diagram involving the interactions between two scalars and a
single gauge boson, the two-point function in momentum space is
\beq
\langle A^{(j,k)}_{\mu} A^{(j',k')}_{\nu} \rangle^{(a)} = 
\left(\frac{1}{4L^{2}}\right)^{2}
\left(\frac{1}{4} \right)^{2}
\sum_{m,l} \sum_{m',l'} G_{\mu,p}^{\alpha(j,k;m,l)}
\left[ \langle A^{(m,l)}_{\alpha} A^{(m',l')}_{\beta} 
\rangle_{\rm{amp}}^{(a)} \right]
G^{\beta(m',l';j',k')}_{\nu, p}~,
\label{Two-point-trilinear}
\eeq
where now we get a factor $(1/4)^{2}$ due to the two vertices, and the
amputated function of Figure~\ref{fig:momenta}~$(a)$ is
\beqa
\langle A^{(m,l)}_{\alpha} A^{(m',l')}_{\beta} \rangle_{\rm{amp}}^{(a)} 
&=& - g_{6}^{2} \, i^{2} \, {\rm{Tr}}(T^{a} T^{b})
\sum_{m_{1},l_{1}} \sum_{m'_{1},l'_{1}}
\int \frac{d^{D}k}{(2\pi)^{D}} (2k + p)_{\alpha} (2k + p)_{\beta}
\nonumber \\ %[0.3em]
& & \hspace{5.5cm} \mbox{} \times
G_{k+p,n}^{(m_{1},l_{1}; m'_{1},l'_{1})} G_{k,n}^{(m'_{2},l'_{2}; m_{2},l_{2})}~,
\label{scalarmattertrilinear}
\eeqa
where it is understood that, due to momentum conservation at each
vertex, $m_{2} = m_{1} - m$, $l_{2} = l_{1} - l$, $m'_{2} = m'_{1} -
m'$ and $l'_{2} = l'_{1} - l'$.  Noting that $M_{m'_{2},l'_{2}}^{2}$
depends on $m'_{2}$, $l'_{2}$ only through $m^{\prime 2}_{2} +
l^{\prime 2}_{2}$, and is therefore invariant under exchange of
$m'_{2}$ and $l'_{2}$ and/or a change in their signs, we have
\beqa
G_{k, n}^{(m'_{2},l'_{2}; m_{2},l_{2})}
= \frac{i}{k^{2} - M_{m'_{2},l'_{2}}^{2}} \, 
\hat{\delta}(m'_{2}, l'_{2}; m', l'; n)
= \frac{i}{k^{2} - M_{m',l'}^{2}} \, 
\hat{\delta}(m'_{2}, l'_{2}; m', l'; n)~,
\eeqa
which shows that the integrand in Eq.~(\ref{scalarmattertrilinear})
depends on $m'_{1}$, $l'_{1}$ only through the generalized
$\delta$-functions defined in Eq.~(\ref{delta}).  We can then do the
sum over $m'_{1}$, $l'_{1}$ using the identity [see
Eq.~(\ref{two-deltas1}) with $n_{1} = n_{2} = n$]
\beqa
&& \hspace{-1.5cm}
\sum_{m'_{1}, l'_{1}} \hat{\delta}(m_{1} ,l_{1}; m'_{1}, l'_{1}; n)
\hat{\delta}(m'_{2}, l'_{2}; m_{2}, l_{2}; n)
\hspace*{6cm}
\nonumber \\ [-0.4em]
&=&
\hat{\delta}(m, l; m',l'; 0) +
e^{i\theta}
\hat{\delta}(m - m_{1} - l_{1}, l - l_{1} + m_{1}; m', l'; 0)
\label{two-deltas}
\\ [.4em]
&& \mbox{} +
e^{2i\theta}
\hat{\delta}(m-2m_{1}, l - 2l_{1}; m', l'; 0) +
e^{3i\theta}
\hat{\delta}(m - m_{1} + l_{1}, l - l_{1} - m_{1}; m', l'; 0)~,
\nonumber
\eeqa
where $\theta = n \pi/2$.  We see that the first term has the same
structure as the tree-level propagator.  It is also independent of the
loop momenta, $m_{1}$, $l_{1}$, so that the corresponding KK sum leads
to a power-law divergence.  The remaining terms give KK violating
transitions that, as we will see, correspond to localized operators.
In order to compare to diagram $(b)$ as computed before, we should
contract with the external propagators and perform the KK sums.  The
KK number preserving term just gives
\beqa
\sum_{m, l} \sum_{m', l'} \hat{\delta}(j,k;m,l;0) 
\hat{\delta}(m, l; m',l'; 0)
\hat{\delta}(m',l';j',k'; 0) &=& 16 \, \hat{\delta}(j,k;j',k';0)~.
\rule{0mm}{5mm}
\nonumber
\eeqa
The contraction of the external propagators with the last three,
KK-number violating terms in Eq.~(\ref{two-deltas}), do not alter the
KK violating structure, and simply give an overall factor of 16.
Thus, unlike for diagrams with the topology of diagram $(b)$, for
diagrams with the topology of diagram $(a)$ it is sufficient to
compute the amputated diagram, while keeping in mind that the
contraction with external propagators produces a factor of 16 that
cancels against some of the factors in the two-point function.  The
net effect is that for these diagrams one should include a factor of
$1/(4L^{2})$ where the length scale $L$ is absorbed by the 4D gauge
coupling via $g_{4}^{2} = g_{6}^{2}/L^{2}$.  A straightforward
computation allows us to express the KK violating contributions to
diagram $(a)$ as in Eq.~(\ref{self-matter}), with scalar coefficients
$A_{S}$ and $B_{S}$ as given in Table~\ref{tableABmatter}.

We see that after adding the two diagrams $(a)$ and $(b)$, the ``gauge
violating'' contribution proportional to $\eta_{\alpha\beta}$ cancels
out.  Therefore, at one-loop order, the scalar gauge interactions give a
contribution to the two point function which is equivalent to the
effect of the localized operator
\beqa
\frac{1}{4} \, \delta_{c_{S}}\!(z) \times \left(-\frac{1}{4} \, \hat{r}_{S} L^{2} 
F_{\mu\nu}^{a} F^{\mu\nu a}\right)~,
\label{local45_2}
\eeqa
where $\delta_{c}(z)$ stands for the Dirac delta-functions at the
conical singularities, as defined in Eq.~(\ref{loc_kkp}), and the
factor of $1/4$ accounts for universal KK wavefunction enhancements.
Here we wrote the dimensionful coefficient in Eq.~(\ref{loc_kkp}) as
$r_{S} = \hat{r}_{S} L^2$, so that the scalar contribution is
\beq
\hat{r}_{S} = \frac{2}{3} \times \frac{g_{4}^2}{16\pi^2} T(\Phi)
\,\Gamma\!\left(\frac{\epsilon}{2}\right) \times
\left\{
\begin{array}{r}
    5/8 \\ [0.4em]
    -1/8 \\ [0.4em]
    -3/8 \\
\end{array}
\right.~,
\hspace{5mm}
c_{S} = 
\left\{
\begin{array}{ccl}
    2/5 & \hspace{5mm} & \rm{for}~n=0 \\ [0.4em]
    2 & & \rm{for}~n=1,3 \\ [0.4em]
    -2/3 & & \rm{for}~n=2 \\
\end{array}
\right.~.
\label{Arc_scalar}
\eeq
This shows explicitly that the logarithmic divergences that appear in
the chiral square compactification renormalize operators precisely at
the three conical singularities.

%%%%%%%%%%%%%%%%%%%%%%%%%%%%%%%%%%%%%
\subsubsection{Fermionic Matter}

We now turn to fermionic matter.  We restrict ourselves to 6D fermions
containing a zero-mode (i.e. we do not consider $n=2$ boundary
conditions).

We use the same labeling conventions as in Figure~\ref{fig:momenta}, 
as we will do throughout this paper. 
As discussed in detail in the previous section, we may concentrate on
the amputated fermion loop diagram, including a factor of $1/4$:
\beqa
- \left( \frac{1}{4} \right) g_{4}^{2} \, i^{2} \, {\rm{Tr}}(T^{a} T^{b})
\sum_{m_{1},l_{1}} \sum_{m'_{1},l'_{1}}
\int \frac{d^{D}k}{(2\pi)^{D}}
{\rm{Tr}}\left\{ \Gamma_{\beta } G_{k+p}^{\pm, (m_{1},l_{1}; m'_{1},l'_{1})} 
\Gamma_{\alpha } G_{k}^{\pm, (m'_{2},l'_{2}; m_{2},l_{2})} P_{\mp} \right\}~,
\label{fermionloop}
\eeqa
where the fermion propagator was given in
Eq.~(\ref{FullFermionPropagator}).  Also, as explained in detail in
the scalar loop calculation described before, we separate the result
into a KK-number preserving contribution that renormalizes bulk
operators and a KK-number violating contribution that renormalize
localized operators.

The trace in Eq.~(\ref{fermionloop}) can be separated into parts
involving the non-compact dimensions
\beqa
(k+p)^{\lambda} k^{\rho} {\rm{Tr}}\left\{ \Gamma_{\beta } \Gamma_{\lambda} 
\Gamma_{\alpha } \Gamma_{\rho} P_{L,R} P_{\mp} \right\} &=& 
2 \left[ (k+p)_{\alpha} k_{\beta} + (k+p)_{\beta} k_{\alpha} - 
k \cdot (k+p) \eta_{\alpha\beta} \right]~,
\label{trace1}
\eeqa
and parts involving the compact dimensions (denoted by indices $i$,
$j$)
\beqa
(k+p)_{i} k_{j} {\rm{Tr}}\left\{ \Gamma_{\beta } \Gamma^{i} 
\Gamma_{\alpha } \Gamma^{j} P_{L} P_{\mp} \right\} &=& 
r_{m_{1},l_{1}} r^{*}_{m'_{2},l'_{2}} M_{m_{1},l_{1}} 
M_{m'_{2},l'_{2}} {\rm{Tr}}\left\{ \Gamma_{\beta } 
\Gamma^{4} \Gamma_{\alpha } \Gamma^{4} P_{L} P_{\mp} \right\} 
\nonumber \\ [0.4em]
&=& 2 \eta_{\alpha\beta} r_{m_{1},l_{1}} r^{*}_{m'_{2},l'_{2}} 
M_{m_{1},l_{1}} M_{m'_{2},l'_{2}}~,
\label{trace2}
\\ [0.4em]
(k+p)_{i} k_{j} {\rm{Tr}}\left\{ \Gamma_{\beta } \Gamma^{i} 
\Gamma_{\alpha } \Gamma^{j} P_{R} P_{\mp} \right\} &=& 
r^{*}_{m_{1},l_{1}} r_{m'_{2},l'_{2}} M_{m_{1},l_{1}} 
M_{m'_{2},l'_{2}} {\rm{Tr}}\left\{ \Gamma_{\beta } 
\Gamma^{4} \Gamma_{\alpha } \Gamma^{4} P_{R} P_{\mp} \right\} 
\nonumber \\ [0.4em]
&=& 2 \eta_{\alpha\beta} r^{*}_{m_{1},l_{1}} r_{m'_{2},l'_{2}} 
M_{m_{1},l_{1}} M_{m'_{2},l'_{2}}~,
\label{trace3}
\eeqa
where we used the useful identities \cite{Dobrescu:2004zi}
\beqa
\label{Pidentities}
\Gamma^5 P_L P_\pm  = \pm i \Gamma^4 P_L P_\pm
 ~,
\nonumber \\ [.3em]
\Gamma^5 P_R P_\pm  = \mp i \Gamma^4 P_R P_\pm  ~,
\eeqa
to replace $\Gamma^{5}$ in favor of $\Gamma^{4}$.  We also used
$k_{4}+p_{4} = -m_{1}/R$, $k_{5} +p_{5} = -l_{1}/R$, $k_{4} =
-m'_{2}/R$ and $k_{5} = -l'_{2}/R$.  The momentum dependent phases,
$r_{j,k}$, were defined in Eq.~(\ref{rjk}).  Notice that the 6D
chirality operator $\overline{\Gamma}$ contained in $P_{\pm}$ gives a
nonvanishing contribution to Eqs.~(\ref{trace2}) and (\ref{trace3}).

We express the result associated with the KK-number violating terms as
\beq
\left( \frac{1}{4L^{2}} \right) G_{\mu,p}^{\alpha(j,k)}
\left\{
- i \frac{g_{4}^2}{16\pi^2} T(\Psi) \delta_{ab} 
\,\Gamma\!\left(\frac{\epsilon}{2}\right) \left[A_{F} (p^2 \eta_{\alpha\beta} - 
p_{\alpha}p_{\beta}) - B_{F} \eta_{\alpha\beta} \right]
\right\}
G^{\beta(j',k')}_{\nu, p}~,
\eeq
and find that for all KK-parity preserving transitions $A_{F} = 4/3$
and $B_{F} = 0$.  As for the scalar loops, the ``gauge violating'' term
proportional to $\eta_{\alpha\beta}$ vanishes.  The equivalent
localized operator is in this case
\beqa
\frac{1}{4} \, \delta_{c_{F}}\!(z) \times \left(-\frac{1}{4} \, \hat{r}_{F} L^{2} 
F_{\mu\nu}^{a} F^{\mu\nu a}\right)~,
\label{local45_3}
\eeqa
where $\delta_{c}(z)$ was defined in Eq.~(\ref{loc_kkp}), the factor
of $1/4$ accounts for universal KK wavefunction enhancements, and
\beq \hat{r}_{F} = \frac{2}{3} \times \frac{g_{4}^2}{16\pi^2} T(\Psi) 
\,\Gamma\!\left(\frac{\epsilon}{2}\right)~,
\hspace{1cm}
c_{F} = 0~.  
\label{Arc_fermion}
\eeq 
Note that the fermion loop does not induce an operator localized at
$(0,L)$, i.e. the three cases defined in (\ref{cases}) give the same
result.

%%%%%%%%%%%%%%%%%%%%%%%%%%%%%%%%%%%%%
\subsubsection{Gauge Self-Interactions}

\FIGURE[t]{
\vspace*{-5mm}
\centerline{
   \resizebox{13cm}{!}{\includegraphics{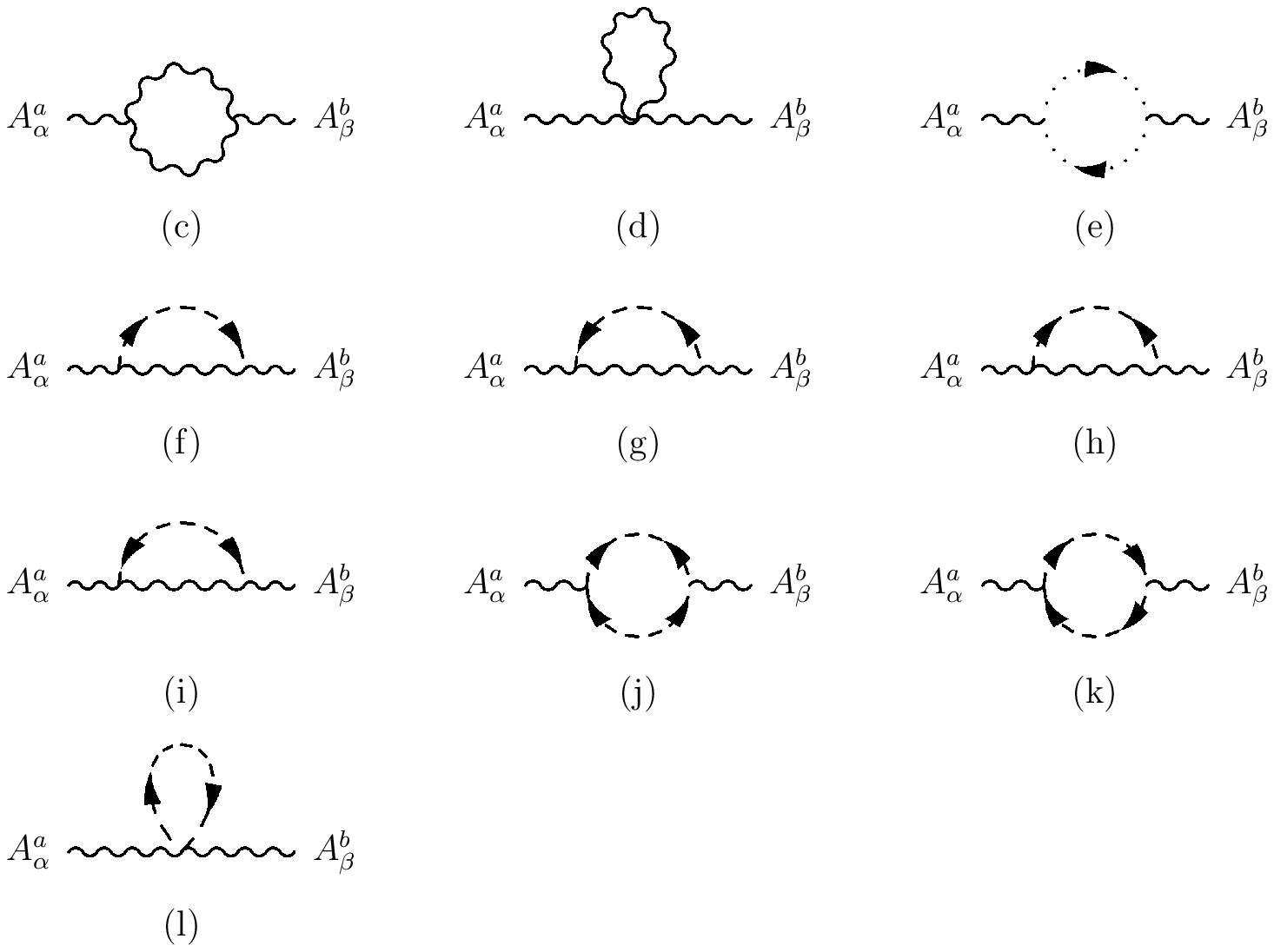}}
}
\caption{Gauge self-energy diagrams at one-loop in the 6D theory:
$(c)$, $(d)$ gauge loops; $(e)$ ghost loop; and $(f)$--$(l)$ loops
associated with the $A_{4}$ and $A_{5}$ components of the 6D gauge
field.  For the latter, the arrows represent the propagation of $A_{+}
= A_{4} + i A_{5}$.}
\label{fig:diagrams}
}

We consider now the one-loop corrections arising from the
self-interactions in a non-abelian gauge theory.  The diagrams are
shown in Figure~\ref{fig:diagrams}.  In terms of the momentum space
propagators for the gauge bosons, the scalar adjoints and the ghost
fields, given in Eqs.~(\ref{GGaugemom}), (\ref{GA4A5mom}) and
(\ref{GGhostmom}), they are:
\beqa
(c) &=& \left( \frac{1}{4} \right) \frac{g_{4}^{2}}{2} f^{adl} f^{bdl}
\sum_{m_{1},l_{1}} \sum_{m'_{1},l'_{1}}
\int \frac{d^{D}k}{(2\pi)^{D}}
\left[ \eta_{\alpha\delta} (p - k)_{\lambda} +
\eta_{\delta\lambda} (2k + p)_{\alpha} -
\eta_{\lambda\alpha} (k + 2 p)_{\delta} \right]
\hspace{7mm}
\nonumber \\ [0.5em]
& & \hspace{-2mm} \mbox{} \times
G^{\lambda\lambda',k+p}_{(m_{1},l_{1}; m'_{1},l'_{1})}
\left[ \eta_{\lambda'\beta} (k + 2p)_{\delta'}
- \eta_{\delta'\lambda'} (2k + p)_{\beta}
+ \eta_{\beta\delta'} (k - p)_{\lambda'} \right]
G^{\delta\delta',k}_{(m'_{2},l'_{2}; m_{2},l_{2})}~,
\label{gaugetri}
\\ [0.75em]
(d) &=& - i \frac{g_{4}^{2}}{2} f^{adl} f^{bdl} \sum_{m_{1},l_{1}}
\int \frac{d^{D}k}{(2\pi)^{D}}
\left[ 2 \eta_{\alpha\beta} \eta_{\delta\delta'}-
\eta_{\alpha\delta} \eta_{\beta\delta'}  -
\eta_{\alpha\delta'} \eta_{\beta\delta} \right]
G^{\delta\delta',k}_{(m_{1},l_{1}; m'_{1},l'_{1})}~,
\label{gaugequartic}
\\ [0.75em]
(e) &=& - \left( \frac{1}{4} \right) g_{4}^{2} f^{ald} f^{bdl}
\sum_{m_{1},l_{1}} \sum_{m'_{1},l'_{1}}
\int \frac{d^{D}k}{(2\pi)^{D}}
(k + p)_{\alpha} G_{\xi,k+p}^{(m_{1},l_{1}; m'_{1},l'_{1})} k_{\beta} \,
G_{\xi,k}^{(m'_{2},l'_{2}; m_{2},l_{2})}~,
\label{ghostloop}
\\ [0.75em]
(f) + (g) &=& \left( \frac{1}{4} \right) \left( \frac{1}{2} \right)^{2}
g_{4}^{2} f^{ald} f^{bdl} \sum_{m_{1},l_{1}} \sum_{m'_{1},l'_{1}}
\int \frac{d^{D}k}{(2\pi)^{D}}
\left[ r_{m',l'} M_{m',l'} - r_{m'_{2},l'_{2}} M_{m'_{2},l'_{2}} \right]
\nonumber \\ [0.5em]
& & \mbox{} \times
G_{++,k+p}^{(m_{1},l_{1}; m'_{1},l'_{1})}
\left[ r^{*}_{m,l} M_{m,l} - r^{*}_{m_{2},l_{2}} M_{m_{2},l_{2}} \right]
G_{\alpha\beta,k}^{(m'_{2},l'_{2}; m_{2},l_{2})} + \mathrm{h.c.}~,
\label{scalarNGB1}
\\ [0.75em]
(h) + (i) &=& \left( \frac{1}{4} \right) \left( \frac{1}{2} \right)^{2}
g_{4}^{2} f^{ald} f^{bdl} \sum_{m_{1},l_{1}} \sum_{m'_{1},l'_{1}}
\int \frac{d^{D}k}{(2\pi)^{D}}
\left[ r_{m',l'} M_{m',l'} - r_{m'_{2},l'_{2}} M_{m'_{2},l'_{2}} \right]
\\ [0.5em]
& & \mbox{} \times
G_{+-,k+p}^{(m_{1},l_{1}; m'_{1},l'_{1})} 
\left[ r_{m,l} M_{m,l} - r_{m_{2},l_{2}} M_{m_{2},l_{2}} \right]
G_{\alpha\beta,k}^{(m'_{2},l'_{2}; m_{2},l_{2})} + \mathrm{h.c.}~,
\label{scalarNGB2}
\nonumber \\ [0.75em]
(j)&=& \left( \frac{1}{4} \right) \frac{g_{4}^{2}}{4} f^{ald} f^{bdl}
\sum_{m_{1},l_{1}} \sum_{m'_{1},l'_{1}}
\int \frac{d^{D}k}{(2\pi)^{D}}
\, (2 k + p)_{\alpha} \,\, (2 k + p)_{\beta} 
\nonumber \\ %[0.3em]
& & \hspace{7.5cm} \mbox{} \times
G_{+-,k+p}^{(m_{1},l_{1}; m'_{1},l'_{1})}
\, G_{-+,k}^{(m'_{2},l'_{2}; m_{2},l_{2})} ~,
\label{scalarlooptri1}
\\ [0.75em]
(k)&=& \left( \frac{1}{4} \right) \frac{g_{4}^{2}}{4} f^{ald} f^{bdl}
\sum_{m_{1},l_{1}} \sum_{m'_{1},l'_{1}}
\int \frac{d^{D}k}{(2\pi)^{D}}
\, (2 k + p)_{\alpha} \, (2 k + p)_{\beta} \, 
\\ [0.5em]
& & \hspace{7.5cm} \mbox{} \times
G_{++,k+p}^{(m_{1},l_{1}; m'_{1},l'_{1})} \, 
G_{++,k}^{(m_{2},l_{2}; m'_{2},l'_{2})}~,
\label{scalarlooptri2}
\\ [0.75em]
(l) &=& i g_{4}^{2} f^{adl} f^{bdl} \, \eta_{\alpha\beta}
\sum_{m_{1},l_{1}}
\int \frac{d^{D}k}{(2\pi)^{D}} G_{++,k}^{(m_{1},l_{1}; m'_{1},l'_{1})}~,
\label{scalarquartic}
\eeqa
where we included a factor $1/4$ for the diagrams involving trilinear
interactions, as explained in subsection~\ref{sec:scalarloop}.  It is
similarly understood that the diagrams involving a quartic interaction
should be contracted with the external propagators in order to obtain
the correct KK-number structure in the two-point function [see
discussion after Eq.~(\ref{allowedKKnumbers})].  The color factors are
\beq
f^{adl} f^{bdl} = C_{2}(A) \delta_{ab}~,
\eeq
where $C_{2}(A) = N$ for a $SU(N)$ group.  A lengthy but
straightforward calculation allows us to write the leading logarithmic
divergences of diagrams $(c)$-$(l)$ as
\beq
- i \frac{g_{4}^2}{16\pi^2} C_2(A) \delta_{ab}
\,\Gamma\!\left(\frac{\epsilon}{2}\right) \left[A_{G} (p^2 \eta^{\mu\nu} - 
p^{\mu}p^{\nu}) - B_{G} \eta^{\mu\nu}\right]~,
\label{self-gauge}
\eeq
where the scalar functions $A_{G}$ and $B_{G}$ are presented in
Table~\ref{tableABgauge}. 

%%%%%%%%%%%%%%%%%%%%%%%%%%%%%%%%%%%%%%
%%% Table for scalar functions A and B for gauge sector
%%%%%%%%%%%%%%%%%%%%%%%%%%%%%%%%%%%%%%
\TABLE[t]{
%\vspace*{-1cm}
%\begin{center}
\begin{tabular}{|c||c|c|}
  \hline
   &  $A_{G}$ & $B_{G}$
\rule{0mm}{5mm} \\ [0.4em]
\hline
\hline
  (c)  & $(\frac{1}{2} \xi - \frac{7}{3}) \times \left\{%
\begin{array}{ll}
    3 \\
    2 \\
    5/2 \\
\end{array}%
\right.$ & $- \frac{1}{4} p^2 \times \left\{%
\begin{array}{ll}
    3 \\
    2 \\
    5/2 \\
\end{array}%
\right. + \frac{3}{4}(4 + \xi+ \xi^2)(M_{m,l}^2+M_{m',l'}^2) \times \left\{%
\begin{array}{ll}
    5/4 \\
    1 \\
    9/8 \\
\end{array}%
\right.$
\rule{0mm}{5mm} \\ [0.4em]
\hline
  (d) & 0&$ - \frac{3}{4} (3+\xi^2) (M_{m,l}^2+M_{m',l'}^2) \times \left\{%
\begin{array}{ll}
    5/4 \\
    1 \\
    9/8 \\
\end{array}%
\right.$ 
\rule{0mm}{5mm} \\ [0.4em]
\hline
  (e) & $\frac{1}{6} \times \left\{%
\begin{array}{ll}
    3 \\
    2 \\
    5/2 \\
\end{array}%
\right.$ & $ \frac{1}{4} p^2 \times \left\{%
\begin{array}{ll}
    3 \\
    2 \\
    5/2 \\
\end{array}%
\right. - \frac{1}{2} \xi(M_{m,l}^2+M_{m',l'}^2) \times \left\{%
\begin{array}{ll}
    5/4 \\
    1 \\
    9/8 \\
\end{array}%
\right.$
\rule{0mm}{5mm} \\ [0.4em]
\hline
  (f) + (g) & 0& $ -\frac{1}{8} (3 + \xi)(M_{m,l}^2+M_{m',l'}^2) \times \left\{%
\begin{array}{ll}
    -7/2 \\
    -2 \\
    -11/4 \\
\end{array}%
\right.$
\rule{0mm}{5mm} \\ [0.4em]
\hline
  (h) + (i)  & 0&0
\rule{0mm}{5mm} \\ [0.4em]
 \hline
  (j) & 0 & 0 
\rule{0mm}{5mm} \\ [0.4em]
\hline
  (k) & $ - \frac{1}{6} \times \left\{%
\begin{array}{ll}
    2 \\
    0 \\
    1 \\
\end{array}%
\right.$ & $ \frac{1}{4}(1+\xi)(M_{m,l}^2+M_{m',l'}^2) \times \left\{%
\begin{array}{cc}
    -1 \\
    0 \\
    -1/2 \\
\end{array}%
\right.$ 
\rule{0mm}{5mm} \\ [0.4em]
\hline
  (l) &  0& $ - \frac{1}{4} (1 + \xi) (M_{m,l}^2+M_{m',l'}^2) \times \left\{%
\begin{array}{cc}
    -1 \\
    0 \\
    -1/2 \\
\end{array}%
\right.$
\rule{0mm}{5mm} \\ [0.4em]
\hline
\end{tabular}
%\end{center}
\caption{Scalar functions $A_{G}$ and $B_{G}$ in the non-abelian gauge
sector, as defined via Eq.~(\ref{self-gauge}), corresponding to the
diagrams $(c)$-$(l)$ of Fig.~\ref{fig:diagrams}.  For each diagram,
the first two cases correspond to even-even mixings with $m-m'$ even
and $m-m'$ odd, and the third to odd-odd mixings, as listed in
Eq.~(\ref{cases}).}
\label{tableABgauge}
}%

Note that the $p^{2}$ terms in $B_{G}$ cancel between the ghost loop
diagram and the diagram involving trilinear gauge self-interactions,
as happens in 4D QCD. Therefore, the ``gauge violating'' terms in
Eq.~(\ref{self-gauge}), proportional to $\eta_{\alpha\beta}$, are
proportional to the KK masses, thus preserving the unbroken 4D gauge
invariance.  Diagrams $(f)$--$(i)$ involving the coupling of a single
scalar to two gauge fields are directly related to the higher
dimensional Higgs mechanism.  Also the divergent parts of diagrams
$(h)$--$(j)$ vanish.  This results from a cancellation between the two
real scalar degrees of freedom in $A_{+}$, as seen in the form of the
propagator $G_{+-,p}^{(m_{1},l_{1}; m'_{1},l'_{1})}$ of
Eq.~(\ref{GA4A5mom}).  Even though for arbitrary $\xi$ the two real
scalar degrees of freedom have different masses, the $\xi$-dependence
cancels out in the infinite terms.  Note also that diagrams $(k)$ and
$(l)$ with $\xi=1$, when the two real degrees of freedom in $A_{+}$
have the same mass, agree with the results derived for a complex
scalar minimally coupled to a gauge field and satisfying $n=3$
boundary conditions, as given in Table~\ref{tableABmatter}.

Adding the results of Table~\ref{tableABgauge} we find that the
contribution to the 6D gauge self-interactions is given by
\beqa
A_{G} =
\frac{1}{2} \, \xi \times
\left\{%
\begin{array}{ll}
    3  \\ [0.4em]
    2  \\ [0.4em]
    5/2  \\ [0.1em]
\end{array}%
\right.
- \left\{%
\begin{array}{ll}
    41/6  \\ [0.4em]
    13/3  \\ [0.4em]
    67/12  \\ [0.1em]
\end{array}%
\right.~,
& \hspace{5mm} &
B_{G} =
\frac{1}{4} (3+\xi) \left( M_{m,l}^2+M_{m',l'}^2 \right)  \times
\left\{%
\begin{array}{ll}
    3 \\ [0.4em]
    2 \\ [0.4em]
    5/2  \\
\end{array}%
\right. ~.
\label{AGBG}
\eeqa
which, together with Eq.~(\ref{self-gauge}), determines the KK-number
violating contribution to the gauge boson two-point function arising
from the gauge self-interactions in non-abelian gauge theories.

%%%%%%%%%%%%%%%%%%%%%%%%%%%%%%%%%%%%%
\subsubsection{Mass Shifts and Localized Operators}
\label{sec:massLocal_gauge}

It should be noted that the parameters $A_{G}$ and $B_{G}$ given in
Eq.~(\ref{AGBG}), are gauge dependent.  In addition, as already
mentioned, we find a ``gauge violating'' term proportional to
$\eta_{\mu\nu}$, i.e. $B_{G} \neq 0$.  This term would arise from the
localized operators [see Eq.~(\ref{loc_ops})]
\beq
{\cal{O}}_{1,2} = A^{\mu} \partial_{+} \partial_{-} A_{\mu} + {\rm h.c.}
\eeq
Although these operators do not break the 4D gauge invariance
associated with the zero-mode gauge field, they are certainly not
invariant under the set of 6D gauge transformations left unbroken by
the chiral square compactification [i.e. those generated by a gauge
transformation parameter satisfying $n=0$ boundary conditions, as does
$A_{\mu}$.]  Nevertheless, given the fact that the kinetic term
renormalization parameter, $A_{G}$, is $\xi$-dependent, such terms are
necessary to ensure that physical quantities such as the mass shifts
of the massive KK modes are gauge invariant.

Indeed, the KK diagonal components in Eq.~(\ref{self-gauge}) give the
leading contribution to the KK mass shifts.  For KK-parity even
states, $(-1)^{m+l} = +1$, case 1 in Eq.~(\ref{AGBG}) gives a
self-energy
\beqa
- i \frac{g_{4}^2}{16\pi^2} C_2(A) \delta_{ab}
\,\Gamma\!\left(\frac{\epsilon}{2}\right) \left[ \left( 
\frac{3}{2} \xi  - \frac{41}{6} \right) (p^2 \eta_{\mu\nu} - 
p_{\mu}p_{\nu}) - \frac{3}{2} (3+\xi) M^{2}_{m,l} \eta_{\mu\nu}\right]~, 
\label{self-gauge-even}
\eeqa
which leads to a mass shift
\beqa
\delta M^{2}_{m,l} &=& \frac{g_{4}^2}{16\pi^2} C_2(A) 
\,\Gamma\!\left(\frac{\epsilon}{2}\right)
\left[ \frac{3}{2} (3+\xi) - \left( \frac{3}{2} \xi - \frac{41}{6} \right) 
\right] M^{2}_{m,l} 
\nonumber \\
&=& \frac{34}{3} \times \frac{g_{4}^2}{16\pi^2} C_2(A) 
\,\Gamma\!\left(\frac{\epsilon}{2}\right) M^{2}_{m,l}~.
\eeqa
Similarly, for KK-parity odd states, $(-1)^{m+l} = -1$, case 3 in
Eq.~(\ref{AGBG}) gives a (diagonal) self-energy
\beqa
- i \frac{g_{4}^2}{16\pi^2} C_2(A) \delta_{ab}
\,\Gamma\!\left(\frac{\epsilon}{2}\right) \left[ \left( 
\frac{5}{4} \xi - \frac{67}{12} \right) (p^2 \eta_{\mu\nu} - 
p_{\mu}p_{\nu}) - \frac{5}{4} (3+\xi) M^{2}_{m,l} \eta_{\mu\nu}\right]~, 
\label{self-gauge-odd}
\eeqa
which leads to a mass shift
\beqa
\delta M^{2}_{m,l} &=& \frac{28}{3} \times \frac{g_{4}^2}{16\pi^2} C_2(A) 
\,\Gamma\!\left(\frac{\epsilon}{2}\right) M^{2}_{m,l}~.
\eeqa
We see that the $\xi$ dependence disappears from the physical mass
shifts, as expected.

The presence of the gauge non-invariant operator associated with
$B_{G}$ and the $\xi$-dependence of $A_{G}$ and $B_{G}$ go hand in
hand.  By an appropriate field redefinition, one should be able to
absorb the $\eta_{\mu\nu}$ term into purely gauge invariant operators,
with gauge independent coefficients.\footnote{We thank H.~C.~Cheng for
discussions on this point.} Notice that the 6D gauge violating term
vanishes for $\xi = -3$.  Therefore, in this particular gauge no
further field redefinition is necessary.  The induced localized
operators are then
\beqa
\frac{1}{4} \, \delta_{c_{G}}\!(z) \times \left(-\frac{1}{4} \, \hat{r}_{G} L^{2} 
F_{\mu\nu}^{a} F^{\mu\nu a}\right)~,
\label{local45_4}
\eeqa
where $\delta_{c}(z)$ was defined in Eq.~(\ref{loc_kkp}), the factor
of $1/4$ accounts for universal KK wavefunction enhancements, and
\beqa
\hat{r}_{G} = - \frac{14}{3} \times \frac{g_{4}^2}{16\pi^2} \, C_{2}(A)
\,\Gamma\!\left(\frac{\epsilon}{2}\right)~,
&\hspace{5mm}&
c_{G} = \frac{3}{7}~.
\label{Arc_gauge}
\eeqa
This should be added to the contributions from scalar and fermion loops
given in Eqs.~(\ref{local45_2}) and (\ref{local45_3}).

Explicit computation of the KK-number violating one-loop corrections
to the trilinear and quartic gauge vertices in the gauge $\xi = -3$
should give precisely the coefficients necessary to provide the
non-abelian completion of the kinetic operator Eq.~(\ref{local45_4}),
but we leave such a check for future work.

%%%%%%%%%%%%%%%%%%%%%%%%%%%%%%%%%%%%%
\subsection{Fermion Two-Point Function}
\label{sec:Fermion_self}

\FIGURE[t]{
\vspace*{-5mm}
\centerline{
   \resizebox{13cm}{!}{\includegraphics{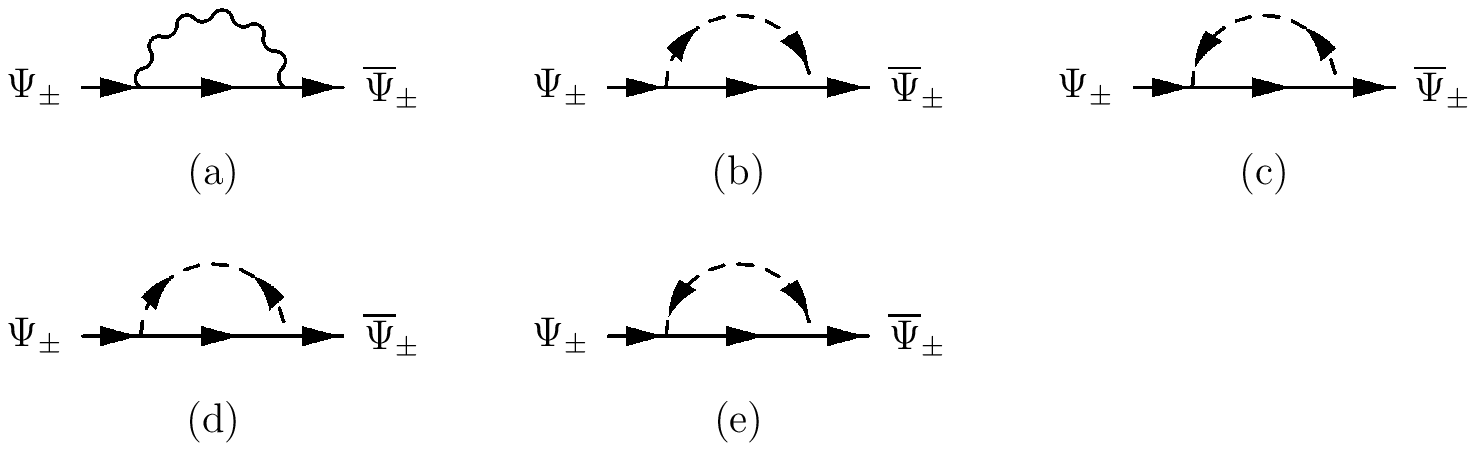}}
}
\caption{One-loop fermion self-energy diagrams.  These diagrams 
arise from the interactions with the 6D gauge sector.  The dashed lines
with two arrows represent the propagation of $A_{+}$.}
\label{fig:Diagramsfermion}
}

We turn now to the one-loop corrections to the fermion two-point
function.  We discuss first the loops arising from the 6D gauge
interactions.  These include interactions with the 4D gauge fields and
interactions with the spinless adjoints.  In
subsection~\ref{sec:Yukawa} we study the corrections arising from
Yukawa interactions.
 
%%%%%%%%%%%%%%%%%%%%%%%%%%%%%%%%%%%%%
\subsubsection{One-loop Diagrams: Spinor Manipulations in 6D}

The diagrams arising from the 6D gauge interactions are shown in
Figure~\ref{fig:Diagramsfermion}.  In terms of the spin-1/2, spin-1
and spin-0 propagators given in Eqs.~(\ref{FullFermionPropagator}),
(\ref{GGaugemom}), (\ref{GA4A5mom}), and according to the vertices
given in Figure~\ref{fig:Feynman1}, they are given by
\beqa
(a) &=& - \left( \frac{1}{4} \right) g_{4}^{2} \, C_{2}(\Psi)
\sum_{m_{1},l_{1}} \sum_{m'_{1},l'_{1}} \Gamma^{\mu}
\int \frac{d^{D}k}{(2\pi)^{D}}
G_{k}^{\pm(m_{2},l_{2}; m'_{2},l'_{2})}
G_{\mu\nu,p-k}^{(m_{1},l_{1}; m'_{1},l'_{1})} \Gamma^{\nu} P_{\pm}~,
\label{fermiongaugeloop}
\\ [0.75em]
(b) &=& - \left( \frac{1}{4} \right) g_{4}^{2} \, C_{2}(\Psi)
\sum_{m_{1},l_{1}} \sum_{m'_{1},l'_{1}} \Gamma^{-}
\int \frac{d^{D}k}{(2\pi)^{D}}
G_{k}^{\pm(m_{2},l_{2}; m'_{2},l'_{2})}
G_{++,p-k}^{(m_{1},l_{1}; m'_{1},l'_{1})} \Gamma^{+} P_{\pm}~,
\label{fermionscalarloop1}
\\ [0.75em]
(c) &=& - \left( \frac{1}{4} \right) g_{4}^{2} \, C_{2}(\Psi)
\sum_{m_{1},l_{1}} \sum_{m'_{1},l'_{1}} \Gamma^{+}
\int \frac{d^{D}k}{(2\pi)^{D}}
G_{k}^{\pm(m_{2},l_{2}; m'_{2},l'_{2})}
G_{--,p-k}^{(m_{1},l_{1}; m'_{1},l'_{1})} \Gamma^{-} P_{\pm}~,
\label{fermionscalarloop2}
\\ [0.75em]
(d) &=& - \left( \frac{1}{4} \right) g_{4}^{2} \, C_{2}(\Psi)
\sum_{m_{1},l_{1}} \sum_{m'_{1},l'_{1}} \Gamma^{+}
\int \frac{d^{D}k}{(2\pi)^{D}}
G_{k}^{\pm(m_{2},l_{2}; m'_{2},l'_{2})}
G_{-+,p-k}^{(m_{1},l_{1}; m'_{1},l'_{1})} \Gamma^{+} P_{\pm}~,
\label{fermionscalarloop3}
\\ [0.75em]
(e) &=& - \left( \frac{1}{4} \right) g_{4}^{2} \, C_{2}(\Psi)
\sum_{m_{1},l_{1}} \sum_{m'_{1},l'_{1}} \Gamma^{-}
\int \frac{d^{D}k}{(2\pi)^{D}}
G_{k}^{\pm(m_{2},l_{2}; m'_{2},l'_{2})}
G_{+-,p-k}^{(m_{1},l_{1}; m'_{1},l'_{1})} \Gamma^{-} P_{\pm}~,
\label{fermionscalarloop4}
\eeqa where a factor of $1/4$ was included in each diagram as
explained in subsection~\ref{sec:scalarloop}, and
\beq
\Gamma^{\pm} = \frac{1}{2} \left( \Gamma^{4} \pm i \Gamma^{5} \right)~.
\label{gammapm}
\eeq
We also chose the directions of the fermion loop momenta $k$,
$(m_{2},l_{2})$ and $(m'_{2},l'_{2})$ opposite to the convention of
Figure~\ref{fig:momenta}, so that they are in the direction of the
fermion number flow.

We concentrate on some useful remarks relevant in the evaluation of
the previous expressions.  In particular, we show how to manipulate
the 6D $\Gamma$-matrices efficiently to understand the spinor
structure, as well as the KK-number violating structure, of the
results.

Starting with diagram $(a)$ and keeping in mind that
$[P_{\pm},P_{L,R}] = 0$, one obtains a term involving only the
momentum along the non-compact dimensions, proportional to
\beq
\Gamma^{\mu} k_{\lambda} \Gamma^{\lambda} \Gamma^{\nu} 
\sum_{m_{1},l_{1}} \sum_{m'_{1},l'_{1}} \, \left[ P_{L} 
\hat{\delta}(m_{2},l_{2}; m'_{2},l'_{2}; n^{\pm}_{L}) +
P_{R} \hat{\delta}(m_{2},l_{2}; m'_{2},l'_{2}; n^{\pm}_{R}) \right]
\hat{\delta}(m_{1},l_{1}; m'_{1},l'_{1};0) P_{\pm} ~,
\label{eqa1}
\eeq
and also a term involving the extra-dimensional momenta that has the
form of~(\ref{eqa1}) with
\beqa
k_{\lambda} \Gamma^{\lambda} P_{L} P_{\pm} &\rightarrow& - r_{m_{2}, 
\pm l_{2}} M_{m_{2}, l_{2}} \Gamma^{4} P_{L} P_{\pm}~,
\nonumber \\ [0.4em]
k_{\lambda} \Gamma^{\lambda} P_{R} P_{\pm} &\rightarrow& - r_{m_{2}, 
\mp l_{2}} M_{m_{2}, l_{2}} \Gamma^{4} P_{R} P_{\pm}~.
\label{eqa2}
\eeqa
To obtain (\ref{eqa2}), we used $k_{4} = - m_{2}/R$, $k_{5} = -
l_{2}/R$, and the identities given in Eq.~(\ref{Pidentities}) to
express $\Gamma^{5}$ in terms of $\Gamma^{4}$.  It is understood that
the factor in Eq.~(\ref{eqa2}) appears under the KK sums.  The
KK-number violating terms in the KK sums of~(\ref{eqa1}) are
nonvanishing only when $n^{\pm}_{L} = 0$, $n^{\pm}_{R} = 0$,
$n^{\pm}_{L} = 2$ or $n^{\pm}_{R} = 2$ [see comments after
Eq.~(\ref{two-deltas1})].  The sums associated with~(\ref{eqa2}),
involving factors of $m_{2} \pm i l_{2}$, are also non-vanishing in
these cases.  The evaluation of the momentum integrals and remaining
$\Gamma$-matrix algebra proceeds exactly as in 4D.

Turning to the evaluation of diagrams $(b)$--$(e)$ with the scalars
$A_{\pm}$ in the loop, which are characteristic of the 6D theory, we
first notice that the definition (\ref{gammapm}) immediately implies 
\beq
\left( \Gamma^{+} \right)^{2} = \left( \Gamma^{-} \right)^{2} = 0~.
\label{Gammaidentity1}
\eeq
Also, from the identities given in Eq.~(\ref{Pidentities}) one can
easily see that
\beq
\Gamma^{\pm} P_{L} P_{\pm} = \Gamma^{\mp} P_{R} P_{\pm} = 0~,
\label{Gammaidentity2}
\eeq
and also
\beqa
\Gamma^{\pm} \Gamma^{\mp} P_{L} P_{\pm} &=& - P_{L} P_{\pm}~,
\nonumber \\
\Gamma^{\mp} \Gamma^{\pm} P_{R} P_{\pm} &=& - P_{R} P_{\pm}~.
\label{Gammaidentity3}
\eeqa
Consider first chirality $+$ fermions.  Using the fermion propagator
given in Eq.~(\ref{FullFermionPropagator}) and $G_{++}$ given in
Eq.~(\ref{GA4A5mom}), as well as
Eqs.~(\ref{Gammaidentity1})--(\ref{Gammaidentity3}), one can see that
diagram $(b)$ is proportional to
\beq
p_{\mu} \Gamma^{\mu} P_{R} P_{+} \sum_{m_{1},l_{1}} 
\sum_{m'_{1},l'_{1}} \, \hat{\delta}(m'_{2},l'_{2}; m_{2},l_{2}; -n^{+}_{L})
\hat{\delta}(m_{1},l_{1}; m'_{1},l'_{1};3)~,
\label{eqb}
\eeq
so that it only renormalizes the 4D kinetic term.  According to the
identity given in Eq.~(\ref{two-deltas1}), the sum in Eq.~(\ref{eqb})
may be written as a sum of generalized $\hat{\delta}$-functions with
boundary conditions given by $n^{+}_{L} + 3$.  As mentioned after
Eq.~(\ref{two-deltas1}), the KK-number violating terms vanish unless
$n^{+}_{L} = 1$, i.e. when $n^{+}_{R} = 0$ [see Eq.~(\ref{nLnR})].

Similarly, diagram $(c)$ is proportional to
\beq
p_{\mu} \Gamma^{\mu} P_{L} P_{+} \sum_{m_{1},l_{1}} 
\sum_{m'_{1},l'_{1}} \, \hat{\delta}(m'_{2},l'_{2}; m_{2},l_{2}; -n^{+}_{R})
\hat{\delta}(m_{1},l_{1}; m'_{1},l'_{1};1)~,
\label{eqc}
\eeq
and the associated KK-number violating terms are non-vanishing only
when $n^{+}_{R} = 1$, i.e. when $n^{+}_{L} = 0$.

For diagrams $(d)$ and $(e)$, it is easy to see that the divergent
parts are $\xi$-independent, since they can at most diverge
logarithmically.  But $G_{+-}$ vanishes for $\xi=1$ [see
Eq.~(\ref{GA4A5mom})], and therefore these diagrams are finite.

Diagrams $(b)$--$(e)$ with chirality $-$ fermions on the external
lines can be treated similarly.

%%%%%%%%%%%%%%%%%%%%%%%%%%%%%%%%%%%%%%
%%% Table for scalar functions A and B for fermion self-energies
%%%%%%%%%%%%%%%%%%%%%%%%%%%%%%%%%%%%%%
\TABLE[t]{
%\vspace*{-1cm}
%\begin{center}
\begin{tabular}{|c||c|c|}
  \hline
   &  $A_{\Psi}$ & $B_{\Psi}$ 
\rule{0mm}{5mm} \\ [0.4em]
\hline
\hline
  (a)  & $\left[1+(\xi-1) \right] \times \left\{%
\begin{array}{ll}
    3 \\
    2 \\
    5/2 \\
\end{array}%
\right.$ & $ \left[ 4+(\xi-1) \right] \times \left\{%
\begin{array}{ll}
    3/2 \\
    1 \\
    5/4 \\
\end{array}%
\right. $ \rule{0mm}{5mm} \\ [0.4em] 
\hline
  (c) &$ 1 \times \left\{%
\begin{array}{ll}
    -1 \\
    0 \\
    -1/2 \\
\end{array}%
\right.$ &0 \rule{0mm}{5mm} \\ [0.4em] 
\hline
\end{tabular}
%\end{center}
\caption{Scalar functions $A_{\Psi}$ and $B_{\Psi}$ for the fermion
self-energies, as defined via Eq.~(\ref{fermion_self}), corresponding
to the diagrams $(a)$ and $(c)$ of Fig.~\ref{fig:Diagramsfermion},
when the zero-mode is left-handed.  Diagrams $(b)$, $(d)$ and $(e)$
are finite in this case.  When the zero-mode is right-handed the
results for diagrams $(b)$ and $(c)$ are exchanged.  For each diagram,
the first two cases correspond to even-even mixings with $m-m'$ even
and $m-m'$ odd, and the third to odd-odd mixings, as listed in
Eq.~(\ref{cases}).}
\label{tableABfermion}
}%

We summarize our results for the fermion self-energies by considering
two cases of phenomenological interest: whether the zero-mode is left-
or right-handed.  If the zero-mode is left-handed, we find that the
KK-number violating contribution to the corresponding fermion
two-point function can be written as
\beq
i \frac{g^{2}_{4}}{16\pi^{2}} \, C_{2}(\Psi) 
\,\Gamma\!\left(\frac{\epsilon}{2}\right) \left\{
A_{\Psi} p_{\mu} \Gamma^{\mu} P_{L} - B_{\Psi} \Gamma^{4}  
\left( r_{m',\mp l'} M_{m',l'} P_{R} + r_{m,\pm l} M_{m,l} P_{L} \right)
\right\} P_{\pm}~,
\label{fermion_self}
\eeq
where $C_{2}(\Psi)$ is the eigenvalue of the quadratic Casimir
operator in the representation of the fermion $\Psi$, and the scalar
function $A_{\Psi}$ and $B_{\Psi}$ are given in
Table~\ref{tableABfermion}.

When the zero-mode is right-handed, we obtain Eq.~(\ref{fermion_self})
with $P_{L} \leftrightarrow P_{R}$.

Since the cases $n_{L} = 2$ or $n_{R} = 2$ do not lead to zero-modes,
and are therefore of less phenomenological interest we do not record
the results here.  However, they can be easily read from
Table~\ref{tableABfermion} and the last identity of
Appendix~\ref{App:MomSpace}.

%%%%%%%%%%%%%%%%%%%%%%%%%%%%%%%%%%%%%
\subsubsection{Mass Shifts and Localized Operators}
\label{sec:massLocal_fermion}

We can read now the self-energies proper [the diagonal entries in
Eq.~(\ref{fermion_self})].  For KK-parity even modes, we get
\beq
i \frac{g^{2}_{4}}{16\pi^{2}} \, C_{2}(\Psi) 
\,\Gamma\!\left(\frac{\epsilon}{2}\right) \left\{
\left( 3 \xi - 1 \right) p_{\mu} \Gamma^{\mu} P_{L} + 
\frac{3}{2} (3+\xi) \left( p_{4} \Gamma^{4} + p_{5} \Gamma^{5}  \right)
\right\} P_{\pm}~,
\label{fermion_self_diag_even}
\eeq
where $p_{4} = -m/R$ and $p_{5} = -l/R$.  Since only the left-handed
fields receive a wavefunction renormalization, after canonical
normalization the mass shift is, to lowest order,
\beqa
\delta M_{m,l} &=& \frac{g_{4}^2}{16\pi^2} \, C_2(\Psi) 
\,\Gamma\!\left(\frac{\epsilon}{2}\right)
\left[ \frac{3}{2} (3+\xi) - \frac{1}{2} \left( 3 \xi - 1 \right) 
\right] M_{m,l} 
\nonumber \\
&=& 5 \times \frac{g_{4}^2}{16\pi^2} \, C_2(\Psi) 
\,\Gamma\!\left(\frac{\epsilon}{2}\right) M_{m,l}~,
\hspace{1cm}{\rm for}~(-1)^{m+l} = +1~,
\eeqa
and we see that the $\xi$-dependence cancels out, as expected.  Notice
also that the phases $r_{m,l}$ in (\ref{fermion_self}) also appear in
the KK mass terms arising from the bulk kinetic term.  They can be
rotated away by a chiral transformation, or equivalently, absorbed in
the definition of the wavefunction profile associated with the
right-handed fermions, as done in \cite{Dobrescu:2004zi}.

For KK-parity odd modes we get the self-energy 
\beq
i \frac{g^{2}_{4}}{16\pi^{2}} \, C_{2}(\Psi) 
\,\Gamma\!\left(\frac{\epsilon}{2}\right) \left\{
\left( \frac{5}{2} \xi - \frac{1}{2} \right) p_{\mu} 
\Gamma^{\mu} P_{L} + \frac{5}{4} (3+\xi) 
\left( p_{4} \Gamma^{4} + p_{5} \Gamma^{5}  \right) 
\right\} P_{\pm}~,
\label{fermion_self_diag_odd}
\eeq
and the corresponding mass shift is 
\beqa
\delta M_{m,l} &=& 4 \times \frac{g_{4}^2}{16\pi^2} \, C_2(\Psi) 
\,\Gamma\!\left(\frac{\epsilon}{2}\right) M_{m,l}~, 
\hspace{1cm}{\rm for}~(-1)^{m+l} = -1~.
\eeqa

It is also interesting to note that Eq.~(\ref{fermion_self})
corresponds to the localized operator
\beqa
& & \frac{1}{4} \, \delta_{c_{\Psi}}\!(z) \times \left( \hat{r}_\Psi L^{2} \,
i \overline{\Psi}_{+ L} \Gamma^{\mu} \partial_{\mu} \Psi_{+ L} \right)
+ \frac{1}{4} \, \delta_{c'_{\Psi}}\!(z) \times \hat{r}^{\prime}_\Psi L^{2}
\left[ i \overline{\Psi}_{+ L} \Gamma^{4} \partial_{-} \Psi_{+ R}
+ {\rm h.c.} \right]~,
\label{localized_fermion}
\eeqa
where $\delta_{c}(z)$ stands for the Dirac delta-functions at the
conical singularities, as defined in Eq.~(\ref{loc_kkp}), and the
factor of $1/4$ accounts for universal KK wavefunction enhancements.
For a chirality minus fermion, $\Psi_{-}$, one should simply make
$\partial_{-} \leftrightarrow \partial_{+}$.  If the zero-mode is
right-handed, one should make $L \leftrightarrow R$ everywhere in
Eq.~(\ref{localized_fermion}).

As noted in subsection~\ref{sec:massLocal_gauge}, the one-loop
computation in $\xi=-3$ gauge automatically gives rise to operators
with a gauge invariant structure, without any additional field
redefinitions.  In this gauge, we obtain
\beqa
\hat{r}_{\Psi} = - 4 \times \frac{g_{4}^2}{16\pi^2} C_{2}(\Psi)
\,\Gamma\!\left(\frac{\epsilon}{2}\right)~,
&\hspace{5mm}&
c_{\Psi} = \frac{1}{2}~,
\label{Frc_gauge}
\eeqa
and 
\beq
\hat{r}^{\prime}_{\Psi} = 0~, 
\eeq
so that no ``mass-terms'' are generated by the gauge interactions.  A
direct calculation of the three-point vertex of two fermions and a
gauge field in the $\xi = -3$ gauge should give the correct
coefficient to provide the gauge invariant completion of the kinetic
term operators of Eq.~(\ref{localized_fermion}).

Not all types of kinetic operators appear in
Eq.~(\ref{localized_fermion}).  This is consistent with the fact that,
when $n^{\pm}_{L} = 0$, so that there is a left-handed zero-mode,
$\Psi_{\pm R}$ vanishes at the fixed points $(0,0)$, $(L,L)$, and
$(L,0)$.  Therefore, the quadratic operators not appearing in
Eq.~(\ref{localized_fermion}) vanish at the singular points [note also
that $\Gamma^{\pm} \Psi_{\pm L} = \Gamma^{\mp} \Psi_{\pm R} = 0$,
according to Eq.~(\ref{Gammaidentity1})].

%%%%%%%%%%%%%%%%%%%%%%%%%%%%%%%%%%%%%
\subsubsection{Yukawa Interactions}
\label{sec:Yukawa}

\FIGURE[t]{
\vspace*{-5mm}
\centerline{
   \resizebox{5cm}{!}{\includegraphics{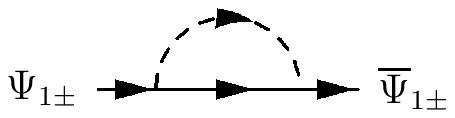}}
}
\caption{One-loop fermion self-energy diagrams associated with the
Yukawa interaction~(\ref{Yukawa}).  The dashed line denotes a complex
scalar satisfying boundary conditions labeled by $n$.}
\label{fig:DiagramsYukawa}
}

We finally consider the effect of Yukawa couplings such as
\beq
{\cal L}_{\rm Yukawa} = \lambda_{6} \Phi \overline{\Psi}_{1 \pm} 
\Psi_{2 \mp} + {\rm h.c.}
\label{Yukawa}
\eeq
Note that the fermions must be different since they must have opposite
6D chiralities in order for the above coupling to be Lorentz
invariant.  $\Phi$ is a complex 6D scalar.  Assume that the boundary
conditions for the scalar are labeled by $n$, while the boundary
conditions for $\Psi_{1\pm}$ and $\Psi_{2\pm}$ are labeled by
$n^{\pm}_{1L}$, $n^{\pm}_{1R}$, $n^{\mp}_{2L}$ and $n^{\mp}_{2R}$,
such that, according to Eq.~(\ref{nLnR}),
\beqa
n^\pm_{1R} &=& n^\pm_{1L} \mp 1 \; {\rm mod} \, 4~, 
\nonumber \\ 
n^\mp_{2R} &=& n^\mp_{2L} \pm 1 \; {\rm mod} \, 4~.
\label{nLnRYukawa}
\eeqa
In addition, 6D Lorentz invariance of the local coupling
(\ref{Yukawa}) also requires \cite{Dobrescu:2004zi}
\beqa
n - n^\pm_{1L} + n^\mp_{2R} &=& 0 \; {\rm mod} \, 4~, 
\label{YukawaInvariance}
\eeqa
which, together with (\ref{nLnRYukawa}) implies $n - n^\pm_{1R} +
n^\mp_{2L} = 0 \; {\rm mod} \, 4$.  To be concrete, let us assume that
$\Psi_{1 \pm}$ has a left-handed zero-mode: $n^{\pm}_{1L} = 0$ and
$n^{\pm}_{1R} = \mp 1$.  Then Eqs.~(\ref{nLnRYukawa}) and
(\ref{YukawaInvariance}) imply
\beqa
n^\mp_{2R} = -n \; {\rm mod} \, 4~, 
\hspace{1cm} 
n^\mp_{2L} = -n \mp 1 \; {\rm mod} \, 4~.
\label{YukawaExample}
\eeqa
The diagram in Figure~\ref{fig:DiagramsYukawa} is then
\beq
- \left( \frac{1}{4} \right) |\lambda_{4}|^{2}
\sum_{m_{1},l_{1}} \sum_{m'_{1},l'_{1}}
\int \frac{d^{D}k}{(2\pi)^{D}}
G_{k, n^{\mp}_{2}}^{\mp(m_{2},l_{2}; m'_{2},l'_{2})}
G_{k+p,n}^{(m_{1},l_{1}; m'_{1},l'_{1})} P_{\pm}~.
\label{fermionHiggsloop}
\eeq
This contains a term that depends on the 4D momentum, with the
structure
\beq
k_{\lambda} \Gamma^{\lambda} \sum_{m_{1},l_{1}} \sum_{m'_{1},l'_{1}} 
\, \left[ P_{R} \hat{\delta}(m_{2},l_{2}; m'_{2},l'_{2}; n^{\mp}_{2L}) +
P_{L} \hat{\delta}(m_{2},l_{2}; m'_{2},l'_{2}; n^{\mp}_{2R}) \right]
\hat{\delta}(m_{1},l_{1}; m'_{1},l'_{1};n) P_{\pm} ~,
\label{eqY1}
\eeq
but given the relations (\ref{YukawaExample}), and the comments
following Eq.~(\ref{two-deltas1}), only the second term gives a
nonvanishing result.  The terms depending on momenta along the extra
dimensions are both non-vanishing, and give rise to the following
structure:
\beqa
\sum_{m_{1},l_{1}} \sum_{m'_{1},l'_{1}} \, r_{m_{2},\mp l_{2}} M_{m_{2},l_{2}}
\hat{\delta}(m_{2},l_{2}; m'_{2},l'_{2}; n^{\mp}_{2L})
\hat{\delta}(m_{1},l_{1}; m'_{1},l'_{1};n) \Gamma^{4} P_{R} P_{\pm} 
&\rightarrow& r_{m',\mp l'} M_{m',l'} \Gamma^{4} P_{R} P_{\pm}~,
\nonumber \\
\sum_{m_{1},l_{1}} \sum_{m'_{1},l'_{1}} \, r_{m_{2},l_{2}} M_{m_{2},l_{2}}
\hat{\delta}(m_{2},l_{2}; m'_{2},l'_{2}; n^{\mp}_{2R})
\hat{\delta}(m_{1},l_{1}; m'_{1},l'_{1};n) \Gamma^{4} P_{L} P_{\pm} 
&\rightarrow& r_{m,\pm l} M_{m,l} \Gamma^{4} P_{L} P_{\pm}~.
\nonumber
\label{eqY2}
\eeqa
%
%%%%%%%%%%%%%%%%%%%%%%%%%%%%%%%%%%%%%%
%%% Table for scalar functions A and B due to Yukawas
%%%%%%%%%%%%%%%%%%%%%%%%%%%%%%%%%%%%%%
\TABLE[t]{
%\begin{center}
\begin{tabular}{|rccc|rcccc|}
  \hline
   &  & $A_{\Psi}$ & & & & $B_{\Psi}$ & &
\rule{0mm}{5mm} \\ [0.4em]
\hline
\hline
& $n=0$ & $n=1,3$ & $n=2$ & & $n=0$ & $n=1$ & $n=2$ & $n=3$
\rule{0mm}{5mm} \\ [0.4em]
\hline
 $\frac{1}{2} \times \left\{%
\begin{array}{ll}
     \\
     \\
     \\
\end{array}%
\right.$ \hspace*{-8mm}
&
$\begin{array}{c}
    3 \\
    2 \\
    5/2 \\
\end{array}$%
&
$\begin{array}{c}
    -1 \\
    0 \\
    -1/2 \\
\end{array}$%
&
$\begin{array}{c}
    -1\\
    -2 \\
    -3/2 \\
\end{array}$%
& $ 1 \times \left\{%
\begin{array}{c}
    \\
    \\
    \\
\end{array}%
\right.$ \hspace*{-8mm}
&
$\begin{array}{c}
    3/2 \\
    1 \\
    5/4 \\
\end{array}$%
&
$\begin{array}{c}
    1/2 \\
    1 \\
    3/4 \\
\end{array}$%
&
$\begin{array}{c}
    -1/2 \\
    -1 \\
    -3/4 \\
\end{array}$%
&
$\begin{array}{c}
    -3/2 \\
    -1 \\
    -5/4 \\
\end{array}$%
\rule{0mm}{5mm} \\ [0.4em]
\hline
\end{tabular}
%\end{center}
\caption{Scalar functions $A_{\Psi}$ and $B_{\Psi}$ associated with
the Yukawa contributions to the fermion self-energies, as defined via
Eq.~(\ref{fermion_self_Yukawa}), corresponding to the diagram of
Fig.~\ref{fig:DiagramsYukawa}.  We assume that the fermion has 6D
chirality $+$, and that it has a zero-mode, i.e. we exclude $n^{+}_{L,R}
= 2$ boundary conditions.  We give the results for scalars satisfying
the four types of boundary conditions, labeled by $n=0,1,2,3$.  For a
given $n$ and for each diagram, there are three possible cases
depending on KK-parity and whether $m-m'$ is even or odd, as listed in
Eq.~(\ref{cases}): the first two cases correspond to even-even mixings
with $m-m'$ even and $m-m'$ odd, and the third case to odd-odd
mixings.}
\label{tableABYukawa}
}%
With the help of Eq.~(\ref{two-deltas1}) and the relations following
it, we can write the diagrams in the form
\beq
i \, \frac{|\lambda_{4}|^{2}}{16\pi^{2}} 
\,\Gamma\!\left(\frac{\epsilon}{2}\right) \left\{
A_{\Psi} p_{\mu} \Gamma^{\mu} P_{L} - B_{\Psi} \Gamma^{4}  
\left( r_{m',\mp l'} M_{m',l'} P_{R} + r_{m,\pm l} M_{m,l} P_{L} \right)
\right\} P_{\pm}~,
\label{fermion_self_Yukawa}
\eeq
where $A_{\psi}$ and $B_{\Psi}$ are given in Table~\ref{tableABYukawa}
for a 6D chirality~$+$ fermion, and for the four allowed values of the
boundary conditions satisfied by the scalar field, $n=0,1,2$ or $3$.
For a 6D chirality~$-$ fermion, the cases $n=1$ and $n=3$ are
interchanged.  Also, when the zero mode of $\Psi_{1\pm}$ is
right-handed, we obtain Eq.~(\ref{fermion_self_Yukawa}) with $P_{L}
\leftrightarrow P_{R}$.

The Yukawa interactions induce localized operators as in
Eq.~(\ref{localized_fermion}).  When the scalar field satisfies $n=0$
boundary conditions we get $\hat{r}_{\Psi} = \hat{r}^{\prime}_{\Psi}$,
$c_{\Psi} = c^{\prime}_{\Psi}$, and
\beqa
\hat{r}_{\Psi} = \frac{5}{8} \times \frac{|\lambda_{4}|^2}{16\pi^2} 
\,\Gamma\!\left(\frac{\epsilon}{2}\right)~,
&\hspace{5mm}&
c_{\Psi} = \frac{2}{5}~.
\label{Frc_Yukawa}
\eeqa
The coefficients of the localized operators for other types of scalar
boundary conditions can be easily worked out from
Table~\ref{tableABYukawa}.

%%%%%%%%%%%%%%%%%%%%%%%%%%%%%%%%%%%%%
\subsection{Two-Point Functions of Scalar Fields}
\label{sec:scalar_self}

In this section we study the new features arising in the calculation
of the one-loop corrections to the two-point function associated with
6D scalars.  Here we encounter the most general KK-number violating
structure arising from operators localized at the three conical
singularities.  In addition, this will serve as a warm up to study the
renormalization associated with the spinless adjoints, $A_{\pm}$.

As is well-known, these diagrams contain quadratic divergences that
tend to lift the zero-mode to the cutoff scale.  This issue only
affects scalars satisfying $n=0$ boundary conditions, and we assume
that a bare contribution is tuned to keep the zero-mode light, if
necessary.  We are interested in the induced KK-number violating
transitions that correspond to operators localized at the conical
singularities.  Notice that for $n=1,3$, the mass term operator
\beq
-M^{2} \Phi^{\dagger} \Phi~,
\label{mass_op}
\eeq
vanishes when evaluated at the singular points $(0,0)$, $(L,L)$ and
$(0,L)$, due to the vanishing of the corresponding KK wavefunctions
given in Eq.~(\ref{KKf}).  This is relevant for the spinless adjoints,
$A_{\pm}$, which satisfy these boundary conditions, and it trivially
shows that there are no divergences corresponding to localized mass
terms for these fields.  For $n=2$ boundary conditions, the mass term
operator~(\ref{mass_op}) is non-vanishing at $(0,L)$.

Here we will simply use dimensional regularization to calculate the
induced localized {\textit{kinetic}} term operators, and will not
worry about potential quadratic divergences.  We show the relevant
diagrams in Figure~\ref{fig:Diagramsscalar}.
\FIGURE[t]{
\vspace*{-5mm}
\centerline{
   \resizebox{14cm}{!}{\includegraphics{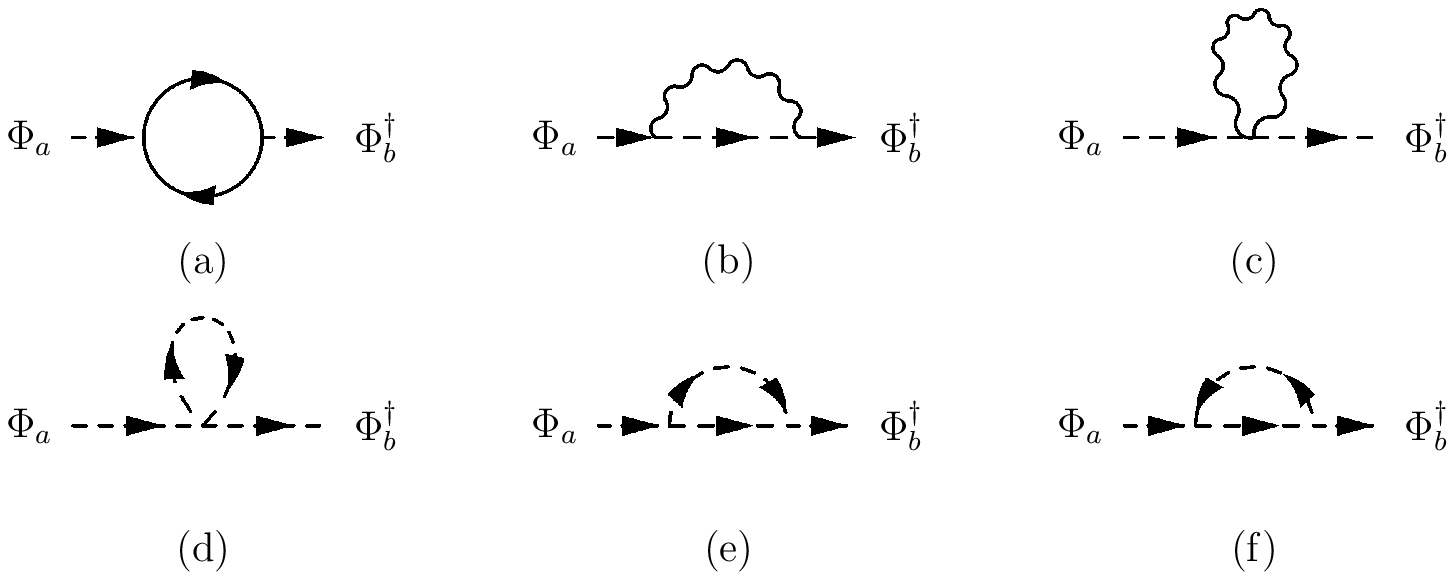}}
}
\caption{One-loop contribution to $\langle \Phi \, \Phi^{\dagger}
\rangle$, where $\Phi$ is a complex scalar (represented by a dashed
line with a single arrow) satisfying any of the four types of boundary
conditions, $n = 0,1,2,3$.  The dashed lines with two arrows
represent the propagation of $A_{+}$.}
\label{fig:Diagramsscalar}
}
%

%%%%%%%%%%%%%%%%%%%%%%%%%%%%%%%%%%%%%
\subsubsection{Gauge Interactions}

Let us consider first the diagrams arising from the gauge
interactions.  The interactions with the spin-1 components of the 6D
gauge field have expressions which are straightforward generalizations
of those applying to 4D scalars minimally coupled to gauge fields (see
Figure~\ref{fig:Feynman3}).  The coupling between two scalars and one
gauge field induces
\beqa
(b) &=& - \left( \frac{1}{4} \right) g_{4}^{2} \, C_{2}(\Phi) \, \delta_{ab}
\sum_{m_{1},l_{1}} \sum_{m'_{1},l'_{1}}
\int \frac{d^{D}k}{(2\pi)^{D}} \, (2p-k)^{\mu} (2p-k)^{\nu}
G_{\mu\nu,k}^{(m_{1},l_{1}; m'_{1},l'_{1})} 
G_{p-k,n}^{(m_{2},l_{2}; m'_{2},l'_{2})}~,
\nonumber \\ [-0.5em]
\label{S_gaugetrilinear}
\eeqa
where $(T^{c} T^{c})_{ab} = C_{2}(\Phi) \delta_{ab}$ is the quadratic
Casimir operator in the representation of $\Phi$.  The coupling with
two gauge bosons induces
\beqa
(c) &=& i g_{4}^{2} \, C_{2}(\Phi) \, \delta_{ab}
\sum_{m_{1},l_{1}}
\int \frac{d^{D}k}{(2\pi)^{D}} \, \eta^{\mu\nu} 
G_{\mu\nu,k}^{(m_{1},l_{1}; m'_{1},l'_{1})}~,
\label{S_gaugequartic}
\eeqa
which, as explained in subsection~\ref{sec:scalarloop} should be
contracted with external propagators to obtain the correct KK-number
violating structure.

\FIGURE[b]{
\vspace*{-5mm}
\centerline{
   \resizebox{15cm}{!}{\includegraphics{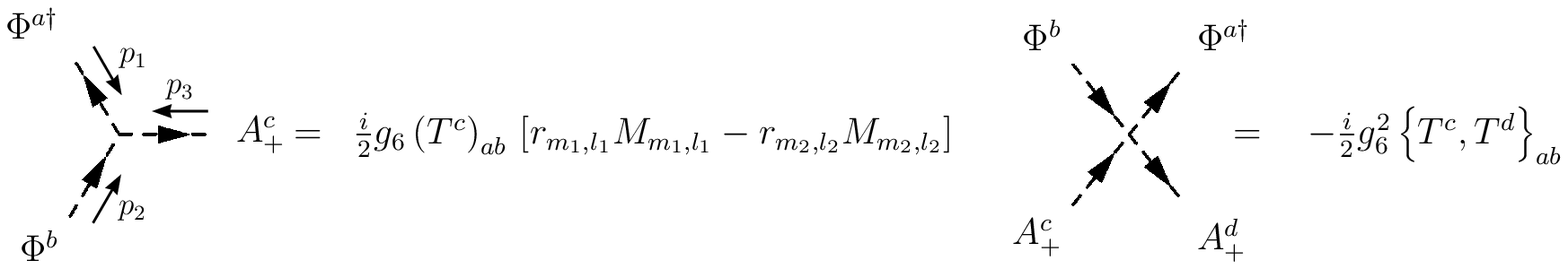}}
}
\caption{Momentum space Feynman rules for the couplings between
scalars and the spinless adjoints, $A_{\pm}$.  There is a trilinear
vertex involving an incoming $A_{+}$ (outgoing $A_{-}$), which can be
obtained from the one shown here by $r_{m,l} \rightarrow r^{*}_{m,l}$,
where the momentum-dependent phases were defined in Eq.~(\ref{rjk}).}
\label{fig:Feynman4}
}
In addition, there are diagrams arising from the gauge couplings to
the spin-0 components of the 6D gauge field.  The corresponding
vertices are shown in Figure~\ref{fig:Feynman4}.  A quartic
interaction induces
\beqa
(d) &=& - i g_{4}^{2} \, C_{2}(\Phi) \, \delta_{ab}
\sum_{m_{1},l_{1}}
\int \frac{d^{D}k}{(2\pi)^{D}} \, G_{++,k}^{(m_{1},l_{1}; m'_{1},l'_{1})}~,
\label{S_scalarquartic}
\eeqa which again requires contraction with external propagators,
while a trilinear interaction gives
\beqa
(e) &=& - \left( \frac{1}{4} \right)^{2} 
g_{4}^{2} \, C_{2}(\Phi) \, \delta_{ab} \sum_{m_{1},l_{1}} \sum_{m'_{1},l'_{1}}
\int \frac{d^{D}k}{(2\pi)^{D}}
\left[ r^{*}_{m,l} M_{m,l} + r^{*}_{m_{2},l_{2}} M_{m_{2},l_{2}} \right]
\nonumber \\ [0.5em]
& & \hspace{3cm} \mbox{} \times
G_{++,p-k}^{(m_{1},l_{1}; m'_{1},l'_{1})}
\left[r_{m'_{2},l'_{2}} M_{m'_{2},l'_{2}} +  r_{m',l'} M_{m',l'} 
\right]
G_{k,n}^{(m_{2},l_{2}; m'_{2},l'_{2})}~,
\label{S_trilinear1}
\eeqa
and a second diagram $(f)$ which is obtained from $(e)$ by taking the
complex conjugates of all phases, $r_{m,l} \rightarrow r^{*}_{m,l}$,
and also $G_{++} \rightarrow G_{--}$.

%%%%%%%%%%%%%%%%%%%%%%%%%%%%%%%%%%%%%%
%%% Table for scalar functions A and B for n=0,2 scalars
%%%%%%%%%%%%%%%%%%%%%%%%%%%%%%%%%%%%%%
\TABLE[t]{
%\begin{center}
\begin{tabular}{|c||rcc|rcc|}
  \hline
&  & $A_{\Phi}$ & & & $B_{\Phi}$ &
\rule{0mm}{5mm} \\ [0.4em]
\hline
\hline
& & $n=0$ & $n=2$ & & $n=0$ & $n=2$
\rule{0mm}{5mm} \\ [0.4em]
\hline
(b) & $(\xi - 3) \times \left\{%
\begin{array}{ll}
     \\
     \\
     \\
\end{array}%
\right.$ \hspace*{-8mm}
&
$\begin{array}{c}
    3 \\
    2 \\
    5/2 \\
\end{array}$%
&
$\begin{array}{c}
    0\\
    1/2 \\
    -1/2 \\
\end{array}$%
& $ (\xi + \xi^{2}) \times \left\{%
\begin{array}{c}
    \\
    \\
    \\
\end{array}%
\right.$ \hspace*{-8mm}
&
$\begin{array}{c}
    5/4 \\
    1 \\
    9/8 \\
\end{array}$%
&
$\begin{array}{c}
    0 \\
    1/8 \\
    -1/8 \\
\end{array}$%
\rule{0mm}{5mm} \\ [0.4em]
\hline
(c) & & 0 & 0 
& $ - (3+\xi^{2}) \times \left\{%
\begin{array}{c}
    \\
    \\
    \\
\end{array}%
\right.$ \hspace*{-8mm}
&
$\begin{array}{c}
    5/4 \\
    1 \\
    9/8 \\
\end{array}$%
&
$\begin{array}{c}
    0 \\
    1/8 \\
    -1/8 \\
\end{array}$%
\rule{0mm}{5mm} \\ [0.4em]
\hline
(d) & & 0 & 0
& $ - (1+\xi) \times \left\{%
\begin{array}{c}
    \\
    \\
    \\
\end{array}%
\right.$ \hspace*{-8mm}
&
$\begin{array}{c}
    -1/4 \\
    0 \\
    -1/8 \\
\end{array}$%
&
$\begin{array}{c}
    0 \\
    -1/8 \\
    1/8 \\
\end{array}$%
\rule{0mm}{5mm} \\ [0.4em]
\hline
(e) + (f) & & 0 & 0
& $ - \frac{1}{2} \times \left\{%
\begin{array}{c}
    \\
    \\
    \\
\end{array}%
\right.$ \hspace*{-8mm}
&
$\begin{array}{c}
    -7/2 \\
    -2 \\
    -11/4 \\
\end{array}$%
&
$\begin{array}{c}
    0 \\
    -3/4 \\
    3/4 \\
\end{array}$%
\rule{0mm}{5mm} \\ [0.4em]
\hline
\end{tabular}
%\end{center}
\caption{Functions $A_{\Phi}$ and $B_{\Phi}$, as defined via
Eq.~(\ref{scalar_self_gauge}), associated with the gauge contributions
to the scalar self-energies, corresponding to diagrams $(b)$--$(f)$ in
Fig.~\ref{fig:Diagramsscalar}.  We give the results for scalars
satisfying $n=0$ and $n=2$ boundary conditions.  For $n=0$, the three
cases in each diagram correspond to cases 1a, 1b and 2 of
Eq.~(\ref{cases}), while for $n=2$, they correspond to cases 1, 2a and
2b of Eq.~(\ref{cases2}).}
\label{tableABscalar_n02}
}%

We wish to consider scalar fields satisfying any of the four possible
boundary conditions, $n=0,1,2$ or $3$.  We start by considering the
cases $n=0$ or $n=2$.  We write the results for diagrams $(b)$--$(f)$
in the form
\beq
i \, \frac{g_{4}^{2}}{16\pi^{2}} \, C_{2}(\Phi) \, \delta_{ab}
\,\Gamma\!\left(\frac{\epsilon}{2}\right) \left\{
A_{\Phi} p^{2} - B_{\Phi}  
\left( M^{2}_{m,l} + M^{2}_{m',l'} \right)
\right\}~.
\label{scalar_self_gauge}
\eeq
A straightforward calculation gives the scalar coefficients $A_{\Phi}$
and $B_{\Phi}$ as summarized in Table~\ref{tableABscalar_n02}.  Adding
the diagrams for the $n=0$ boundary conditions we get
\beqa
A_{\Phi} = (\xi - 3) \times \left\{%
\begin{array}{c}
    3 \\
    2 \\
    5/2 \\
\end{array}
\right.~,
\hspace{1cm}
B_{\Phi} =  \xi  \times \left\{%
\begin{array}{l}
    3/2 \\
    1 \\
    5/4\\
\end{array}
\right. 
- \mbox{} \left\{%
\begin{array}{l}
    7/4 \\
    2 \\
    15/8 \\
\end{array}
\right.~,
\label{ABphi_gauge0}
\eeqa
where the three cases correspond to those listed in Eq.~(\ref{cases}).

For scalars satisfying  $n=2$ boundary conditions we get instead
\beqa
A_{\Phi} = (\xi - 3) \times \left\{%
\begin{array}{r}
    0 \\
    1/2 \\
    -1/2 \\
\end{array}
\right.~,
\hspace{1cm}
B_{\Phi} =  \xi  \times \left\{%
\begin{array}{r}
    0 \\
    1/4 \\
    -1/4\\
\end{array}
\right. 
+ \mbox{} \left\{%
\begin{array}{r}
    0 \\
    1/8 \\
    -1/8 \\
\end{array}
\right.~,
\label{ABphi_gauge2}
\eeqa
where now the three cases correspond to those listed in
Eq.~(\ref{cases2}).  Notice that, to lowest order, the corresponding
mass shifts are proportional to $2B_{\Phi} - A_{\Phi}$, and that the
$\xi$ dependence cancels out in the difference, both in
Eqs.~(\ref{ABphi_gauge0}) and (\ref{ABphi_gauge2}).

%%%%%%%%%%%%%%%%%%%%%%%%%%%%%%%%%%%%%%
%%% Table for scalar functions B and B' for n=1,3 scalars
%%%%%%%%%%%%%%%%%%%%%%%%%%%%%%%%%%%%%%
\TABLE[t]{
%\begin{center}
\begin{tabular}{|c||c|c|}
  \hline
& $B_{\Phi}$ & $B'_{\Phi}$
\rule{0mm}{5mm} \\ [0.4em]
\hline
\hline
(b) & $ \xi \times \left\{%
\begin{array}{ll}
    1/4 \\
    0 \\
    1/8 \\
    1/8 \\
\end{array}%
\right. 
\mbox{} - \xi^{2} \times \left\{%
\begin{array}{ll}
    5/4 \\
    1 \\
    9/8 \\
    9/8 \\
\end{array}%
\right.$ 
& $ \xi \times \left\{%
\begin{array}{ll}
    0 \\
    0 \\
    1/8 \\
    -1/8 \\
\end{array}%
\right. 
\mbox{} - \xi^{2} \times \left\{%
\begin{array}{ll}
    0 \\
    0 \\
    1/8 \\
    -1/8 \\
\end{array}%
\right.$ 
\rule{0mm}{5mm} \\ [0.4em]
\hline
(c) & $ (3 + \xi^{2}) \times \left\{%
\begin{array}{ll}
    5/4 \\
    1 \\
    9/8 \\
    9/8 \\
\end{array}%
\right.$ 
& $ (3 + \xi^{2}) \times \left\{%
\begin{array}{ll}
    0 \\
    0 \\
    1/8 \\
    -1/8 \\
\end{array}%
\right.$ 
\rule{0mm}{5mm} \\ [0.4em]
\hline
(d) & $ - (1 + \xi) \times \left\{%
\begin{array}{ll}
    1/4 \\
    0 \\
    1/8 \\
    1/8 \\
\end{array}%
\right.$ 
& $ - (1 + \xi) \times \left\{%
\begin{array}{ll}
    0 \\
    0 \\
    1/8 \\
    -1/8 \\
\end{array}%
\right.$ 
\rule{0mm}{5mm} \\ [0.4em]
\hline
(e) + (f) & $ - \frac{1}{2} \times \left\{%
\begin{array}{ll}
    -1/2 \\
    2 \\
    3/4 \\
    3/4 \\
\end{array}%
\right.$ 
& $ - \frac{1}{2} \times \left\{%
\begin{array}{ll}
    0 \\
    0 \\
    -5/4 \\
    5/4 \\
\end{array}%
\right.$ 
\rule{0mm}{5mm} \\ [0.4em]
\hline
\end{tabular}
%\end{center}
\caption{Functions $B_{\Phi}$ and $B'_{\Phi}$, as defined via
Eq.~(\ref{scalar_self_gauge13}), associated with the gauge
contributions to the scalar self-energies, corresponding to diagrams
$(b)$--$(f)$ in Fig.~\ref{fig:Diagramsscalar}, in the case where the
scalars satisfy $n=3$ boundary conditions.  The four lines in each
diagram correspond to cases 1a, 1b, 2a and 2b of Eqs.~(\ref{cases})
and (\ref{cases2}).  For $n=1$, the roles of $B_{\Phi}$ and
$B'_{\Phi}$ are interchanged.}
\label{tableABscalar_n13}
}%
It remains to consider the gauge contributions to scalars satisfying
$n=1$ or $n=3$ boundary conditions.  In these cases we write the
results for diagrams $(b)$--$(f)$ in the form
\beq
- i \, \frac{g_{4}^{2}}{16\pi^{2}} \, C_{2}(\Phi) \, \delta_{ab}
\,\Gamma\!\left(\frac{\epsilon}{2}\right) M_{m,l} M_{m',l'} \left\{
B_{\Phi} r_{m,l} r^{*}_{m',l'} + B'_{\Phi} r^{*}_{m,l} r_{m',l'} 
\right\}~.
\label{scalar_self_gauge13}
\eeq
Notice that there is no $p^{2}$ term, as expected from the fact that
the 4D-like kinetic operator for fields satisfying these boundary
conditions vanishes at the conical singularities.  A straightforward
calculation gives the scalar coefficients $B_{\Phi}$ and $B'_{\Phi}$
as summarized in Table~\ref{tableABscalar_n13} for the $n=3$ case.
Adding the various contributions, we obtain
\beqa
B_{\Phi} = \left\{%
\begin{array}{c}
    15/4 \\
    2 \\
    23/8 \\
    23/8 \\
\end{array}
\right. ~,
\hspace{1cm}
B'_{\Phi} = \left\{%
\begin{array}{c}
    0 \\
    0 \\
    7/8 \\
    -7/8\\
\end{array}
\right.~,
\label{ABphi_gauge13}
\eeqa
where the four lines correspond to cases 1a, 1b, 2a and 2b in
Eqs.~(\ref{cases}) and (\ref{cases2}).  We see that the
$\xi$-dependence cancels out.

For $n=1$ boundary conditions, the roles of $B_{\Phi}$ and $B'_{\Phi}$
in Eq.~(\ref{ABphi_gauge13}) are interchanged.

%%%%%%%%%%%%%%%%%%%%%%%%%%%%%%%%%%%%%
\subsubsection{Localized Operators}

Comparing to the general results of subsection~\ref{sec:localized}, we
see that the above expressions correspond to localized operators as in
Eq.~(\ref{localOp}).  Here we summarize the induced localized
operators in terms of the coefficients $r_{i} = \hat{r}_{i} L^{2}$,
$c_{i}$, $r'_{3} = \hat{r}'_{3} L^{2}$ and $c'_{3}$ of
Eq.~(\ref{localOp}), for the four possible boundary conditions
defining the scalar field:
\begin{itemize}
\item
For $n=0$:
\beqa
\hat{r}_{1} &=& \frac{5}{4} (\xi-3) \times \frac{g_{4}^{2}}{16\pi^{2}} 
\, C_{2}(\Phi) 
\,\Gamma\!\left(\frac{\epsilon}{2}\right)~, 
\hspace{1.2cm}
c_{1} = \frac{2}{5}~, 
\nonumber \\
\hat{r}_{2} &=& \frac{5}{16} (2\xi-3) \times \frac{g_{4}^{2}}{16\pi^{2}} 
\, C_{2}(\Phi) 
\,\Gamma\!\left(\frac{\epsilon}{2}\right)~, 
\hspace{1cm}
c_{2} = \frac{2}{5} \, \frac{(2\xi + 1)}{(2\xi - 3)}~, 
\label{rcn0}
\eeqa
while $\hat{r}_{3} = \hat{r}'_{3} = 0$.

\item
For $n=2$:
\beqa
c_{1} \hat{r}_{1} &=& \frac{1}{2} (\xi-3) \times \frac{g_{4}^{2}}{16\pi^{2}} 
\, C_{2}(\Phi) 
\,\Gamma\!\left(\frac{\epsilon}{2}\right)~, 
\nonumber \\
c_{2} \hat{r}_{2} &=& \frac{1}{8} (2\xi+1) \times \frac{g_{4}^{2}}{16\pi^{2}} 
\, C_{2}(\Phi) 
\,\Gamma\!\left(\frac{\epsilon}{2}\right)~, 
\label{rcn2}
\eeqa
while $\hat{r}_{3} = \hat{r}'_{3} = 0$.  In this case, no operators at
$(0,0)$ or $(L,L)$ are generated.

\item
For $n=3$:
\beqa
\hat{r}_{3} &=& \frac{23}{16} \times \frac{g_{4}^{2}}{16\pi^{2}} 
\, C_{2}(\Phi) 
\,\Gamma\!\left(\frac{\epsilon}{2}\right)~, 
\hspace{1cm}
c_{3} = \frac{14}{23}~, 
\nonumber \\
c'_{3} \hat{r}'_{3} &=& \frac{7}{8} \times \frac{g_{4}^{2}}{16\pi^{2}} 
\, C_{2}(\Phi) 
\,\Gamma\!\left(\frac{\epsilon}{2}\right)~, 
\label{rcn3}
\eeqa
while $\hat{r}_{1} = \hat{r}_{2} = 0$.  Notice that the operators
associated with $r'_{3}$ are generated only at $(0,L)$, not at $(0,0)$
or $(L,L)$.

\item
For $n=1$:
\beqa
c_{3} \hat{r}_{3} &=& \frac{7}{8} \times \frac{g_{4}^{2}}{16\pi^{2}} 
\, C_{2}(\Phi) 
\,\Gamma\!\left(\frac{\epsilon}{2}\right)~, 
\nonumber \\
\hat{r}'_{3} &=& \frac{23}{16} \times \frac{g_{4}^{2}}{16\pi^{2}} 
\, C_{2}(\Phi) 
\,\Gamma\!\left(\frac{\epsilon}{2}\right)~, 
\hspace{1cm}
c'_{3} = \frac{14}{23}~, 
\label{rcn1}
\eeqa
while $\hat{r}_{1} = \hat{r}_{2} = 0$.  Notice that the operators
associated with $r_{3}$ are generated only at $(0,L)$, not at $(0,0)$
or $(L,L)$.

\end{itemize}

%%%%%%%%%%%%%%%%%%%%%%%%%%%%%%%%%%%%%
\subsubsection{Yukawa Interactions}
\label{sec:ScalarYukawa}

We end this section by considering the effects of Yukawa interactions,
which as explained in Eq.~(\ref{Yukawa}), involve two 6D Weyl fermions
of opposite 6D chiralities.  We have 
\beqa
(a) &=& \left( \frac{1}{4} \right) |\lambda_{4}|^{2} 
\, \delta_{ab}
\sum_{m_{1},l_{1}} \sum_{m'_{1},l'_{1}}
\int \frac{d^{D}k}{(2\pi)^{D}}
{\rm{Tr}}\left\{ G_{k+p}^{\pm, (m_{1},l_{1}; m'_{1},l'_{1})} 
G_{k}^{\mp, (m'_{2},l'_{2}; m_{2},l_{2})} P_{\mp} \right\}~,
\hspace{8mm}
\label{S_fermionloop}
\eeqa
which can be easily evaluated using the identities (\ref{Pidentities})
and (\ref{Gammaidentity1})--(\ref{Gammaidentity3}) to simplify the
trace.

We restrict to the case where at least one of the 6D fermions running
in the loop has a zero-mode.  If the zero-mode is left-handed the
boundary conditions for the fermions in the loop are related to those
of the scalar by Eqs.~(\ref{YukawaExample}).  Analogous relations hold
if the zero-mode is right-handed. The boundary conditions obeyed by 
the scalar field are labeled by $n$.

Independently of the chirality of the fermion zero-mode, when $n=0$ or
$n=2$ we can write the result of the diagram as
\beq
i \, \frac{|\lambda_{4}|^{2}}{16\pi^{2}} \, \delta_{ab}
\,\Gamma\!\left(\frac{\epsilon}{2}\right) \left\{
A_{\Phi} p^{2} - B_{\Phi}  
\left( M^{2}_{m,l} + M^{2}_{m',l'} \right)
\right\}~,
\label{scalar_self_Yukawa}
\eeq
where the scalar coefficients $A_{\Phi}$ and $B_{\Phi}$ are given by
\beqa
\begin{array}{lclcl}
A_{\Phi} = 2~, & \hspace{5mm} & B_{\Phi} = 2~, & \hspace{5mm} & {\rm{for}}~n=0~,
\\ [0.3em]
A_{\Phi} = 0~, & & B_{\Phi} = 0~, & & {\rm{for}}~n=2~.
\end{array}
\eeqa
Notice that for $n=0$, this corresponds to localized operators as in 
Eq.~(\ref{localOp}), with 
\beqa
\hat{r}_{1} = \hat{r}_{2} &=& \frac{g_{4}^{2}}{16\pi^{2}} 
\, C_{2}(\Phi) 
\,\Gamma\!\left(\frac{\epsilon}{2}\right)~, 
\hspace{1.2cm}
c_{1} = c_{2} = 0~, 
\label{rcn0_Yukawa}
\eeqa
and $\hat{r}_{3} = \hat{r}'_{3} = 0$.

When $n=1$ or $n=3$ the result depends on whether \textit{both} 6D
fermions in the loop give rise to a 4D chiral zero-mode, or
\textit{only one} of them.  We write the results in the form
\beq
- i \, \frac{|\lambda_{4}|^{2}}{16\pi^{2}} \, \delta_{ab}
\,\Gamma\!\left(\frac{\epsilon}{2}\right) M_{m,l} M_{m',l'} \left\{
B_{\Phi} r_{m,l} r^{*}_{m',l'} + B'_{\Phi} r^{*}_{m,l} r_{m',l'} 
\right\}~.
\label{scalar_self_Yukawa13}
\eeq
Let us label the boundary conditions associated with the two fermions
participating in the Yukawa interaction, Eq.~(\ref{Yukawa}), by
$n^{\pm}_{iL}$ and $n^{\mp}_{iR}$, with $i=1,2$, as in
subsection~\ref{sec:Yukawa}.  For $n=3$, both fermions in the loop
have a zero mode when either $n^{+}_{1L} = n^{-}_{2L}=0$ or
$n^{-}_{1R} = n^{+}_{2R}=0$.  In these instances, we get
\beqa
\begin{array}{rclcrclcl}
B_{\Phi} &=&  \left\{%
\begin{array}{c}
    1 \\
    0 \\
    1/2 \\
    1/2 \\
\end{array}
\right.~, 
& \hspace{5mm} & 
B'_{\Phi} &=&  \left\{%
\begin{array}{c}
    0 \\
    0 \\
    -1/2 \\
    1/2 \\
\end{array}
\right.~, 
\end{array}
\hspace{1cm} {\rm for}~n=3~{\rm and~two~zero~modes}~,
\label{BBp3}
\eeqa
where the four lines correspond to cases 1a, 1b, 2a and 2b in
Eqs.~(\ref{cases}) and (\ref{cases2}).  For $n=1$, the two fermions in
the loop can have zero-modes simultaneously when $n^{+}_{1R} =
n^{-}_{2R}=0$ or $n^{-}_{1L} = n^{+}_{2L}=0$.  In such instances, one
obtains Eq.~(\ref{BBp3}) with $B_{\Phi}$ and $B'_{\Phi}$ interchanged.

The second category occurs when only one of the fermions has a
zero-mode.  For $n=3$, this happens when $n^{+}_{1R}=0$ or $n^{-}_{1L}
= 0$ or $n^{+}_{2L}=0$ or $n^{-}_{2R}=0$, while for $n=1$, it happens
when $n^{+}_{1L}=0$ or $n^{-}_{1R} = 0$ or $n^{+}_{2R}=0$ or
$n^{-}_{2L}=0$.  In all such instances, one gets the \textit{opposite}
sign of the contributions when two fermion zero-modes are
simultaneously present.

%%%%%%%%%%%%%%%%%%%%%%%%%%%%%%%%%%%%%
\subsection{Two-Point Function of the Spinless Adjoints }
\label{sec:Spinless_self}

We end our exploration of the one loop corrections in the chiral
square background by studying the two-point functions associated with
the spinless adjoints, $A_{\pm}$.

We need to consider two types of two-point functions: $\langle A_{+}
A_{+}^{\dagger} \rangle$ and $\langle A_{+} A_{+} \rangle$, which
together with their complex conjugates determine the structure of
localized counterterms needed to absorb the logarithmic divergences
that appear at one-loop.  As mentioned in the previous section, the
fact that the spinless adjoints satisfy $n=1$ or $n=3$ boundary
conditions ensures that no localized mass terms are generated.
However, one can generate a localized tadpole term, proportional to 
$F_{45}$, the field strength with indices in the extra dimensions 
\cite{Csaki:2002ur}.
The coefficient is quadratically divergent, but is proportional to
the trace of the generators associated with the field in the loop.
Since for the standard model gauge group and field content
all such traces vanish, we do not compute this tadpole in what 
follows. It is easy to do with the technology we have developed.

%%%%%%%%%%%%%%%%%%%%%%%%%%%%%%%%%%%%%
\subsubsection{Gauge Interactions}
\label{sec:Spinless_self_gauge}

\FIGURE[t]{
\vspace*{-5mm}
\centerline{
   \resizebox{14cm}{!}{\includegraphics{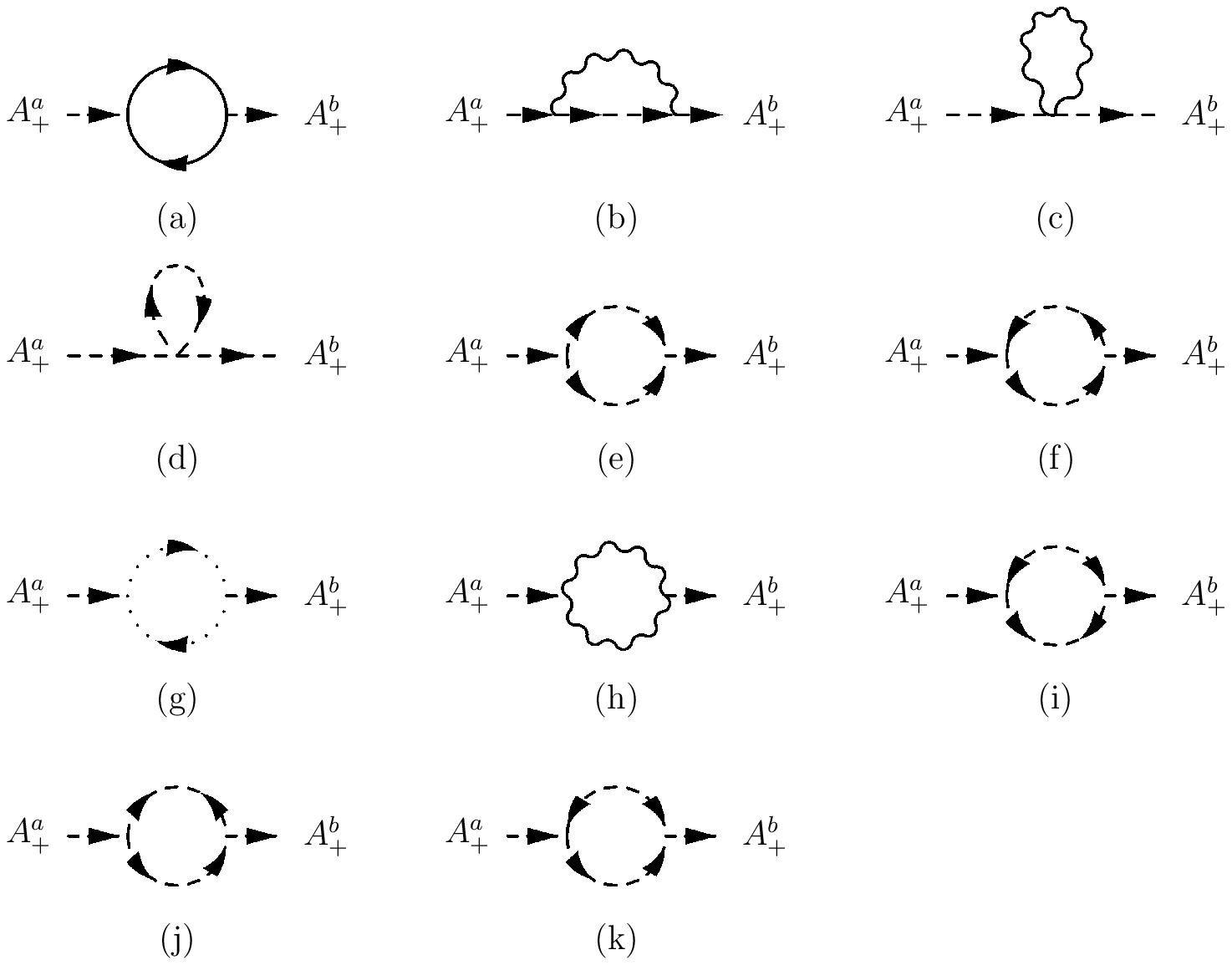}}
}
\caption{One-loop contribution to $\langle A_{+} A_{+}^{\dagger}
\rangle$.  The dashed lines with two arrows represent the propagation
of $A_{+}$.}
\label{fig:Diagramsscalar1}
}
We show the diagrams contributing to the two point function $\langle
A_{+} A_{+}^{\dagger} \rangle$ in Figure~\ref{fig:Diagramsscalar1},
and those contributing to $\langle A_{+} A_{+} \rangle$ in
Figure~\ref{fig:Diagramsscalar2}.  We start with the diagrams in
Figure~\ref{fig:Diagramsscalar1}.

Diagrams $(b)$--$(d)$ are trivially related to the corresponding
diagrams encountered in our study of general scalar fields in
section~\ref{sec:scalar_self}.  Diagram $(b)$ in
Figure~\ref{fig:Diagramsscalar1} is given by $(1/2)^{2}$ times
expression (\ref{S_gaugetrilinear}) with $G_{n} \rightarrow G_{++}$,
and $C_{2}(\Phi) \rightarrow C_{2}(A)$.  Similarly, diagram $(c)$ is
given by $1/2$ times expression (\ref{S_gaugequartic}) with
$C_{2}(\Phi) \rightarrow C_{2}(A)$.  Diagram $(d)$ is given by $1/4$
of expression (\ref{S_scalarquartic}).  Diagrams $(e)$ and $(f)$ are
given by
\beqa
(e) &=& - \left( \frac{1}{4} \right)^{3} \frac{g_{4}^{2}}{2} \, C_{2}(A) \, \delta_{ab} 
\sum_{m_{1},l_{1}} \sum_{m'_{1},l'_{1}}
\int \frac{d^{D}k}{(2\pi)^{D}}
\left[ r^{*}_{m_{1},l_{1}} M_{m_{1},l_{1}} + 
r^{*}_{m_{2},l_{2}} M_{m_{2},l_{2}} \right]
\nonumber \\ [0.5em]
& & \mbox{} \times
G_{++,k+p}^{(m_{1},l_{1}; m'_{1},l'_{1})}
\left[ r_{m'_{1},l'_{1}} M_{m'_{1},l'_{1}} + 
r_{m'_{2},l'_{2}} M_{m'_{2},l'_{2}} \right]
G_{++,k}^{(m_{2},l_{2}; m'_{2},l'_{2})}~,
\label{Ap_trilinear1}
\\ [0.75em]
(f) &=& - \left( \frac{1}{4} \right)^{3} g_{4}^{2} \, C_{2}(A) \, \delta_{ab} 
\sum_{m_{1},l_{1}} \sum_{m'_{1},l'_{1}}
\int \frac{d^{D}k}{(2\pi)^{D}}
\left[ r_{m,l} M_{m,l} + r_{m_{1},l_{1}} M_{m_{1},l_{1}} \right]
\nonumber \\ [0.5em]
& & \mbox{} \times
G_{--,k+p}^{(m_{1},l_{1}; m'_{1},l'_{1})}
\left[ r^{*}_{m',l'} M_{m',l'} + r^{*}_{m'_{1},l'_{1}} M_{m'_{1},l'_{1}} 
\right]
G_{++,k}^{(m_{2},l_{2}; m'_{2},l'_{2})}~.
\label{Ap_trilinear2}
\eeqa
Finally, there is a diagram involving the ghost fields,
\beqa
(g) &=& \left( \frac{1}{4} \right)^{2} \xi^{2} g_{4}^{2} \, C_{2}(A) \, \delta_{ab}
\sum_{m_{1},l_{1}} \sum_{m'_{1},l'_{1}}
r_{m_{2},l_{2}} M_{m_{2},l_{2}} r^{*}_{m'_{1},l'_{1}} 
M_{m'_{1},l'_{1}}
\nonumber \\ [0.1em]
& & \hspace{5cm} \mbox{} \times
\int \frac{d^{D}k}{(2\pi)^{D}}
G_{\xi,k+p}^{(m_{1},l_{1}; m'_{1},l'_{1})}
G_{\xi,k}^{(m'_{2},l'_{2}; m_{2},l_{2})}~,
\label{Ap_ghostloop}
\eeqa
and one involving the interaction between two gauge fields and a
single spinless adjoint given by
\beqa
(h) &=& - \left( \frac{1}{4} \right)^{2} \frac{g_{4}^{2}}{2} \, C_{2}(A) \, \delta_{ab} 
\sum_{m_{1},l_{1}} \sum_{m'_{1},l'_{1}}
\int \frac{d^{D}k}{(2\pi)^{D}}
\left[ r_{m_{1},l_{1}} M_{m_{1},l_{1}} + r_{m_{2},l_{2}} 
M_{m_{2},l_{2}} \right]
\nonumber \\ [0.3em]
& & \hspace{2cm} \mbox{} \times
G_{k+p}^{\mu\nu(m_{1},l_{1}; m'_{1},l'_{1})}
\left[ r^{*}_{m'_{1},l'_{1}} M_{m'_{1},l'_{1}} + 
r^{*}_{m'_{2},l'_{2}} M_{m'_{2},l'_{2}} \right]
G_{\mu\nu,k}^{(m'_{2},l'_{2}; m_{2},l_{2})}~.
\label{Ap_gaugeNGB}
\eeqa
These last two diagrams are present to account for the would-be
Goldstone modes contained in $A_{+}$ that realize the Higgs mechanism
at each spin-1 KK level.

Notice that diagrams $(i)$, $(j)$ and $(k)$ in
Figure~\ref{fig:Diagramsscalar1}, involving the propagators $G_{+-}$
or $G_{-+}$, are finite as a result of a cancellation between the two
real degrees of freedom in $A_{+}$, so we need not consider them.

%%%%%%%%%%%%%%%%%%%%%%%%%%%%%%%%%%%%%%
%%% Table for scalar functions B and B' for A_+ A_-
%%%%%%%%%%%%%%%%%%%%%%%%%%%%%%%%%%%%%%
\TABLE[t]{
%\begin{center}
\begin{tabular}{|c||c|c|}
  \hline
& $B_{+}$ & $B'_{+}$
\rule{0mm}{5mm} \\ [0.4em]
\hline
\hline
(b) & $ \frac{1}{4} \xi (1 + \xi) \times \left\{%
\begin{array}{ll}
    1/4 \\
    0 \\
    1/8 \\
    1/8 \\
\end{array}%
\right. 
\mbox{} - \frac{1}{2} \xi^{2} \times \left\{%
\begin{array}{ll}
    5/4 \\
    1 \\
    9/8 \\
    9/8 \\
\end{array}%
\right.$ 
& $ \frac{1}{4} \xi (1 + \xi) \times \left\{%
\begin{array}{ll}
    0 \\
    0 \\
    1/8 \\
    -1/8 \\
\end{array}%
\right. 
\mbox{} - \frac{1}{2} \xi^{2} \times \left\{%
\begin{array}{ll}
    0 \\
    0 \\
    1/8 \\
    -1/8 \\
\end{array}%
\right.$ 
\rule{0mm}{5mm} \\ [0.4em]
\hline
(c) & $ \frac{1}{2} (3 + \xi^{2}) \times \left\{%
\begin{array}{ll}
    5/4 \\
    1 \\
    9/8 \\
    9/8 \\
\end{array}%
\right.$ 
& $ \frac{1}{2} (3 + \xi^{2}) \times \left\{%
\begin{array}{ll}
    0 \\
    0 \\
    1/8 \\
    -1/8 \\
\end{array}%
\right.$ 
\rule{0mm}{5mm} \\ [0.4em]
\hline
(d) & $ - \frac{1}{4} (1 + \xi) \times \left\{%
\begin{array}{ll}
    1/4 \\
    0 \\
    1/8 \\
    1/8 \\
\end{array}%
\right.$ 
& $ - \frac{1}{4} (1 + \xi) \times \left\{%
\begin{array}{ll}
    0 \\
    0 \\
    1/8 \\
    -1/8 \\
\end{array}%
\right.$ 
\rule{0mm}{5mm} \\ [0.4em]
\hline
(e) & $ - \frac{1}{8} \times \left\{%
\begin{array}{ll}
    -5 \\
    -4 \\
    -9/2 \\
    -9/2 \\
\end{array}%
\right.$ 
& $0$ 
\rule{0mm}{5mm} \\ [0.4em]
\hline
(f) & $ - \frac{1}{4} \times \left\{%
\begin{array}{ll}
    -21/4 \\
    -3 \\
    -33/8 \\
    -33/8 \\
\end{array}%
\right.$ 
& $ - \frac{1}{4} \times \left\{%
\begin{array}{ll}
    0 \\
    0 \\
    -1/8 \\
    1/8 \\
\end{array}%
\right.$ 
\rule{0mm}{5mm} \\ [0.4em]
\hline
(g) & $ \frac{1}{4} \xi^{2} \times \left\{%
\begin{array}{ll}
    -5/4 \\
    -1 \\
    -9/8 \\
    -9/8 \\
\end{array}%
\right.$ 
& $ \frac{1}{4} \xi^{2} \times \left\{%
\begin{array}{ll}
    0 \\
    0 \\
    1/8 \\
    -1/8 \\
\end{array}%
\right.$ 
\rule{0mm}{5mm} \\ [0.4em]
\hline
(h) & $ - \frac{1}{8} \left( 3 + \xi^{2} \right) \times \left\{%
\begin{array}{ll}
    -2 \\
    -2 \\
    -2 \\
    -2 \\
\end{array}%
\right.$ 
& $ - \frac{1}{8} \left( 3 + \xi^{2} \right)  \times \left\{%
\begin{array}{ll}
    0 \\
    0 \\
    1/2 \\
    -1/2 \\
\end{array}%
\right.$ 
\rule{0mm}{5mm} \\ [0.4em]
\hline
\end{tabular}
%\end{center}
\caption{Functions $B_{+}$ and $B'_{+}$, as defined via
Eq.~(\ref{Ap_self_gauge}), associated with the gauge contributions to
the two-point function $\langle A_{+} A_{+}^{\dagger} \rangle$,
corresponding to diagrams $(b)$--$(h)$ in
Fig.~\ref{fig:Diagramsscalar1}.  The four lines in each diagram
correspond to cases 1a, 1b, 2a and 2b of Eqs.~(\ref{cases}) and
(\ref{cases2}).}
\label{tableBBpspinless1}
}%

We write the results for diagrams $(b)$--$(h)$ in the form
\beq
- i \, \frac{g_{4}^{2}}{16\pi^{2}} \, C_{2}(A) \, \delta_{ab}
\,\Gamma\!\left(\frac{\epsilon}{2}\right) M_{m,l} M_{m',l'} \left\{
B_{+} r_{m,l} r^{*}_{m',l'} + B'_{+} r^{*}_{m,l} r_{m',l'} 
\right\}~.
\label{Ap_self_gauge}
\eeq
Notice that there is no $p^{2}$ term, as expected from the fact that
the 4D kinetic operator for fields satisfying $n=3$ boundary
conditions vanishes at the conical singularities.  A straightforward
calculation gives the scalar coefficients $B_{+}$ and $B'_{+}$ as
summarized in Table~\ref{tableBBpspinless1}.  Adding the various
contributions, we obtain
\beqa
B_{+} = \left\{%
\begin{array}{c}
    9/2 \\
    7/2 \\
    4 \\
\end{array}
\right. ~,
\hspace{1cm}
B'_{+} = 0~,
\label{ABspinless_gauge}
\eeqa
where the three lines correspond to the cases listed in
Eq.~(\ref{cases}).  We see that the $\xi$-dependence disappears and
that $B'_{+}$ vanishes.  Notice that the vanishing of $B'_{+}$ is
essential to get a rotationally invariant (in the plane of the extra
dimensions) structure for the induced localized operators: a
nonvanishing result would lead to operators of the form $(\partial_{+}
A_{+})(\partial_{-} A_{-})$.  $B_{+}$, on the other hand, leads to
localized operators of the form $(\partial_{-} A_{+})(\partial_{+}
A_{-})$, which are rotationally invariant.

Consider now the diagrams in Figure~\ref{fig:Diagramsscalar2} for the
$\langle A_{+} A_{+} \rangle$ two-point function.  Diagram $(m)$ can
be obtained from diagram $(b)$ in Figure~\ref{fig:Diagramsscalar1},
with the replacement $G_{++} \rightarrow - G_{-+}$, the minus sign
coming from the ordering of the 4D momenta.  Diagram $(n)$ can be
obtained from Eq.~(\ref{Ap_ghostloop}) with $r_{m_{2},l_{2}}
\rightarrow r^{*}_{m_{2},l_{2}}$.  Similarly, diagram $(o)$ can be
obtained from Eq.~(\ref{Ap_gaugeNGB}) by making the replacements
$r_{m_{1},l_{1}} \rightarrow r^{*}_{m_{1},l_{1}}$ and $r_{m_{2},l_{2}}
\rightarrow r^{*}_{m_{2},l_{2}}$.  Diagram $(p)$ can be obtained from
diagram $(d)$ in Figure~\ref{fig:Diagramsscalar1}, with the
replacement $G_{++} \rightarrow G_{-+}$.  Finally, diagram $(q)$ can
be obtained from Eq.~(\ref{Ap_trilinear2}) by making the replacements
$r_{m,l} \rightarrow - r^{*}_{m,l}$ and $r_{m_{1},l_{1}} \rightarrow
r^{*}_{m_{2},l_{2}}$.  It is easy to see that diagrams $(r)$, $(s)$
and $(t)$ in Figure~\ref{fig:Diagramsscalar2} are finite as a result
of a cancellation between the two real scalar degrees of freedom in
$A_{+}$, and we do not consider them in the following.

\FIGURE[t]{
\vspace*{-5mm}
\centerline{
   \resizebox{14cm}{!}{\includegraphics{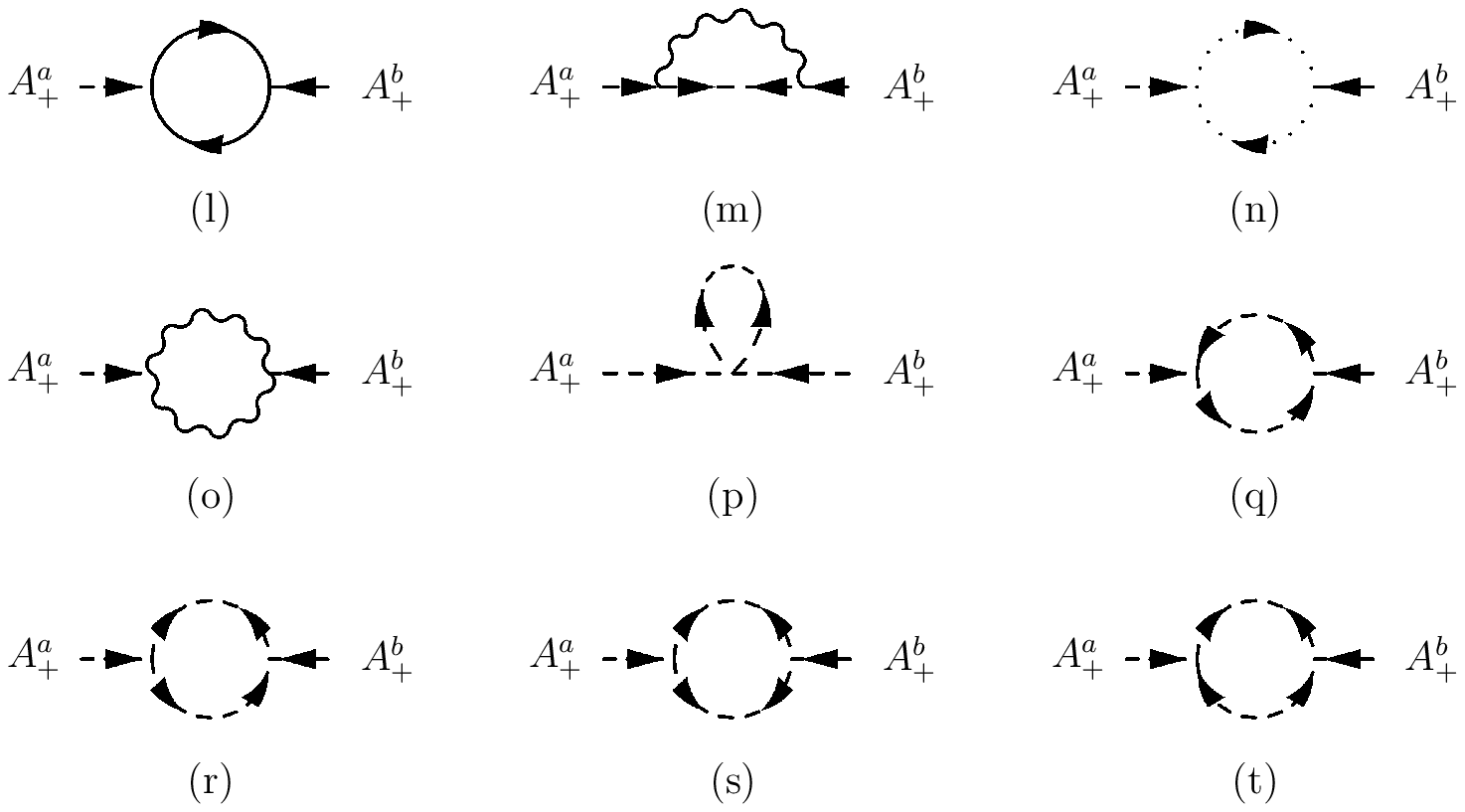}}
}
\caption{One-loop contribution to $\langle A_{+} A_{+} \rangle$.  The
dashed lines with two arrows represent the propagation of $A_{+}$.}
\label{fig:Diagramsscalar2}
}

We write the results for diagrams $(m)$--$(q)$ in the form
\beq
- i \, \frac{g_{4}^{2}}{16\pi^{2}} \, C_{2}(A) \, \delta_{ab}
\,\Gamma\!\left(\frac{\epsilon}{2}\right) 
r^{*}_{m,l} r^{*}_{m',l'} M_{m,l} M_{m',l'} \tilde{B}_{+} ~.
\label{ApAp_self_gauge}
\eeq
A straightforward calculation gives the scalar coefficients
$\tilde{B}_{+}$ as summarized in Table~\ref{tableABspinles2}.  Adding
the various contributions, we obtain
\beqa
\tilde{B}_{+} = \left\{%
\begin{array}{c}
    -9/2 \\
    -7/2 \\
    -4 \\
\end{array}
\right. ~,
\label{ABspinless_gauge2}
\eeqa
where the three lines correspond to the cases listed in
Eq.~(\ref{cases}).  Again, the $\xi$ dependence cancels out, as it
should.  Furthermore, we see from Eqs.~(\ref{ABspinless_gauge}) and
(\ref{ABspinless_gauge2}) that $\tilde{B}_{+} = - B_{+}$, which is
essential to obtain a gauge invariant structure for the induced
localized operators, as we discuss in
subsection~\ref{sec:Spinless_localized}.

%%%%%%%%%%%%%%%%%%%%%%%%%%%%%%%%%%%%%%
%%% Table for scalar functions B and B' for A_+ A_+
%%%%%%%%%%%%%%%%%%%%%%%%%%%%%%%%%%%%%%
\TABLE[t]{
%\begin{center}
\begin{tabular}{ccccc}
\begin{tabular}[t]{|c||c|}
  \hline
& $\tilde{B}_{+}$
\rule{0mm}{5mm} \\ [0.4em]
\hline
\hline
(m) & $ - \frac{1}{4} \xi (1 - \xi) \times \left\{%
\begin{array}{ll}
    5/2 \\
    2 \\
    9/4 \\
\end{array}%
\right.$ 
\rule{0mm}{5mm} \\ [0.4em]
\hline
(p) & $ - \frac{1}{4} (1 - \xi) \times \left\{%
\begin{array}{ll}
    5/2 \\
    2 \\
    9/4 \\
\end{array}%
\right.$ 
\rule{0mm}{5mm} \\ [0.4em]
\hline
\end{tabular}
& \hspace{-5mm} &
\begin{tabular}[t]{|c||c|}
  \hline
& $\tilde{B}_{+}$
\rule{0mm}{5mm} \\ [0.4em]
\hline
\hline
(n) & $ \frac{1}{4} \xi^{2} \times \left\{%
\begin{array}{ll}
    1 \\
    1 \\
    1 \\
\end{array}%
\right.$ 
\rule{0mm}{5mm} \\ [0.4em]
\hline
(q) & $ \frac{1}{4} \times \left\{%
\begin{array}{ll}
    -5 \\
    -3 \\
    -4 \\
\end{array}%
\right.$ 
\rule{0mm}{5mm} \\ [0.4em]
\hline
\end{tabular}
& \hspace{-5mm} &
\begin{tabular}[t]{|c||c|}
  \hline
& $\tilde{B}_{+}$
\rule{0mm}{5mm} \\ [0.4em]
\hline
\hline
(o) & $ - \frac{1}{8} \left( 3 + \xi^{2} \right) \times \left\{%
\begin{array}{ll}
    7 \\
    6 \\
    13/2 \\
\end{array}%
\right.$ 
\rule{0mm}{5mm} \\ [0.4em]
\hline
\end{tabular}
\end{tabular}
%\end{center}
\caption{Scalar functions $\tilde{B}_{+}$, as defined via
Eq.~(\ref{ApAp_self_gauge}), associated with the gauge contributions
to the two-point function $\langle A_{+} A_{+} \rangle$, corresponding
to diagrams $(m)$--$(q)$ in Fig.~\ref{fig:Diagramsscalar2}.  The three
lines in each diagram correspond to the cases listed in
Eq.~(\ref{cases}).}
\label{tableABspinles2}
}%
%

%%%%%%%%%%%%%%%%%%%%%%%%%%%%%%%%%%%%%
\subsubsection{Fermions and Spinless Adjoints}
\label{sec:Spinless_self_fermion}

Before studying the structure of localized KK-number violating terms
associated with the spinless adjoints, we compute the effect of the
interactions of the spinless adjoints with scalar and fermionic
matter.  In this subsection we give the result of the fermion loops,
and in the next we consider scalar loops.

The contribution due to fermions to the two-point function $\langle
A_{+} A^{\dagger}_{+} \rangle$ is given by (see
Fig.~\ref{fig:Diagramsscalar1})
\beqa
(a) &=& - \left( \frac{1}{4} \right) g_{4}^{2} \, i^{2} 
\, T(\Psi) \delta_{ab}
\sum_{m_{1},l_{1}} \sum_{m'_{1},l'_{1}}
\int \frac{d^{D}k}{(2\pi)^{D}}
{\rm{Tr}}\left\{ \Gamma^{+} G_{k+p}^{\pm, (m_{1},l_{1}; m'_{1},l'_{1})} 
\Gamma^{-} G_{k}^{\pm, (m'_{2},l'_{2}; m_{2},l_{2})} P_{\mp} \right\}~,
\nonumber \\ [-0.4em]
\label{Ap_fermionloop}
\eeqa
where ${\rm{Tr}}(T^{a} T^{b}) = T(\Psi) \delta_{ab}$.  Assuming that
the fermion has a zero-mode (of any 4D chirality), we obtain
\beq
- i \, \frac{g_{4}^{2}}{16\pi^{2}} \, T(\Psi) \, \delta_{ab}
\,\Gamma\!\left(\frac{\epsilon}{2}\right) 
r_{m,l} r^{*}_{m',l'} M_{m,l} M_{m',l'} \, B_{+}~,
\label{Ap_self_fermionLoop}
\eeq
with $B_{+} = -2$. 

The fermion loop $(l)$ in Fig.~\ref{fig:Diagramsscalar2} is given by
Eq.~(\ref{Ap_fermionloop}) with $\Gamma^{+} \rightarrow \Gamma^{-}$.
Assuming that the fermion has a zero-mode (of any 4D chirality), we
obtain
\beq
- i \, \frac{g_{4}^{2}}{16\pi^{2}} \, T(\Psi) \, \delta_{ab}
\,\Gamma\!\left(\frac{\epsilon}{2}\right) 
r^{*}_{m,l} r^{*}_{m',l'} M_{m,l} M_{m',l'} \tilde{B}_{+} ~,
\label{ApAp_self_fermionLoop}
\eeq
with $\tilde{B}_{+} = 2$.  As for the diagrams arising from the gauge
self-interactions, the fermion loop gives $\tilde{B}_{+} = - B_{+}$,
which implies a gauge invariant counterterm.

\FIGURE[b]{
\vspace*{-8mm}
\centerline{
   \resizebox{13cm}{!}{\includegraphics{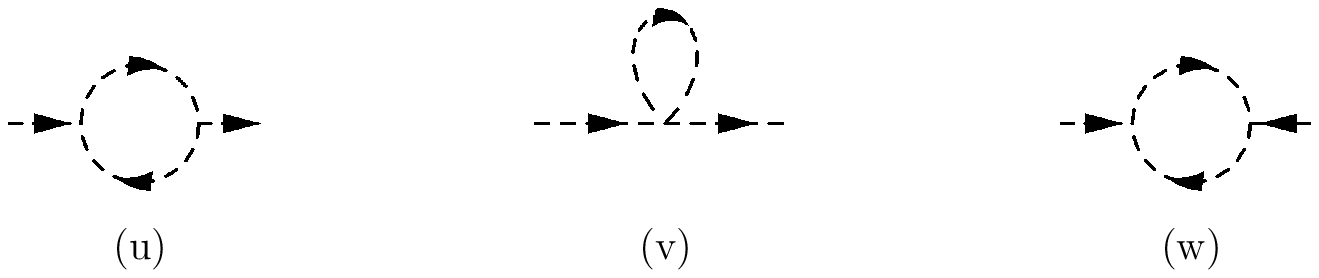}}
}
\caption{One-loop contribution to $\langle A_{+} A_{+}^{\dagger}
\rangle$ and $\langle A_{+} A_{+} \rangle$ due to scalars interacting
with the spinless adjoints via the 6D gauge interactions of
Figure~\ref{fig:Feynman4}.}
\label{fig:Diagramsscalar3}
}
%

%%%%%%%%%%%%%%%%%%%%%%%%%%%%%%%%%%%%%
\subsubsection{Scalars and Spinless Adjoints}
\label{sec:Spinless_self_scalars}

We finally consider the effect of scalar fields interacting with
spinless adjoints via the vertices of Figure~\ref{fig:Feynman4}.  As
shown in Figure~\ref{fig:Diagramsscalar3}, there is a diagram
involving a trilinear interaction,
\beqa
(u) &=& - \left( \frac{1}{4} \right)^{2} g_{4}^{2} \, T(\Phi) \, \delta_{ab} 
\sum_{m_{1},l_{1}} \sum_{m'_{1},l'_{1}}
\int \frac{d^{D}k}{(2\pi)^{D}}
\left[ r_{m_{1},l_{1}} M_{m_{1},l_{1}} + 
r_{m_{2},l_{2}} M_{m_{2},l_{2}} \right]
\nonumber \\ [0.5em]
& & \mbox{} \times
G_{k+p,n}^{(m_{1},l_{1}; m'_{1},l'_{1})}
\left[ r^{*}_{m'_{1},l'_{1}} M_{m'_{1},l'_{1}} + 
r^{*}_{m'_{2},l'_{2}} M_{m'_{2},l'_{2}} \right]
G_{k,n}^{(m'_{2},l'_{2}; m_{2},l_{2})}~,
\label{Ap_self_scalartrilinear}
\eeqa
and one involving a quartic interaction:
\beqa
(v) &=& - i g_{4}^{2} \, T(\Phi) \, \delta_{ab}
\sum_{m_{1},l_{1}}
\int \frac{d^{D}k}{(2\pi)^{D}} \, G_{k,n}^{(m_{1},l_{1}; m'_{1},l'_{1})}~,
\label{Ap_self_scalarquartic}
\eeqa
where ${\rm{Tr}}(T^{a} T^{b}) = T(\Phi) \delta_{ab}$.
%
%%%%%%%%%%%%%%%%%%%%%%%%%%%%%%%%%%%%%%
%%% Table for scalar functions B and B' for A_+ A_- with scalar matter
%%%%%%%%%%%%%%%%%%%%%%%%%%%%%%%%%%%%%%
\TABLE[t]{
%\begin{center}
\begin{tabular}{ccc}
\begin{tabular}{|c||rccc|rccc|}
  \hline
   &  &  & $B_{+}$ & & & & $B'_{+}$ &
\rule{0mm}{5mm} \\ [0.4em]
\hline
\hline
& & $n=0$ & $n=1,3$ & $n=2$ & & $n=0$ & $n=1,3$ & $n=2$
\rule{0mm}{5mm} \\ [0.4em]
\hline
  (u) & $ - \frac{1}{4} \times \left\{%
\begin{array}{ll}
     \\
     \\
     \\
     \\
\end{array}%
\right.$ \hspace*{-8mm}
&
$\begin{array}{c}
    -2 \\
    -2 \\
    -2 \\
    -2 \\
\end{array}$%
&
$\begin{array}{c}
    0 \\
    0 \\
    0 \\
    0 \\
\end{array}$%
&
$\begin{array}{c}
     2 \\
     2 \\
     2 \\
     2 \\
\end{array}$%
& $ - \frac{1}{4} \times \left\{%
\begin{array}{c}
     \\
    \\
    \\
    \\
\end{array}%
\right.$ \hspace*{-8mm}
&
$\begin{array}{c}
    0 \\
    0 \\
    1/2 \\
    -1/2 \\
\end{array}$%
&
$\begin{array}{c}
    0 \\
    0 \\
    -1/2 \\
    1/2 \\
\end{array}$%
&
$\begin{array}{c}
    0 \\
    0 \\
    1/2 \\
    -1/2 \\
\end{array}$%
\rule{0mm}{5mm} \\ [0.4em]
\hline
  (v) & $ (-1) \times \left\{%
\begin{array}{ll}
     \\
     \\
     \\
     \\
\end{array}%
\right.$ \hspace*{-8mm}
&
$\begin{array}{c}
    -5/4 \\
    -1 \\
    -9/8 \\
    -9/8 \\
\end{array}$%
&
$\begin{array}{c}
    1/4 \\
    0 \\
    1/8 \\
    1/8 \\
\end{array}$%
&
$\begin{array}{c}
    3/4 \\
    1\\
    7/8 \\
    7/8 \\
\end{array}$%
& $ (-1) \times \left\{%
\begin{array}{c}
    \\
    \\
    \\
    \\
\end{array}%
\right.$ \hspace*{-8mm}
&
$\begin{array}{c}
    0 \\
    0 \\
    -1/8 \\
    1/8 \\
\end{array}$%
&
$\begin{array}{c}
    0 \\
    0 \\
    1/8 \\
    -1/8 \\
\end{array}$%
&
$\begin{array}{c}
    0 \\
    0 \\
    -1/8 \\
    1/8 \\
\end{array}$%
\rule{0mm}{5mm} \\ [0.4em]
\hline
\end{tabular}
\end{tabular}
%\end{center}
%
\caption{Functions $B_{+}$ and $B'_{+}$, as defined via
Eq.~(\ref{Ap_self_scalar}), associated with the scalar contributions
to the two-point function $\langle A_{+} A_{+}^{\dagger} \rangle$,
corresponding to diagrams $(u)$ and $(v)$ in
Fig.~\ref{fig:Diagramsscalar3}.  We give the results for scalars
satisfying any of the four types of boundary conditions, $n=0,1,2$ or
$3$.  The four lines in each diagram correspond to cases 1a, 1b, 2a
and 2b of Eqs.~(\ref{cases}) and (\ref{cases2}).}
\label{tableBBpspinless2}
}%
We can write these diagrams as
\beq
- i \, \frac{g_{4}^{2}}{16\pi^{2}} \, T(\Phi) \, \delta_{ab}
\,\Gamma\!\left(\frac{\epsilon}{2}\right) M_{m,l} M_{m',l'} \left\{
B_{+} r_{m,l} r^{*}_{m',l'} + B'_{+} r^{*}_{m,l} r_{m',l'} 
\right\}~,
\label{Ap_self_scalar}
\eeq
with the scalar coefficients $B_{+}$, $B'_{+}$ given in
Table~\ref{tableBBpspinless2}.  Adding the diagrams, we get for
scalars satisfying any of the four types of boundary conditions,
$n=0,1,2$ or $3$:
\beqa
\begin{array}{cccccccc}
& $n=0$ & & $n=1,3$ & & $n=2$
\rule{0mm}{5mm} \\ [0.4em]
B_{+} = 
\left\{%
\begin{array}{ll}
     \\
     \\
     \\
\end{array}%
\right. \hspace*{-3mm}
&
\begin{array}{c}
    7/4 \\
    3/2 \\
    13/8 \\
\end{array}
& &
\begin{array}{c}
    -1/4 \\
    0 \\
    -1/8 \\
\end{array}
& &
\begin{array}{c}
    -5/4 \\
    -3/2 \\
    -11/8 \\
\end{array}
&~, \hspace{1cm} & 
B'_{+} = 0~,
\end{array}
\label{BBpspinless_scalar}
\eeqa
so that again the rotationally violating contribution proportional to
$B'_{+}$ vanishes.

The contribution to the two-point function $\langle A_{+} A_{+}
\rangle$ from diagram $(w)$ in Figure~\ref{fig:Diagramsscalar3} is
given by Eq.~(\ref{Ap_self_scalartrilinear}) with $r_{m_{1},l_{1}}
\rightarrow r^{*}_{m_{1},l_{1}}$ and $r_{m_{2},l_{2}} \rightarrow
r^{*}_{m_{2},l_{2}}$, and the result can be written as
\beq
- i \, \frac{g_{4}^{2}}{16\pi^{2}} \, T(\Phi) \, \delta_{ab}
\,\Gamma\!\left(\frac{\epsilon}{2}\right) 
r^{*}_{m,l} r^{*}_{m',l'} M_{m,l} M_{m',l'} \tilde{B}_{+}~,
\label{ApAp_self_scalarloop}
\eeq
with $\tilde{B}_{+} = - B_{+}$, and $B_{+}$ given in
Eq.~(\ref{BBpspinless_scalar}).

%%%%%%%%%%%%%%%%%%%%%%%%%%%%%%%%%%%%%
\subsubsection{Localized operators}
\label{sec:Spinless_localized}

We computed in the previous subsections the contributions to the
spinless adjoint two-point functions due to the gauge
self-interactions, as well as to the interactions with fermions and
scalars arising from the 6D gauge interactions.  We showed that one
obtains the structure
\beqa
\langle A^{(m',l')}_{+} A^{(m,l)\dagger}_{+} \rangle & \sim &
B_{+} r_{m,l} r^{*}_{m',l'} M_{m,l} M_{m',l'}~,
\nonumber \\ [0.5em]
\langle A^{(m',l')}_{+} A^{(m,l)}_{+} \rangle & \sim &
\tilde{B}_{+} r^{*}_{m,l} r^{*}_{m',l'} M_{m,l} M_{m',l'}~,
\label{Ap_self_summary}
\eeqa
with $\tilde{B}_{+} = - B_{+}$.  It follows from this and the
hermitian conjugates of relations (\ref{Ap_self_summary}), that the
logarithmic divergences require the gauge invariant localized
counterterm
\beq
\frac{1}{4} \, \delta_{c_{+}}\!(z) \times \left(- \frac{1}{2} \, \hat{r}_{+} L^{2} 
\, F^{a}_{45} F^{a}_{45} \right)~,
\label{F45localized}
\eeq
where $\delta_{c}(z)$ was defined in Eq.~(\ref{loc_kkp}), the factor
of $1/4$ accounts for universal KK wavefunction enhancements, and
\beqa
- \frac{1}{2} F^{2}_{45} &=& \frac{1}{8} F^{2}_{+-} 
\nonumber \\
&=&  \frac{1}{8} \left[ \partial_{+} A_{-} - \partial_{-} A_{+} \right]^{2}~.
\eeqa
Recalling the KK decompositions $A^{j,k}_{+}(x,z) \sim -
A^{j,k}_{+}(x) f^{j,k}_{3}(z)$, $A^{j,k}_{-}(x,z) \sim A^{j,k}_{-}(x)
f^{j,k}_{1}(z)$, and using Eqs.~(\ref{delf}), one can see that the
tree-level contribution to $\langle A^{(m',l')}_{+}
A^{(m,l)\dagger}_{+} \rangle$ associated with the vertex
(\ref{F45localized}) is
\beq
- \frac{i}{4} \, r_{m,l} r^{*}_{m',l'}  M_{m,l} M_{m',l'}
\times \left\{%
\begin{array}{c}
    (2+c_{+}) \, \hat{r}_{+}  \\
    (2-c_{+}) \, \hat{r}_{+}  \\
    2 \, \hat{r}_{+}  \\
\end{array}%
\right.~, 
\label{F45rule}
\eeq
where the three lines correspond to the cases listed in
Eq.~(\ref{cases}).  From this we can read the one-loop contributions
to $\hat{r}_{+}$ from the results of the previous subsections.  The
gauge self-interactions, Eqs.~(\ref{Ap_self_gauge}),
(\ref{ABspinless_gauge}), (\ref{ApAp_self_gauge}) and
(\ref{ABspinless_gauge2}), give a contribution
\beqa
\hat{r}_{+} &=& 8 \times \frac{g_{4}^{2}}{16\pi^{2}} \, C_{2}(A)
\,\Gamma\!\left(\frac{\epsilon}{2}\right)~,
\hspace{1cm}
c_{+} = \frac{1}{4}~,
\label{rcSpinless_gauge}
\eeqa
the fermions [see Eqs.~(\ref{Ap_self_fermionLoop}),
and (\ref{ApAp_self_fermionLoop})] give
\beqa
\hat{r}_{+} = - 4 \times \frac{g_{4}^{2}}{16\pi^{2}} \, T(\Psi)
\,\Gamma\!\left(\frac{\epsilon}{2}\right)~,
\hspace{1cm}
c_{+} = 0~,
\label{rcSpinless_fermion}
\eeqa
and the scalars [see Eqs.~(\ref{Ap_self_scalar}),
(\ref{BBpspinless_scalar}) and (\ref{ApAp_self_scalarloop})] give
\beqa
\hat{r}_{+} = \frac{13}{4} \times \frac{g_{4}^{2}}{16\pi^{2}} \, T(\Phi)
\,\Gamma\!\left(\frac{\epsilon}{2}\right)~,
\hspace{1cm}
c_{+} = \frac{2}{13}~.
\label{rcSpinless_scalar}
\eeqa
The previous results for the localized operators assume that the
fermion and scalar fields include a zero-mode in their KK towers.  The
coefficients of the localized operators induced by scalar fields
satisfying other boundary conditions can be easily read from the
results presented in subsection~\ref{sec:Spinless_self_scalars}.

%%%%%%%%%%%%%%%%%%%%%%%%%%%%%%%%%%%%%
\section{Summary and Conclusions}
\label{sec:conclusions}

We considered the one-loop structure of general field theories in six
dimensions, with two of the dimensions compactified on the ``chiral
square'' of Ref.~\cite{Dobrescu:2004zi}.  This compactification has a
very simple geometric description: start from a square region in the
plane and identify adjacent sides of the square.  This
compactification to four dimensions has the desirable property of
leading to a chiral four-dimensional theory, and is therefore
appropriate for phenomenological applications.  In fact, as shown in
\cite{Dobrescu:2004zi}, the construction is equivalent to a
$T^{2}/Z_{4}$ orbifold.  The geometric construction makes it clear
that there are three singular points with a conical symmetry.  Our
explicit one-loop computation shows that there are logarithmic
divergences that require the introduction of counterterms precisely at
these three points.  It also shows that the localized counterterms
obey a rotational symmetry, as expected from the conical nature of the
singularities.

We derived the propagators for spin-0, spin-1/2 and spin-1 fields in
momentum space and showed how to take into account the ``boundary
conditions'' that define the propagation of these fields on the chiral
square background.  Putting the information about the compactification
in the form of the propagators permits the use of vertices that
conserve momentum in the extra dimensions, and therefore allows us to
consider arbitrary interactions in a universal manner.

We also considered the 4D spin-0 fields that arise from the extra
dimensional components of 6D gauge fields.  Their interactions among
themselves and with other fermion and scalar fields are constrained by
the 6D gauge invariance.  These states are characteristic of the
present class of six-dimensional theories.  We find that the
self-interactions give a positive mass shift, while the gauge
interactions with fermions give a negative mass shift.  This is
similar to their spin-1 counterparts.  However, the numerical
coefficients are different.  When applied to the standard model field
content, one finds that the lightest KK particle is the spinless
adjoint associated with the hypercharge interactions. Thus, these 
scenarios give rise to a scalar dark matter candidate, with Yukawa-like 
couplings determined by the gauge interactions.

Our results can be summarized succinctly by giving the coefficients of
the various quadratic operators involving the given fields.  In order
to do so, we define the following shorthand notation for the various
kinetic term operators.  For the 6D gauge system we have two types of
kinetic terms:
\beqa
{\cal{O}}_{A} = -\frac{1}{4} F_{\mu\nu}^{a} F^{\mu\nu a}~,
& \hspace{5mm} &
{\cal{O}}_{45} = - \frac{1}{2} F_{45}^{a} F_{45}^{a}~.
\label{kineticOpsGauge} 
\eeqa 
For 6D Weyl fermions, $\Psi_{\pm}$, with a left-handed zero-mode, the
kinetic terms generated at the singularities have the form
\beqa
{\cal{O}}_{\Psi_{L}} = i \overline{\Psi}_{\pm} \Gamma^{\mu} P_{L} 
\partial_{\mu} \Psi_{\pm}~,
& \hspace{5mm} &
{\cal{O}}^{M}_{\Psi_{L}} = 
i \overline{\Psi}_{\pm} \Gamma^{\pm} P_{R} \partial_{\mp} \Psi_{\pm} 
+ {\rm h.c.}~,
\label{kineticOpsFermion} 
\eeqa 
while for 6D Weyl fermions with a right-handed zero-mode they have an
analogous structure with $P_{L} \leftrightarrow P_{R}$.  For scalar
fields there are four types of localized kinetic term operators as
shown in Eq.~(\ref{localOp}).  However, for scalars having a
zero-mode, i.e. satisfying $n=0$ boundary conditions, only two types
of kinetic terms are generated:
\beqa
{\cal{O}}_{\Phi} = \partial_{\mu} \Phi^{\dagger} \partial^{\mu} \Phi~,
& \hspace{5mm} &
{\cal{O}}^{M}_{\Phi} = 
\Phi^{\dagger}(\partial_{+} \partial_{-} \Phi) + {\rm{h.c.}}~,
\label{kineticOpsScalar} 
\eeqa 
In this case, there are also induced mass terms, $\Phi^{\dagger} \Phi$.

Assuming that the zero-mode fermion is left-handed, we found that the
quantum effects induce localized kinetic term operators at the points
$(0,0)$ and $(L,L)$, which we write as
\beqa
& & \frac{L^{2}}{4} \left[ \delta(x^4) \delta(x^5) + \delta(L - x^4)
\delta(L - x^5) \right] \times \left\{ \hat{r}^{A}_{1} {\cal{O}}_{A} 
+ \hat{r}^{45}_{1} {\cal{O}}_{45} 
\right.
\nonumber \\ [0.4em]
& & \hspace{6cm} \left. \mbox{}
+ \hat{r}^{\Psi}_{1} {\cal{O}}_{\Psi_{L}} + 
\hat{r}^{\Psi,M}_{1} {\cal{O}}^{M}_{\Psi_{L}}
+ \hat{r}^{\Phi}_{1} {\cal{O}}_{\Phi}
+ \hat{r}^{\Phi,M}_{1} {\cal{O}}^{M}_{\Phi}
\right\}~,
\label{localOps1_summary}
\eeqa
where, for convenience, we wrote an explicit factor of $1/4$ to
account for the KK wavefunction enhancements arising from
Eq.~(\ref{KKf}) evaluated at the singular points.  The coefficients of
the operators at these two conical singularities are found to be
identical, as required by KK-Parity.  If bare contributions at the
cutoff scale $\Lambda$ can be neglected, RG evolution induces
coefficients that can be read from Eqs.~(\ref{Arc_scalar}),
(\ref{Arc_fermion}), (\ref{Arc_gauge}), (\ref{Frc_gauge}),
(\ref{Frc_Yukawa}), (\ref{rcn0}), (\ref{rcn0_Yukawa}),
(\ref{rcSpinless_gauge}), (\ref{rcSpinless_fermion}) and
(\ref{rcSpinless_scalar}):
\beqa
\hat{r}^{A}_{1} &=& \frac{g^{2}_{4}}{16\pi^{2}} 
\ln \frac{\Lambda^2}{\mu^2} \left[ -\frac{14}{3} C_{2}(A) + 
\frac{2}{3} \sum_{\Psi} T(\Psi) + \frac{5}{12} \sum_{\Phi} T(\Phi) \right]~,
\nonumber \\ [0.4em]
\hat{r}^{45}_{1} &=& \frac{g^{2}_{4}}{16\pi^{2}} 
\ln \frac{\Lambda^2}{\mu^2} \left[ 8 \, C_{2}(A) - 
4 \sum_{\Psi} T(\Psi) + \frac{13}{4} \sum_{\Phi} T(\Phi) \right]~,
\nonumber \\ [0.4em]
\hat{r}^{\Psi}_{1} &=& \frac{1}{16\pi^{2}} 
\ln \frac{\Lambda^2}{\mu^2} \left[ - 4 \sum_{{\rm gauge}} g^{2}_{4} C_{2}(\Psi) 
+ \frac{5}{8} \sum_{i} \lambda^{2}_{4,i} \right]~,
\nonumber \\ [0.4em]
\hat{r}^{\Psi,M}_{1} &=& \frac{1}{16\pi^{2}} 
\ln \frac{\Lambda^2}{\mu^2} \left[ \frac{5}{8} \sum_{i} \lambda^{2}_{4,i} \right]~,
\label{r1} \\ [0.4em]
\hat{r}^{\Phi}_{1} &=& \frac{1}{16\pi^{2}} 
\ln \frac{\Lambda^2}{\mu^2} \left[ - \frac{15}{2} \sum_{{\rm gauge}} g^{2}_{4} C_{2}(\Phi) 
+ \sum_{i} \lambda^{2}_{4,i} \right]~,
\nonumber \\ [0.4em]
\hat{r}^{\Phi,M}_{1} &=& \frac{1}{16\pi^{2}} 
\ln \frac{\Lambda^2}{\mu^2} \left[ - \frac{45}{16} \sum_{{\rm gauge}} g^{2}_{4} C_{2}(\Phi) 
+ \sum_{i} \lambda^{2}_{4,i} \right]~,
\nonumber 
\eeqa
where $\mu$ is the renormalization scale, and $g_{4}$ and
$\lambda_{4,i}$ are the 4-dimensional gauge and Yukawa couplings,
respectively.  In the equations for $\hat{r}^{A}_{1}$ and
$\hat{r}^{45}_{1}$ the first sum runs over 6D Weyl fermions, while the
second runs over 6D {\textit{complex}} scalars satisfying $n=0$
boundary conditions.  $C_{2}(F)$ is the Casimir eigenvalue in the
representation of the fields $F = A_{\mu}$, $\Psi$ or $\Phi$, while
${\rm{Tr}}(T^{a} T^{b}) = T(F) \delta_{ab}$, where $T^{a}$ are the
generators in the representation of the field $F$.  The terms
proportional to $C_{2}(A)$ include the contributions of the complete
6D gauge multiplet, i.e. both the 4D spin-1 components, as well as the
spinless adjoints.  The sums in the expression for
$\hat{r}^{\Psi}_{1}$ and $r^{\Psi,M}_{1}$ run over its gauge
interactions, as well as the Yukawa interactions with complex scalars
satisfying $n=0$ boundary conditions.  We also derived relations for
scalars satisfying more general boundary conditions in
Section~\ref{sec:scalar_self}.

In addition, one finds operators at a third singular point with
coordinates $(0,L)$:
\beqa
& & \frac{L^{2}}{4} \, \delta(x^4) \delta(L - x^5) \left( \hat{r}^{A}_{2} {\cal{O}}_{A} 
+ \hat{r}^{45}_{2} {\cal{O}}_{45} 
+ \hat{r}^{\Psi}_{2} {\cal{O}}_{\Psi_{L}} + 
\hat{r}^{\Psi,M}_{2} {\cal{O}}^{M}_{\Psi_{L}}
+ \hat{r}^{\Phi}_{2} {\cal{O}}_{\Phi}
+ \hat{r}^{\Phi,M}_{2} {\cal{O}}^{M}_{\Phi}
\right)~,
\label{localOps2_summary}
\eeqa
where the coefficients are in general independent from those in
Eq.~(\ref{localOps1_summary}).  The contribution due to physics below
the cutoff scale $\Lambda$ was found to be
\beqa
\hat{r}^{A}_{2} &=& \frac{g^{2}_{4}}{16\pi^{2}} 
\ln \frac{\Lambda^2}{\mu^2} \left[ - 2 \, C_{2}(A) + 
\frac{1}{6} \sum_{\Phi} T(\Phi) \right]~,
\nonumber \\ [0.4em]
\hat{r}^{45}_{2} &=& \frac{g^{2}_{4}}{16\pi^{2}} 
\ln \frac{\Lambda^2}{\mu^2} \left[ 2 \, C_{2}(A) 
+ \frac{1}{2} \sum_{\Phi} T(\Phi)\right]~,
\nonumber \\ [0.4em]
\hat{r}^{\Psi}_{2} &=& \frac{1}{16\pi^{2}} 
\ln \frac{\Lambda^2}{\mu^2} \left[ 
- 2 \sum_{{\rm gauge}} g^{2}_{4} C_{2}(\Psi) + 
\frac{1}{4} \sum_{i} \lambda^{2}_{4,i} \right]~,
\nonumber \\ [0.4em]
\hat{r}^{\Psi,M}_{2} &=& \frac{1}{16\pi^{2}} 
\ln \frac{\Lambda^2}{\mu^2} \left[ 
\frac{1}{4} \sum_{i} \lambda^{2}_{4,i} \right]~.
\label{r2} \\ [0.4em]
\hat{r}^{\Phi}_{2} &=& \frac{1}{16\pi^{2}} 
\ln \frac{\Lambda^2}{\mu^2} \left[ - 3 \sum_{{\rm gauge}} 
g^{2}_{4} C_{2}(\Phi) \right]~,
\nonumber \\ [0.4em]
\hat{r}^{\Phi,M}_{2} &=& \frac{1}{16\pi^{2}} 
\ln \frac{\Lambda^2}{\mu^2} \left[ - \frac{5}{8} 
\sum_{{\rm gauge}} g^{2}_{4} C_{2}(\Phi) \right]~.
\nonumber 
\eeqa

It should be noted that the above coefficients were obtained in an
$R_{\xi}$-gauge with $\xi = -3$ (the expressions for the various
two-point functions in an arbitrary gauge are given in the main text).
This is a convenient gauge since all induced operators have
automatically a gauge invariant structure.  To see the gauge
invariance explicitly for other choices of the gauge parameter, a
further field redefinition is required.  Of course, physical
quantities that may be calculated from the two point functions, such
as the mass shifts, are $\xi$-independent. See the discussion in
subsection~\ref{sec:massLocal_gauge}.

Although in this paper we did not compute explicitly the
renormalization of KK-number violating gauge interactions, the result
of such a computation in $\xi = -3$ gauge should give rise to
operators with the precise coefficients necessary to provide the gauge
invariant completions of the kinetic operators in
Eqs.~(\ref{kineticOpsGauge})--(\ref{kineticOpsScalar}), according to
the standard prescription $\partial_{M} \rightarrow D_{M} =
\partial_{M} - i A_{M}$.  Thus, the operators given in
Eqs.~(\ref{localOps1_summary}), (\ref{r1}), (\ref{localOps2_summary})
and (\ref{r2}) provide a very convenient summary of the one-loop
results computed in this paper, allowing a straightforward
determination of the induced mass-shifts, or of any KK-number
violating gauge interactions of interest.

For example, the leading corrections to the gauge boson masses can be
obtained from
\beqa
M_{A^{(j,k)}} &=& M_{j,k} \left( 1 - \frac{1}{2} \ZZ_{A^{(j,k)}} \right)~,
\label{A_mass-shifts}
\eeqa
where $\ZZ_{A^{(j,k)}}$ is the wavefunction renormalization of
$A^{(j,k)}_{\mu}$ coming from the localized operator ${\cal{O}}_{A}$
in Eq.~(\ref{kineticOpsGauge}), and the tree-level mass of the
$(j,k)$-th level is given by $M_{j,k} = \sqrt{j^{2}+k^{2}}/R$.  The
spinless adjoints, on the other hand, receive only a ``mass''
renormalization associated with ${\cal{O}}_{45}$ in
Eq.~(\ref{kineticOpsGauge}), since the 4D kinetic term operators
vanish at the singular points.  In fact, ${\cal{O}}_{45}$, when
expanded in KK modes, contains precisely the gauge invariant linear
combination of $A_{4}$ and $A_{5}$ that is orthogonal to the eaten
would-be Nambu-Goldstone modes.  Thus, only this physical degree of
freedom receives a mass shift from the localized operators, given by
\beqa
M_{A^{(j,k)}_{S\!A}} &=& M_{j,k} \left( 1 + \frac{1}{2} 
\ZZ_{A^{(j,k)}_{S\!A}} \right)~.
\label{Ap_mass-shifts}
\eeqa
For fermions, one finds to first order in perturbation theory,
\beqa
M_{\Psi^{(j,k)}} &=& M_{j,k} \left( 1 - \frac{1}{2} \ZZ_{\Psi^{(j,k)}} + 
\frac{1}{2} \ZZ^{\prime}_{\Psi^{(j,k)}} \right)~,
\label{Psi_mass-shifts}
\eeqa
where $\ZZ_{\Psi^{(j,k)}}$ is the localized 4D kinetic term
renormalization constant, and $\ZZ^{\prime}_{\Psi^{(j,k)}}$ the
renormalization of the kinetic terms with transverse derivatives, i.e.
mass renormalization in a KK language.  Notice that only one of the 4D
chiralities receives a wavefunction renormalization due to localized
operators.  For scalars, one similarly has
\beqa
M^{2}_{\Phi^{(j,k)}} &=& 4 m^{2}_{0} + M^{2}_{j,k} \left( 1 - \ZZ_{\Phi^{(j,k)}} + 
\ZZ^{\prime}_{\Phi^{(j,k)}} \right)~,
\label{Phi_mass-shifts}
\eeqa
where $m_{0}$ is the mass of the zero-mode, and the factor of four
arises from the normalization of the heavy KK states relative to the
zero-mode.

For KK-parity even states, the $\ZZ$'s are related to the coefficients
of the localized kinetic term operators of
Eqs.~(\ref{localOps1_summary}) and (\ref{localOps2_summary}) by
\beqa
\begin{array}{rclcrcl}
\ZZ_{A^{(j,k)}} &=& 2 \, \hat{r}^{A}_{1} + \hat{r}^{A}_{2}~,
& \hspace{2mm} &
\ZZ_{A^{(j,k)}_{S\!A}} &=& 2 \, \hat{r}^{45}_{1} + \hat{r}^{45}_{2}~,
\\[0.4em]
\ZZ_{\Psi^{(j,k)}} &=& 2 \, \hat{r}^{\Psi}_{1} + \hat{r}^{\Psi}_{2}~, 
& &
\ZZ_{\Phi^{(j,k)}} &=& 2 \, \hat{r}^{\Phi}_{1} + \hat{r}^{\Phi}_{2}~, 
\\[0.4em]  
\ZZ^{\prime}_{\Psi^{(j,k)}} &=& 2 \,{\rm Re} 
\left( 2 \, \hat{r}^{\Psi,M}_{1} + \hat{r}^{\Psi,M}_{2} \right)~,
& &
\ZZ^{\prime}_{\Phi^{(j,k)}} &=& 2 \,{\rm Re} 
\left( 2 \, \hat{r}^{\Phi,M}_{1} + \hat{r}^{\Phi,M}_{2} \right)~.
\end{array}
\label{deltaZ}
\eeqa

For KK-parity odd states, the $\ZZ$'s are as in Eqs.~(\ref{deltaZ})
except that the $r_{2}$'s do not contribute, since the corresponding
KK wavefunctions vanish at $(x^{4},x^{5})=(0,L)$.  The explicit mass
formulae for both KK-parity even and KK-parity odd states were given
in Eqs.~(\ref{OddGaugeMass})--(\ref{EvenScalarMass}) of the
Introduction, where we also included the results for scalars
satisfying boundary conditions other than $n=0$.

As mentioned before, the localized operators summarized in
Eqs.~(\ref{localOps1_summary})--(\ref{r2}) contain much more
information than the mass shifts.  They also encode information about
KK transitions, as well as new interactions with the massive gauge
fields.  As an important example of KK-number violating couplings, we
consider those between zero-mode fermions, $\psi$, and massive
KK-parity even gauge bosons, $A_\mu^{(j,k)}$.  We write the effective
4D coupling as
\beq
g_{4} C^{\Psi A}_{j,k} \, \overline{\psi} \gamma^\mu A_\mu^{(j,k)} \psi~,
\label{coupling}
\eeq
where the dimensionless parameters $C^{\Psi A}_{j,k}$ are determined,
to lowest order in perturbation theory, by the coefficients defined in
Eqs.~(\ref{r1}) and (\ref{r2}), as
\beqa
C^{\Psi A}_{j,k} = - \frac{1}{2} \ZZb_{A^{(j,k)}} + \frac{1}{2} 
\ZZb_{\Psi^{(j,k)}} - 
\frac{1}{2} \ZZb^{\prime}_{\Psi^{(j,k)}} ~,
\label{interactions}
\eeqa
where now
\beqa
\label{interactions-more}
\ZZb_{A^{(j,k)}} &=& 
2 \, \hat{r}^{A}_{2} + (-1)^{j} \, \hat{r}^{A}_{1}~,
\nonumber \\ [0.4em]
\ZZb_{\Psi^{(j,k)}} &=& 
2 \, \hat{r}^{\Psi}_{2} + (-1)^{j} \, \hat{r}^{\Psi}_{1}~,
 \\ [0.4em] 
\ZZb^{\prime}_{\Psi^{(j,k)}} &=& 2
\,{\rm Re}\left[ 2 \, \hat{r}^{\Psi,M}_{2} + (-1)^{j} \, 
\hat{r}^{\Psi,M}_{1} \right]~. 
\nonumber
\eeqa
Notice that when $j$ is even, the KK-number violating couplings,
$C^{\Psi A}_{j,k}$, are simply related to the mass shifts of the heavy
states involved.  However, when $j$ is odd, $C^{\Psi A}_{j,k}$ depends
on a different linear combination of $\hat{r}_{1}$ and $\hat{r}_{2}$
than the one appearing in the mass shifts, e.g. Eq.~(\ref{deltaZ}).
These couplings may play a crucial role in discriminating these
scenarios from other kinds of new physics \cite{PhenoPaper}.

%%%%%%%%%%%%%%%%%%%%%%%%%%%%%%%%%%%%%
\bigskip

{\bf Acknowledgements:} \ We would like to thank Gustavo Burdman,
H.~C.~Cheng and Bogdan Dobrescu for interesting discussions.
This work was supported by DOE under contract DE-FG02-92ER-40699.

\appendix
%%%%%%%%%%%%%%%%%%%%%%%%%%%%%%%%%%%%%
\renewcommand{\theequation}{A.\arabic{equation}}
\section{Kaluza-Klein Number versus Momentum Space
Representations}
\label{App:MomSpace}

In this Appendix we derive in detail the general relation between the
KK-number and momentum space representations of a generic two-point
function.  We derive results that are sufficiently general to cover
the cases arising in the treatment of fermion and gauge fields.  In
particular, we allow for two-point functions connecting fields that
satisfy different boundary conditions, labeled by integers $n_{1}$ and
$n_{2}$.

The procedure is straightforward: starting from the propagator in
configuration space, $G(p;z,z')$, satisfying the appropriate boundary
conditions, one can either project on the KK wavefunctions,
$f^{(j,k)}(z)$, given in Eq.~(\ref{KKf}), or on the momentum space
wavefunctions, $h^{(m,l)}(z)$, given in Eq.~(\ref{planewaves}).
However, one must exert some care since the two sets of functions form
complete sets on different spacetime regions, and the quantum numbers
$(j,k)$ and $(m,l)$ cover different ranges.
  
%%%%%%%%%%%%%%%%%%%%%%%%%%%%%%%%%%%%%
\subsection{Diagonal Propagators}
\label{App:Diagonal}

We start with propagators that preserve Kaluza-Klein number and
postpone the analysis of Kaluza-Klein number violation to the next
subsection.  Using Eq.~(\ref{rotationmatrix}) we can derive, for any
two integers $n_{1}, n_{2} \in \{0,1,2,3\}$ and arbitrary expansion
coefficients $\tilde{g}_{j,k}$, the identity
\beqa
& & \hspace{-1.5cm} \frac{1}{4 L^{2}} \int_{-L}^{L} d^{2}z \, d^{2}z^{\prime} \,
\left[ h^{(m,l)}(z) \right]^{*} \,
\left( \frac{1}{4 L^{2}} \sum_{j,k} \, \tilde{g}_{j,k} \,
f_{n_{1}}^{(j,k)}(z) \left[f_{n_{2}}^{(j,k)}(z^{\prime})\right]^{*}
\right) h^{(m',l')}(z^{\prime})
\hspace*{2cm}
\nonumber \\
&=& \sum_{j,k}
\frac{1}{2\left[1+\delta_{j,0}\delta_{k,0} \right]}
\hat{\delta}(m,l;j,k;n_{1}) \,
\tilde{g}_{j,k} \,
\frac{1}{2 \left[1+\delta_{m',0}\delta_{l',0} \right]}
\hat{\delta}(j,k;m',l';n_{2})
\nonumber \\
&=& \frac{1}{4} \left[ \tilde{g}_{m,l} +
e^{ i (\theta_{1} - \theta_{2})} \tilde{g}_{-l,m} +
e^{ 2 i (\theta_{1} - \theta_{2})} \tilde{g}_{-m,-l} +
e^{ 3 i (\theta_{1} - \theta_{2})} \tilde{g}_{l,-m} \right]
\frac{\hat{\delta}(m,l;m',l';n_{2})}{\left[1+\delta_{m,0}\delta_{l,0}
\right]^{2}}~,
\label{convtemp}
\eeqa
where $\theta_{i} = n_{i} \pi/2$.  The significance of the ``tilde''
notation in $\tilde{g}_{j,k}$ will become clear in the following
paragraphs.

In Eq.~(\ref{convtemp}) it was necessary to assume that the sums over
$j$ and $k$ run unrestricted over all integer values.  To use this
identity, the simplest way to proceed is to extend the restricted sums
one naturally encounters when working in the KK-number representation
[see comments after Eq.~(\ref{planewaves})] to the whole range of
integers.  This can be achieved by noting that the KK wavefunctions
defined in Eq.~(\ref{KKf}) satisfy the relations
\beqa
\begin{array}{lclcl}
f_{n}^{(-j,k)}(z) = e^{i\theta} f_{n}^{(k,j)}(z)~,
& \hspace{0mm}
&
f_{n}^{(-j,-k)}(z) = e^{2i\theta} f_{n}^{(j,k)}(z)~,
& \hspace{0mm}
&
f_{n}^{(j,-k)}(z) = e^{3i\theta} f_{n}^{(k,j)}(z)~.
\end{array}
\label{continuation}
\eeqa
Therefore, for arbitrary coefficients $g_{j,k}$, one can write
\beq
{\sum_{j,k}}' \, g_{j,k} \, f_{n_{1}}^{(j,k)}(z)
\left[ f_{n_{2}}^{(j,k)}(z^{\prime}) \right]^{*}
=
\frac{1}{4}
\sum_{j,k} \, \tilde{g}_{j,k} \,
f_{n_{1}}^{(j,k)}(z)
\left[ f_{n_{2}}^{j,k}(z^{\prime}) \right]^{*}~,
\label{extension1}
\eeq
where, following our convention, the $'$ superscript in the sum on the
left-hand-side indicates that it runs over the restricted range $j >
0$, $k \geq 0$ and $j=k=0$, while the sum on the right-hand-side
stands for a double sum over {\textit{all}} integers.  The ``tilded''
quantities $\tilde{g}_{j,k}$ are defined in terms of $g_{j,k}$ as
follows:
\beq
\tilde{g}_{j,k} = \left\{
\begin{array}{ll} g_{j,k}  &
    \hspace{1cm} {\textrm{for $j > 0$, $k \geq 0$}}
    \\ [.5em]   e^{ - i (\theta_{1} - \theta_{2})} g_{k,-j}  &
    \hspace{1cm} {\textrm{for $j \leq 0$, $k > 0$}}
    \\ [.5em]   e^{ - 2 i (\theta_{1} - \theta_{2})}g_{-j,-k}  &
    \hspace{1cm} {\textrm{for $j < 0$, $k \leq 0$}}
    \\ [.5em]   e^{ - 3 i (\theta_{1} - \theta_{2})}g_{-k,j}  &
    \hspace{1cm} {\textrm{for $j \geq 0$, $k < 0$}}
\end{array}
\right. ~,
\label{gtilde}
\eeq
and $\tilde{g}_{0,0} = 4 g_{0,0}$.  Then, using Eqs.~(\ref{convtemp}),
(\ref{extension1}) and the definition~(\ref{gtilde}) one obtains
\beqa
\frac{1}{4 L^{2}} \int_{-L}^{L} d^{2}z \, d^{2}z^{\prime} \,
\left[ h^{(m,l)}(z) \right]^{*} \,
\left( \frac{1}{L^{2}} {\sum_{j,k}}' \, g_{j,k} \,
f_{n_{1}}^{(j,k)}(z) \left[f_{n_{2}}^{(j,k)}(z^{\prime})\right]^{*}
\right) h^{(m',l')}(z^{\prime})
\hspace*{1.5cm}
\nonumber \\
= \tilde{g}_{m,l} \, \frac{\hat{\delta}(m,l;m',l';n_{2})}{\left[1+
\delta_{m,0}\delta_{l,0} \right]^{2}}~.
\label{conv}
\eeqa

It follows that for a propagator with the general representation in
KK-number space,
\beq
G_{n_{1},n_{2}}(p; z; z^{\prime}) = \frac{1}{L^{2}} \, {\sum_{j,k}}' 
\, g_{j,k} \,
f_{n_{1}}^{(j,k)}(z) \left[f_{n_{2}}^{(j,k)}(z^{\prime})\right]^{*}~,
\label{GenProp_App}
\eeq
and using Eq.~(\ref{conv}), as well as the
completeness relation
\beq
\frac{1}{4 L^2}
\sum_{m,l} h^{(m,l)}(z) \left[ h^{(m,l)}(z^{\prime}) \right]^*
= \delta^{(2)}(z - z^{\prime})~,
\label{completenessExp}
\eeq
we can write
\beqa
G_{n_{1},n_{2}}(p; z; z^{\prime}) &=& \int^{L}_{-L} d^{2}y  
\, d^{2}y^{\prime}
\delta^{(2)}(z - y) G_{n_{1},n_{2}}(p; y; y^{\prime})
\delta^{(2)}(y^{\prime} - z^{\prime})
\nonumber \\
&=&
\frac{1}{4 L^{2}} \sum_{m,l} \sum_{m',l'} G_{p, n_{1},n_{2}}^{(m,l; m',l')}
h^{(m,l)}(z) \left[ h^{(m',l')}(z^{\prime}) \right]^{*}~,
\label{GMomRep_App}
\eeqa
with $G_{p, n_{1},n_{2}}^{(m,l; m',l')}$, as defined in Eq.~(\ref{Gp4p5}),
explicitly given by
\beqa
G_{p, n_{1},n_{2}}^{(m,l; m',l')} = \tilde{g}_{m,l} \,
\frac{\hat{\delta}(m,l;m',l';n_{2})}{\left[1+\delta_{m,0}\delta_{l,0} 
\right]^{2}}
&=& \tilde{g}_{m',l'} \,
\frac{\hat{\delta}(m,l;m',l';n_{1})}{\left[1+\delta_{m,0}\delta_{l,0} 
\right]^{2}}~.
\label{Gml_App}
\eeqa
To obtain the second equality we used the relations 
\beq
\tilde{g}_{l',-m'} = e^{i(\theta_{1} - \theta_{2})} \tilde{g}_{m',l'}~,
\hspace{1cm}
\tilde{g}_{-m',-l'} = e^{2i(\theta_{1} - \theta_{2})} \tilde{g}_{m',l'}~,
\hspace{1cm}
\tilde{g}_{-l',m'} = e^{3i(\theta_{1} - \theta_{2})} \tilde{g}_{m',l'}~,
\nonumber
\eeq
which follow from the definitions (\ref{gtilde}).  

Specializing Eq~(\ref{Gml_App}) to the scalar case with $n_{1} = n_{2}
= n$, Eqs.~(\ref{DiagGenProp}) and (\ref{GKKScalar}), one obtains the
scalar result of Eq.~(\ref{Gml}).  Recall that $\tilde{g}_{0,0} = 4
g_{0,0}$, as stated after Eq.~(\ref{gtilde}).

%%%%%%%%%%%%%%%%%%%%%%%%%%%%%%%%%%%%%
\subsection{Kaluza-Klein Mixing}
\label{App:KKmixing}

Now we consider two-point functions with an arbitrary KK-number
violating structure, as in Eq.~(\ref{connection}):
\beqa
G(p; z; z^{\prime}) &=&
\frac{1}{L^{2}} \, {\sum_{j,k}}' {\sum_{j',k'}}' \, g_{(j,k);(j',k')} \,
f_{n_{1}}^{(j,k)}(z) \left[f_{n_{2}}^{(j',k')}(z^{\prime})\right]^{*}
\nonumber \\
&=&
\frac{1}{4 L^{2}} \sum_{m,l} \sum_{m',l'} G_{n_{1}, n_{2}}^{(m,l; m',l')}
h^{(m,l)}(z)
\left[ h^{(m',l')}(z^{\prime}) \right]^{*}~.
\label{connection_App}
\eeqa
As was done in Eq.~(\ref{extension1}), we may extend the definition of
the coefficients $g_{(j,k);(j',k')}$ in such a way that the summations
over KK number can be taken over an unrestricted
range:~\footnote{Although $\tilde{g}_{(j,k);(j',k')}$ depends on
$n_{1}$ and $n_{2}$, we do not indicate this dependence to avoid
further notational cluttering.}
\beq
{\sum_{j,k}}' {\sum_{j',k'}}' \, g_{(j,k);(j',k')} \, f_{n_{1}}^{(j,k)}(z)
\left[ f_{n_{2}}^{(j',k')}(z^{\prime}) \right]^{*}
= \frac{1}{16} \sum_{j,k} \sum_{j',k'} \, \tilde{g}_{(j,k);(j',k')}
\, f_{n_{1}}^{(j,k)}(z)
\left[ f_{n_{2}}^{(j',k')}(z^{\prime}) \right]^{*}~.
\label{extension2}
\eeq

In order to write in a compact form the required extension
$\tilde{g}_{(j,k);(j',k')}$, we define a ``reordering'' function
\beq
R(j,k) = \left\{
\begin{array}{ll} (j,k) &
    \hspace{1cm} {\textrm{if}}~(j,k) \in S_{0} = \{j > 0, k \geq 0\}
    \\ [.5em]   (k,-j)  &
    \hspace{1cm} {\textrm{if}}~(j,k) \in S_{1} = \{j \leq 0, k > 0\}
    \\ [.5em]   (-j,-k)  &
    \hspace{1cm} {\textrm{if}}~(j,k) \in S_{2} = \{j < 0, k \leq 0\}
    \\ [.5em]   (-k,j)  &
    \hspace{1cm} {\textrm{if}}~(j,k) \in S_{3} = \{j \geq 0, k < 0\}
\end{array}
\right. ~,
\label{reordering}
\eeq
and also $P(j,k) = \omega$ if $(j,k) \in S_{\omega}$, giving the
quadrant to which $(j,k)$ belongs.  In terms of these auxiliary
functions the ``tilde'' operation is given by
\beqa
\begin{array}{rclcl}
\tilde{g}_{(j,k);(j',k')} &=& g_{R(j,k);R(j'k')} e^{-i P(j,k) \theta_{1} +
i P(j',k') \theta_{2}}~,
& \hspace{3mm}
& \textrm{if}~(j,k)~\&~(j',k') \neq (0,0)
\\ [.5em]
\tilde{g}_{(j,k);(0,0)} &=& 4 g_{R(j,k);(0,0)} e^{-i P(j,k) \theta_{1}}~,
& \hspace{3mm}
& \textrm{if}~(j,k) \neq (0,0)
\\ [.5em]
\tilde{g}_{(0,0);(j',k')} &=& 4 g_{(0,0);R(j'k')} e^{i P(j',k') \theta_{2}}~,
& \hspace{3mm}
& \textrm{if}~(j',k') \neq (0,0)
\\ [.5em]
\tilde{g}_{(0,0);(0,0)} &=& 16 g_{(0,0);(0,0)} ~,
\end{array}
\label{gtildeKKviol0}
\eeqa
where $\theta_{i} = n_{i} \pi/2$.

To relate the expansion coefficients $g_{(j,k);(j',k')}$ and
$G_{n_{1}, n_{2}}^{(m,l; m',l')}$ (KK-number and momentum bases,
respectively) in the KK-number violating case, we can project
Eq.~(\ref{connection_App}) on momentum space, as in Eq.~(\ref{Gp4p5}).
With the help of Eqs.~(\ref{rotationmatrix}) and (\ref{delta}), and
following a procedure similar to the one used to derive
Eqs.~(\ref{convtemp}) and (\ref{conv}), we find
\beqa
G_{n_{1}, n_{2}}^{(m,l; m',l')}
&=& \frac{1}{16} \sum_{j,k} \sum_{j',k'}
\frac{1}{[1+\delta_{j,0}\delta_{k,0}]} \hat{\delta}(m,l;j,k;n_{1}) \,
\tilde{g}_{(j,k);(j',k')} \,
\frac{1}{[1+\delta_{m',0}\delta_{l',0}]} \hat{\delta}(j',k';m',l';n_{2})
\nonumber \\
&=& \frac{1}{[1+\delta_{m,0}\delta_{l,0}][1+\delta_{m',0}\delta_{l',0}]}
\, \tilde{g}_{(m,l);(m',l')}~.
\label{KKmom}
\eeqa
Note that Eq.~(\ref{Gml_App}) is a subcase of the previous relation.

Using the fact that $g_{(j,k);(j',k')}$ and
$\tilde{g}_{(j,k);(j',k')}$ coincide when $j > 0$, $k \geq 0$, one
immediately obtains Eq.~(\ref{momKK1}).  Eqs.~(\ref{momKK2}),
involving zero-modes, are also immediately derived from
Eq.~(\ref{KKmom}) and the definitions (\ref{gtildeKKviol0}) and
(\ref{reordering}).

%%%%%%%%%%%%%%%%%%%%%%%%%%%%%%%%%%%%%
\subsection{Useful Identities for the Generalized Functions
$\hat{\delta}(m,l;m',l';n)$}
\label{App:Identities}

Here we record some useful relations involving the
$\hat{\delta}$-function introduced in Eq.~(\ref{delta}):
\beqa
\hat{\delta}(m_{1}, l_{1}; m_{2}, l_{2}; n) &=&
\hat{\delta}(m_{2}, l_{2}; m_{1}, l_{1}; -n)~,
\nonumber\\
\hat{\delta}(-m_{1}, -l_{1}; m_{2}, l_{2}; n) &=&
\hat{\delta}(m_{1}, l_{1}; -m_{2}, -l_{2}; n)
\nonumber\\
&=& \hat{\delta}(m_{1}, l_{1}; m_{2}, l_{2}; -n)~,
\nonumber\\
\hat{\delta}(m_{1}, l_{1}; l_{2}, -m_{2}; n) &=&
e^{-i n \pi/2} \hat{\delta}(m_{1}, l_{1}; m_{2}, l_{2}; n)~,
\\
\hat{\delta}(m_{1}, l_{1}; -m_{2}, -l_{2}; n) &=&
e^{-i n \pi} \hat{\delta}(m_{1}, l_{1}; m_{2}, l_{2}; n)~,
\nonumber\\
\hat{\delta}(m_{1}, l_{1}; - l_{2}, m_{2}; n) &=&
e^{-3i n \pi/2} \hat{\delta}(m_{1}, l_{1}; m_{2}, l_{2}; n)~.
\nonumber
\eeqa
We also note that the products of generalized $\hat{\delta}$'s that
appear in the diagrams (see Figure~\ref{fig:momenta}) can be
simplified using
\beqa
\sum_{m'_{1}, l'_{1}} \hat{\delta}(m_{1} ,l_{1}; m'_{1}, l'_{1}; n_{1})
\hat{\delta}(m'_{2}, l'_{2}; m_{2}, l_{2}; n_{2})
&=&
\hat{\delta}(m, l; m',l'; n_{1} - n_{2}) 
\label{two-deltas1} \\ [-0.8em]
& &  \hspace{-1.5cm} \mbox{} + 
e^{i\theta_{1}}
\hat{\delta}(m - m_{1} - l_{1}, l - l_{1} + m_{1}; m', l'; n_{1} - n_{2})
\nonumber \\ [.4em]
&& \hspace{-1.5cm} \mbox{} +
e^{2i\theta_{1}}
\hat{\delta}(m-2m_{1}, l - 2l_{1}; m', l'; n_{1} - n_{2})
\nonumber \\ [.4em]
& &  \hspace{-1.5cm} \mbox{} + 
e^{3i\theta_{1}}
\hat{\delta_{1}}(m - m_{1} + l_{1}, l - l_{1} - m_{1}; m', l'; n_{1} - n_{2})~,
\nonumber
\eeqa
where $\theta_{1} = n_{1} \pi/2$, and $m_{2} = m_{1} - m$, $l_{2} =
l_{1} - l$, $m'_{2} = m'_{1} - m'$ and $l'_{2} = l'_{1} - l'$.

One can check that the last three terms in Eq.~(\ref{two-deltas1}),
when summed over $(m_{1},l_{1})$ vanish, except when KK-parity is
preserved and
\begin{itemize}
\item $n_{1} = n_{2} = 0$:
\beqa
\sum_{m_{1}, l_{1}} \left(
\sum_{m'_{1}, l'_{1}} \hat{\delta}(m_{1} ,l_{1}; m'_{1}, l'_{1}; 0)
\hat{\delta}(m'_{2}, l'_{2}; m_{2}, l_{2}; 0) -
\hat{\delta}(m, l; m',l'; 0) \right)
&=&
4 \times \left\{%
\begin{array}{ll}
    3 \\
    2 \\
    5/2 \\
\end{array}%
\right.~,
\nonumber
\eeqa
\item $n_{1} = n_{2} = \pm1$:
\beqa
\sum_{m_{1}, l_{1}} \left(
\sum_{m'_{1}, l'_{1}} \hat{\delta}(m_{1} ,l_{1}; m'_{1}, l'_{1}; \pm 1)
\hat{\delta}(m'_{2}, l'_{2}; m_{2}, l_{2}; \pm 1) -
\hat{\delta}(m, l; m',l'; 0) \right)
&=&
4 \times \left\{%
\begin{array}{ll}
    -1 \\
    0 \\
    -1/2 \\
\end{array}%
\right.~,
\nonumber
\eeqa
\item $n_{1} = n_{2} = 2$:
\beqa
\sum_{m_{1}, l_{1}} \left(
\sum_{m'_{1}, l'_{1}} \hat{\delta}(m_{1} ,l_{1}; m'_{1}, l'_{1}; 2)
\hat{\delta}(m'_{2}, l'_{2}; m_{2}, l_{2}; 2) -
\hat{\delta}(m, l; m',l'; 0) \right)
&=&
4 \times \left\{%
\begin{array}{ll}
    -1 \\
    -2 \\
    -3/2 \\
\end{array}%
\right.~,
\nonumber
\eeqa
\end{itemize}
where the three cases in each of the bullets are as defined in
Eq.~(\ref{cases}), which we quote here again for convenience:
\beqa
\begin{tabular}{lcl}
  Case 1a: & $(-1)^{m+l} = (-1)^{m'+l'} = +1$, & $m-m'$ even, \\ [0.3em]
  Case 1b: & $(-1)^{m+l} = (-1)^{m'+l'} = +1$, & $m-m'$ odd, \\ [0.3em]
  Case 2: & $(-1)^{m+l} = (-1)^{m'+l'} = -1$, \\
\end{tabular}
\eeqa
or
\begin{itemize}
\item $n_{1} = 0$, $n_{2} = 2$ or $n_{1} = 2$, $n_{2} = 0$:
\beqa
\sum_{m_{1}, l_{1}} \left(
\sum_{m'_{1}, l'_{1}} \hat{\delta}(m_{1} ,l_{1}; m'_{1}, l'_{1}; 0)
\hat{\delta}(m'_{2}, l'_{2}; m_{2}, l_{2}; 2) -
\hat{\delta}(m, l; m',l'; 2) \right)
&=&
4 \times \left\{%
\begin{array}{ll}
    0 \\
    1/2 \\
    -1/2 \\
\end{array}%
\right.~,
\nonumber
\eeqa
\end{itemize}
where in this latter bullet the three cases are different from the
previous ones, as given in Eq.~(\ref{cases2}):
\beqa
\begin{tabular}{lcl}
  Case 1: & $(-1)^{m+l} = (-1)^{m'+l'} = +1$, \\ [0.3em]
  Case 2a: & $(-1)^{m+l} = (-1)^{m'+l'} = -1$, & $m-m'$ even, \\ [0.3em]
  Case 2b: & $(-1)^{m+l} = (-1)^{m'+l'} = -1$, & $m-m'$ odd. \\ [0.3em]
\end{tabular}
\eeqa
%

%%%%%%%%%%%%%%%%%%%%%%%%%%%%%%%%%%%%%
\setcounter{equation}{0}
\renewcommand{\theequation}{B.\arabic{equation}}
\section{Tree-level Propagators on the Chiral Square}
\label{App:Propagators}

In this Appendix we derive the propagators for fields of various spins
on the ``chiral square'' background of
\cite{Dobrescu:2004zi,Burdman:2005sr}.  We follow the general strategy
of deriving the propagator in the mixed position and momentum space
representation, making use of the KK wavefunctions in Eq.~(\ref{KKf})
to take care of the appropriate boundary conditions, after which it is
a simple matter to find the corresponding momentum space expressions
using the formulae derived in Appendix~\ref{App:MomSpace}.

%%%%%%%%%%%%%%%%%%%%%%%%%%%%%%%%%%%%%
\subsection{Chiral Fermions}

We start by computing the fermion propagator in mixed position and
momentum space.  We need to solve
\beq
i \left( - i \Gamma^{\mu} p_{\mu} + \Gamma^{4} \partial_{4} + \Gamma^{5}
\partial_{5} \right)
G^{\pm}(p; z; z^{\prime}) = i P_\mp \delta^{(2)}(z - z^{\prime})~,
\label{GFermionEqn}
\eeq
where the $\pm$ superscript in $G^{\pm}(p; z; z^{\prime})$ refers to
the two possible 6D chiralities defined by
\beq
P_\pm = \frac{1}{2} \left( 1 \pm \overline{\Gamma} \right) ~,
\eeq
and the six-dimensional chirality operator is $\overline{\Gamma} =
\Gamma^0\Gamma^1\Gamma^2\Gamma^3\Gamma^4\Gamma^5$. 
A convenient representation of the $8 \times 8$ $\Gamma$-matrices is
\beq
\Gamma^\mu = \gamma^\mu \otimes \sigma^0 \; \; , \; \; \;
\Gamma^{4,5} = i \gamma_5 \otimes \sigma^{1,2} ~,
\eeq
where $\gamma^{\mu}$ are the 4D $\gamma$-matrices, $\gamma_{5}=
i\gamma^{0}\gamma^{1}\gamma^{2}\gamma^{3}$ is the 4D chirality
operator, $\sigma^0$ is the $2\times 2$ unit matrix and $\sigma^i$ are
the Pauli matrices.  In this representation, $\overline{\Gamma} =
-\gamma_5 \otimes \sigma^3$.

Each 6D chiral fermion contains both left- and right-handed
4-dimensional chiralities.  Since the folded square identifications
require them to obey different boundary conditions, it is useful to
treat them separately by using the $8 \times 8$ 4D chirality projectors
\beq
P_{L,R} = \frac{1}{2} \left( 1 \mp i\Gamma^0\Gamma^1\Gamma^2\Gamma^3
\right)~.
\label{LeftRightProjectors}
\eeq
Defining
\beqa
G^{\pm}_{LL}(p; z,z') = P_{L} \, G^{\pm}(p; z,z') P_{R} &=&
\int d^{4}x \, e^{ipx}
\langle \Psi_{\pm_{L}}(x, z) \overline{\Psi_{\pm_{L}}}(0, z') \rangle~,
\nonumber \\ [.4em]
G^{\pm}_{RL}(p; z,z') = P_{R} \, G^{\pm}(p; z,z') P_{R} &=&
\int d^{4}x \, e^{ipx}
\langle \Psi_{\pm_{R}}(x, z) \overline{\Psi_{\pm_{L}}}(0, z') \rangle~,
\nonumber \\ [.4em]
G^{\pm}_{LR}(p; z,z') = P_{L} \, G^{\pm}(p; z,z') P_{L} &=&
\int d^{4}x \, e^{ipx}
\langle \Psi_{\pm_{L}}(x, z) \overline{\Psi_{\pm_{R}}}(0, z') \rangle~,
\\ [.4em]
G^{\pm}_{RR}(p; z,z') = P_{R} \, G^{\pm}(p; z,z') P_{L} &=&
\int d^{4}x \, e^{ipx}
\langle \Psi_{\pm_{R}}(x, z) \overline{\Psi_{\pm_{R}}}(0, z') \rangle~,
\nonumber
\eeqa
we can derive the equations obeyed by $G^{\pm}_{LL}$, $G^{\pm}_{RL}$,
etc.  as follows.  Applying the differential operator $i \left( - i
\Gamma^{\mu} p_{\mu} + \Gamma^{4} \partial_{4} + \Gamma^{5}
\partial_{5} \right)$ to Eq.~(\ref{GFermionEqn}), and then projecting
by $P_{L}$ on the left and by $P_{R}$ on the right, we can obtain a
differential equation for $G^{\pm}_{LL}$:
\beq
\left( p^{2} + \partial^{2}_{4} + \partial^{2}_{5} \right)
G^{\pm}_{LL}(p; z, z') = i  P_{L} P_\pm \Gamma^{\mu} p_{\mu}
\delta^{(2)}(z - z^{\prime})~,
\label{LRfirstProjection}
\eeq
where we used the fact that $\Gamma^{4}$ and $\Gamma^{5}$ commute with
$P_{L,R}$.  We see that this equation is solved by
\beq
G^{\pm}_{LL}(p; z, z') = P_{L} P_\pm \Gamma^{\mu} p_{\mu}
G_{n^{\pm}_{L}}(p; z, z')~,
\label{LLfermionScalarRelation}
\eeq
where $G_{n^{\pm}_{L}}(p; z, z')$ is the 6-dimensional scalar
propagator satisfying Eq.~(\ref{GScalarEqn}). It is given
explicitly by
\beq
G_{n^{\pm}_{L}}(p; z, z') = \frac{1}{L^{2}} \, {\sum_{j,k}}' \, g_{S}^{j,k}
f_{n^{\pm}_{L}}^{(j,k)}(z) \left[ f_{n^{\pm}_{L}}^{(j,k)}(z^{\prime})
\right]^{*}~,
\label{GLLKKsoln}
\eeq
where $g_{S}^{j,k}$ is the 4-dimensional scalar propagator defined in
Eq.~(\ref{GKKScalar}).  The integers $n^{\pm}_{L}$ label the boundary
conditions obeyed by the left-handed components of the 6D fermion in
question.

We can find $G^{\pm}_{RL}$ from the solution $G^{\pm}_{LL}$ above by
projecting directly Eq.~(\ref{GFermionEqn}), $P_{L} \cdots P_{R}$, to
obtain
\beq
\Gamma^{\mu} p_{\mu} G^{\pm}_{RL} + i \left( \Gamma^{4} \partial_{4} +
\Gamma^{5} \partial_{5} \right) G^{\pm}_{LL} = 0~.
\label{LRsecondProjection}
\eeq
Using the identities Eqs.~(\ref{Pidentities}), we find
\beqa
G^{\pm}_{RL} &=&  - \frac{i}{p^{2}} p_{\mu} \Gamma^{\mu}
\Gamma^{4} \partial_{\pm} G^{\pm}_{LL}
\nonumber \\
&=& P_{R} P_\pm \Gamma^{4} \left( i \partial_{\pm}
G_{n^{\pm}_{L}} \right)~,
\label{GLRsoln}
\eeqa
where $\partial_{\pm}$ were defined in Eq.~(\ref{partialpm}), and
$G_{n^{\pm}_{L}}$ is given in Eq.~(\ref{GLLKKsoln}).

Proceeding in an analogous fashion (i.e. projecting by $P_{R}$ on the
left and by $P_{L}$ on the right) we see that $G^{\pm}_{RR}$ is given
by
\beq
G^{\pm}_{RR}(p; z, z') = P_{R} P_\pm \Gamma^{\mu} p_{\mu}
G_{n^{\pm}_{R}}(p; z, z')~,
\label{RRfermionScalarRelation}
\eeq
where $G_{n^{\pm}_{R}}(p; z, z')$ is given by Eq.~(\ref{GLLKKsoln})
with $n^{\pm}_{L} \rightarrow n^{\pm}_{R}$, and also that
$G^{\pm}_{LR}$ is given by
\beqa
G^{\pm}_{LR} &=& P_{L} P_\pm \Gamma^{4}
\left( i \partial_{\mp} G_{n^{\pm}_{R}} \right)~.
\label{GRLsoln}
\eeqa
The integers $n^\pm_L$ and
$n^\pm_R$ associated with a given 6D fermion are related by Eq.~(\ref{nLnR}).
Using Eqs.~(\ref{delf}) and (\ref{GLLKKsoln}) we can easily calculate
the partial derivatives needed in Eqs.~(\ref{GLRsoln}) and
(\ref{GRLsoln}):
\beq
i \, \partial_{\pm} G_{n}(p; z, z') = - \frac{1}{L^{2}} \,
{\sum_{j,k}}' \, r_{j,\pm k} M_{j,k} \, g_{S}^{j,k}
f_{n \mp 1}^{(j,k)}(z) \left[ f_{n}^{(j,k)}(z^{\prime}) \right]^{*}~,
\label{GRLKKsoln}
\eeq
where the phases $r_{j,k}$ were defined in Eq.~(\ref{rjk}).

Putting the results (\ref{LLfermionScalarRelation}),
(\ref{GLRsoln}), (\ref{RRfermionScalarRelation}) and (\ref{GRLsoln})
together, we see that the fermion propagator has the representation
\beqa
G^{\pm}_{LL}(p; z, z') &=& P_{L} P_\pm
\frac{1}{L^{2}} \, {\sum_{j,k}}' \, \Gamma^{\mu} p_{\mu} \, g_{S}^{j,k} \,
 \left[ f_{n^{\pm}_{L}}^{(j,k)}(z) \right] 
 \left[ f_{n^{\pm}_{L}}^{(j,k)}(z^{\prime}) \right]^*~,
\nonumber \\ [.3em]
G^{\pm}_{RL}(p; z, z') &=& - P_{R} P_\pm
\frac{1}{L^{2}} \, {\sum_{j,k}}' \, \Gamma^{4} M_{j,k} \, g_{S}^{j,k} \,
\left[ r_{j,\pm k} f_{n^{\pm}_{R}}^{(j,k)}(z) \right]
\left[ f_{n^{\pm}_{L}}^{(j,k)}(z^{\prime}) \right]^*~,
\nonumber \\ [.3em]
G^{\pm}_{LR}(p; z, z') &=& - P_{L} P_\pm
\frac{1}{L^{2}} \, {\sum_{j,k}}' \, \Gamma^{4} M_{j,k} \, g_{S}^{j,k} \,
\left[ f_{n^{\pm}_{L}}^{(j,k)}(z) \right]
\left[ r_{j,\pm k} f_{n^{\pm}_{R}}^{(j,k)}(z^{\prime}) \right]^*~,
\label{GFermionKK} \\ [.3em]
G^{\pm}_{RR}(p; z, z') &=& P_{R} P_\pm
\frac{1}{L^{2}} \, {\sum_{j,k}}' \, \Gamma^{\mu} p_{\mu} \, g_{S}^{j,k} \,
\left[ r_{j,\pm k} f_{n^{\pm}_{R}}^{(j,k)}(z) \right]
\left[ r_{j,\pm k} f_{n^{\pm}_{R}}^{(j,k)}(z^{\prime}) \right]^*~,
\nonumber
\eeqa
where $n_{L}^{\pm}$ and $n_{R}^{\pm}$ are related as in Eq.~(\ref{nLnR}).

We may now project Eqs.~(\ref{GFermionKK})
on momentum space, as in Eq.~(\ref{Gp4p5}).  Using Eqs.~(\ref{gtilde})
and (\ref{conv}), we obtain
\beqa
G_{LL,p}^{\pm, (m,l; m',l')} &=& P_{L} P_\pm \Gamma^{\mu} p_{\mu}
\, g_{S}^{m,l} \, \hat{\delta}(m,l; m',l';n^{\pm}_{L})~,
\nonumber \\ [.3em]
G_{RL,p}^{\pm, (m,l; m',l')} &=& - P_{R} P_\pm \Gamma^{4} r_{m,\pm l}
M_{m,l}
\, g_{S}^{m,l} \, \hat{\delta}(m,l; m',l';n^{\pm}_{L})~,
\nonumber \\ [.3em]
G_{LR,p}^{\pm, (m,l; m',l')} &=& - P_{L} P_\pm \Gamma^{4} r_{m,\mp l}
M_{m,l}
\, g_{S}^{m,l} \, \hat{\delta}(m,l; m',l';n^{\pm}_{R})~,
\label{GFermionmom} \\ [.3em]
G_{RR,p}^{\pm, (m,l; m',l')} &=& P_{R} P_\pm \Gamma^{\mu} p_{\mu}
\, g_{S}^{m,l} \, \hat{\delta}(m,l; m',l';n^{\pm}_{R})~.
\nonumber
\eeqa
Note that by using the relations~(\ref{nLnR}) the ``tilde'' operation
defined in Eq.~(\ref{gtilde}) simplifies considerably, and there is no
need to distinguish among the four possible sign assignments of the
momenta $m,l$.

By adding the four results in Eq.~(\ref{GFermionmom}), and using again
the identities~(\ref{Pidentities}), we can write the fermion
propagator in the more compact form given in
Eq.~(\ref{FullFermionPropagator}).  Recall that the extra-dimensional
momenta with lower indices are given by $p_{4} = -m/R$ and $p_{5} =
-l/R$.

%%%%%%%%%%%%%%%%%%%%%%%%%%%%%%%%%%%%%
\subsection{Gauge Fields: The Spin-1 Components}

After integration by parts, the terms involving $A^{\mu}$ in
Eq.~(\ref{s1}), with the gauge fixing Eq.~(\ref{gf}), can be written
as
\beq
 - \frac{1}{2} A^{\mu} \left[ \left(p^{2} + \partial_{4}^{2} +
 \partial_{5}^{2}\right) \eta_{\mu\nu} -
 \left( 1-\frac{1}{\xi} \right) p_{\mu} p_{\nu}
\right] A^{\nu}~,
\eeq
where we went to the momentum space associated with the non-compact
dimensions.

The spin-1 propagator in the mixed representation is defined by
\beq
\left[ (p^{2} + \partial_{4}^{2} + \partial_{5}^{2}) \eta_{\mu\lambda} -
\left( 1-\frac{1}{\xi} \right) p_{\mu} p_{\lambda} \right]
G^{\lambda\nu}(p; z; z^{\prime}) = - i \, \delta_{\mu}^{\nu}
\delta^{(2)}(z - z^{\prime})~,
\label{GGaugeEqn}
\eeq
and the solution satisfying the boundary conditions is
\beq
G_{\mu\nu}(p; z; z^{\prime}) = \frac{1}{L^{2}} \, {\sum_{j,k}}' \,
g_{\mu\nu}^{j,k} f_{0}^{(j,k)}(z)
\left[ f_{0}^{(j,k)}(z^{\prime}) \right]^{*}~,
\label{GGaugesoln}
\eeq
with $g_{\mu\nu}^{j,k}$ as given in Eq.~(\ref{GKKGauge}).

Going to momentum space in the compactified dimensions, as in
Eq.~(\ref{Gp4p5}), and using Eq.~(\ref{Gml_App}) with $n_{1} = n_{2} =
0$, we can immediately derive Eq.~(\ref{GGaugemom}).

%%%%%%%%%%%%%%%%%%%%%%%%%%%%%%%%%%%%%
\subsection{Gauge Fields: The Spin-0 Components}

In this subsection, we concentrate on the slightly trickier issues
associated with the scalar degrees of freedom contained in the
six-dimensional gauge field, $A_{M}$.  Up to surface terms that do not
contribute as a result of the boundary conditions discussed in
Ref.~\cite{Burdman:2005sr}, the terms in Eq.~(\ref{s1}) [with the
gauge fixing Eq.~(\ref{gf})] quadratic in $A_{4}$, $A_{5}$ can be
written as
\beqa
{\cal L} &\supset& \frac{1}{2} \sum_{i,j = 4,5} A^{i} \left[ (p^{2} + 
\partial_{4}^{2} +
\partial_{5}^{2}) \delta_{ij} - (1-\xi) \partial_{i} \partial_{j}
\right] A^{j}
\nonumber \\
&=& \frac{1}{4} (A^{*}_{+}\, , \, A^{*}_{-})
\left(
\begin{array}{cc}  p^{2} + \frac{1}{2} (1+\xi) \partial_{+} \partial_{-} &
   - \frac{1}{2} (1-\xi) \partial^{2}_{+} \\ [0.5em]
   - \frac{1}{2} (1-\xi) \partial^{2}_{-} & p^{2} + \frac{1}{2} (1+\xi) 
   \partial_{+} \partial_{-}
\end{array} \right)
\left(
\begin{array}{c} A_{+} \\ [.5em] A_{-}
\end{array} \right) ~,
\label{A4A5action}
\eeqa
where $\partial_{\pm}$ were defined in Eq.~(\ref{partialpm}) and we
wrote the second line in the $A_{\pm}$ basis as defined in
Eq.~(\ref{pmdefns}).  The fields $A_{\pm}$ are convenient since they
satisfy well defined boundary conditions \cite{Burdman:2005sr}:
\beq
A_{\pm}(x^{\mu}; z) = \mp \frac{1}{L} \, {\sum_{j,k}}' \,
A_{\pm}^{(j,k)}(x^{\mu}) f_{3,1}^{(j,k)}(z)~.
\label{ApmKK}
\eeq
We define the propagator for the $A_{\pm}$ system by
\beq
\label{prop45}
\frac{1}{2}
\left(
\begin{array}{cc}  p^{2} + \frac{1}{2} (1+\xi) \partial_{+} \partial_{-} &
    - \frac{1}{2} (1-\xi) \partial^{2}_{+}
\\  [0.5em]
- \frac{1}{2} (1-\xi) \partial^{2}_{-} & p^{2} +
\frac{1}{2} (1+\xi) \partial_{+} \partial_{-}
\end{array} \right)
\left(
\begin{array}{cc} G_{++} & G_{+-} \\ [.5em] G_{-+} & G_{--}
\end{array} \right) = i \, \delta^{(2)}(z - z^{\prime})~.
\eeq

To find the solution to Eq.~(\ref{prop45}) we make the following ansatz
\beqa
G_{++}(p; z; z^{\prime}) &=& \frac{1}{L^{2}} \, {\sum_{j,k}}' \, g^{j,k}_{++}
\, f_{3}^{(j,k)}(z) \left[ f_{3}^{(j,k)}(z^{\prime}) \right]^{*}~,
\nonumber \\
G_{+-}(p; z; z^{\prime}) &=& \frac{1}{L^{2}} \, {\sum_{j,k}}' \, g^{j,k}_{+-}
\, f_{3}^{(j,k)}(z) \left[ f_{1}^{(j,k)}(z^{\prime}) \right]^{*}~,
\nonumber \\
G_{-+}(p; z; z^{\prime}) &=& \frac{1}{L^{2}} \, {\sum_{j,k}}' \, g^{j,k}_{-+}
\, f_{1}^{(j,k)}(z) \left[ f_{3}^{(j,k)}(z^{\prime}) \right]^{*}~,
\label{Gsoln_A4A5} \\
G_{--}(p; z; z^{\prime}) &=& \frac{1}{L^{2}} \, {\sum_{j,k}}' \, g^{j,k}_{--}
\, f_{1}^{(j,k)}(z) \left[ f_{1}^{(j,k)}(z^{\prime}) \right]^{*}~,
\nonumber
\eeqa
which satisfies the boundary conditions implied by Eq.~(\ref{ApmKK}).
If we further use Eq.~(\ref{delf}) it is easy to see that the ansatz
(\ref{Gsoln_A4A5}) solves Eq.~(\ref{prop45}) provided
\beq
\label{propg}
\frac{1}{2}
\left(
\begin{array}{cc}  p^{2} - \frac{1}{2} (1+\xi) M^{2}_{j,k} &
    \frac{1}{2} (1-\xi) r_{j,k}^{2} M^{2}_{j,k}
    \\ [.5em] \frac{1}{2} (1-\xi) r_{j,-k}^{2} M^{2}_{j,k} &
    p^{2} - \frac{1}{2} (1+\xi) M^{2}_{j,k}
\end{array} \right)
\left(
\begin{array}{cc} g^{j,k}_{++} &   g^{j,k}_{+-}
    \\ [.5em] g^{j,k}_{-+} &   g^{j,k}_{--}
\end{array} \right) =
i \left( \begin{array}{cc}  1 & 0  \\ [0.4em] 0 & 1 \end{array} \right)~.
\eeq
The solution to this system is
\beqa
\left(
\begin{array}{cc} g^{j,k}_{++} & g^{j,k}_{+-}
    \\ [.5em]  g^{j,k}_{-+} & g^{j,k}_{--}
\end{array} \right)
&=& \frac{2 i }{(p^{2} - M^{2}_{j,k}) (p^{2} - \xi M^{2}_{j,k})}
\left(
\begin{array}{cc}  p^{2} - \frac{1}{2} (1+\xi) M^{2}_{j,k} &
    - \frac{1}{2} (1-\xi) r_{j,k}^{2} M^{2}_{j,k}
    \\ [.5em] - \frac{1}{2} (1-\xi) r_{j,-k}^{2} M^{2}_{j,k} &
    p^{2} - \frac{1}{2} (1+\xi) M^{2}_{j,k}
\end{array} \right)
\nonumber \\ [.6em]
&=&
\left(
\begin{array}{cc}  g^{j,k}_{h} + g^{j,k}_{\phi}  &
    - r_{j,k}^{2} \, \left( g^{j,k}_{h} - g^{j,k}_{\phi} \right)
    \\ [.5em] - r_{j,-k}^{2} \, \left( g^{j,k}_{h} - g^{j,k}_{\phi}
    \right) &
    g^{j,k}_{h} + g^{j,k}_{\phi}
\end{array} \right)
\label{gsoln}
\\ [0.6em]
&=&
\left(
\begin{array}{cc} r_{j,k} & r_{j,k}
    \\ [.5em]   -r^{*}_{j,k} &   r^{*}_{j,k}
\end{array} \right)
\left(
\begin{array}{cc} g^{j,k}_{h} & 0
    \\ [.5em]  0 &  g^{j,k}_{\phi}
\end{array} \right)
\left(
\begin{array}{cc} r^{*}_{j,k} & - r_{j,k}
    \\ [.5em]   r^{*}_{j,k} &   r_{j,k}
\end{array} \right)~,
\nonumber
\eeqa
where we defined
\beq
g^{j,k}_{h} = \frac{i}{p^{2} - M^{2}_{j,k}}~, \hspace{1cm}
g^{j,k}_{\phi} = \frac{i}{p^{2} - \xi M^{2}_{j,k}}~.
\label{ghphi_App}
\eeq

Eqs.~(\ref{Gsoln_A4A5}), (\ref{gsoln}) and (\ref{ghphi_App}) completely
specify the propagator (in mixed position/momentum space) associated
with the two degrees of freedom $A_{4}$ and $A_{5}$.

However, one must be careful on how the above propagator, defined as
the inverse of the quadratic operator in Eq.~(\ref{A4A5action}),
should be used, since one must impose the constraint $A_{+} =
A_{-}^{\dagger}$.  Starting with the path integral, one can see that
the relation between the various components $G_{++}$, $G_{+-}$,
$G_{-+}$ and $G_{--}$ defined in Eq.~(\ref{Gsoln_A4A5}) and the
tree-level two-point functions that appear in the Feynman rules are
\beqa
\int d^{4}x \, e^{ipx} \langle A_{+}(x,z) A_{+}^{\dagger}(0,z') \rangle
&=& \frac{1}{2} \left[ G_{++}(p;z,z') + G_{--}(p;z',z) \right]~,
\nonumber \\
\int d^{4}x \, e^{ipx} \langle A_{+}(x,z) A_{+}(0,z') \rangle
&=& \frac{1}{2} \left[ G_{+-}(p;z,z') + G_{+-}(p;z',z) \right]~,
\nonumber \\
\int d^{4}x \, e^{ipx} \langle A_{+}^{\dagger}(x,z) A_{+}^{\dagger}(0,z') \rangle
&=& \frac{1}{2} \left[ G_{-+}(p;z,z') + G_{-+}(p;z',z) \right]~,
\label{Two-point_Prop}
\\
\int d^{4}x \, e^{ipx} \langle A_{+}^{\dagger}(x,z) A_{+}(0,z') \rangle
&=& \frac{1}{2} \left[ G_{--}(p;z,z') +
G_{++}(p;z',z) \right]~.
\nonumber
\eeqa
Here we wrote all correlators in terms of $A_{+}$ by using $A_{-} =
A_{+}^{\dagger}$.  The last two relations are simply the complex
conjugates of the first two.  Projecting, for example, the second
relation on the momentum space wavefunctions (\ref{planewaves}), one
gets (we indicate only the dependence on the extra dimensional
momenta)
\beqa
\langle A_{+}^{m,l} A_{+}^{m',l'} \rangle
&=& \frac{1}{2} \left[ G_{+-}^{(m,l;m',l')} + G_{+-}^{(-m',-l';-m,-l)} \right]
\nonumber \\
&=& \frac{1}{2} \left[ \tilde{g}^{m,l}_{+-} \, \hat{\delta}(m,l;m',l';1) + 
\tilde{g}^{-m,-l}_{+-} \, \hat{\delta}(-m,-l;-m',-l';3) \right]
\nonumber \\
&=& \frac{1}{2} \left[ \tilde{g}^{m,l}_{+-} + \tilde{g}^{-m,-l}_{+-} \right]  
\, \hat{\delta}(m,l;m',l';1)~,
\label{ApApexample}
\eeqa
where we used Eq.~(\ref{Gml_App}).  The remaining relations in
Eq.~(\ref{Two-point_Prop}) can be similarly expressed in terms of
\beqa
G_{++}^{(m,l; m',l')} &=& \left( g^{m,l}_{h} + g^{m,l}_{\phi} \right)
\hat{\delta}(m,l; m',l';3)~,
\nonumber \\
G_{+-}^{(m,l; m',l')} &=& - r_{m,l}^{2} \, \left( g^{m,l}_{h} - g^{m,l}_{\phi}
\right)
\, \hat{\delta}(m,l; m',l';1)~,
\nonumber \\
G_{-+}^{(m,l; m',l')} &=& - r_{m,l}^{*2} \, \left( g^{m,l}_{h} - g^{m,l}_{\phi}
\right)  \,
\hat{\delta}(m,l; m',l';3)~,
\label{GA4A5mom_App} \\
G_{--}^{(m,l; m',l')} &=& \left( g^{m,l}_{h} + g^{m,l}_{\phi} \right)
\hat{\delta}(m,l; m',l';1)~,
\nonumber
\eeqa
which follow from Eq.~(\ref{Gml_App}), noting that from the definition
of $\tilde{g}^{m,l}_{+-}$, Eq.~(\ref{gtilde}) with $n_{1} = 3$ and
$n_{2} = 1$, and the explicit expression for $g^{m,l}_{+-}$ in
Eq.~(\ref{gsoln}), one finds $\tilde{g}^{m,l}_{+-} = g^{m,l}_{+-}$ and
similarly $\tilde{g}^{m,l}_{-+} = g^{m,l}_{-+}$, $\tilde{g}^{m,l}_{++}
= g^{m,l}_{++}$ and $\tilde{g}^{m,l}_{--} = g^{m,l}_{--}$.  In
Eq.~(\ref{GA4A5mom_App}) we also used the definitions
(\ref{ghphi_App}).

%%%%%%%%%%%%%%%%%%%%%%%%%%%%%%%%%%%%%
\subsection{Feynman Rules for Gauge Interactions}

We finally present the Feynman rules in momentum space for the
interactions among fermions and gauge fields in six dimensions.  These
can be read directly from the vertices derived in
\cite{Burdman:2005sr}, and we simply present them diagramaticaly in
Figures~\ref{fig:Feynman1} and \ref{fig:Feynman2}.

\FIGURE{
\vspace*{-5mm}
\centerline{
   \resizebox{15.5cm}{!}{\includegraphics{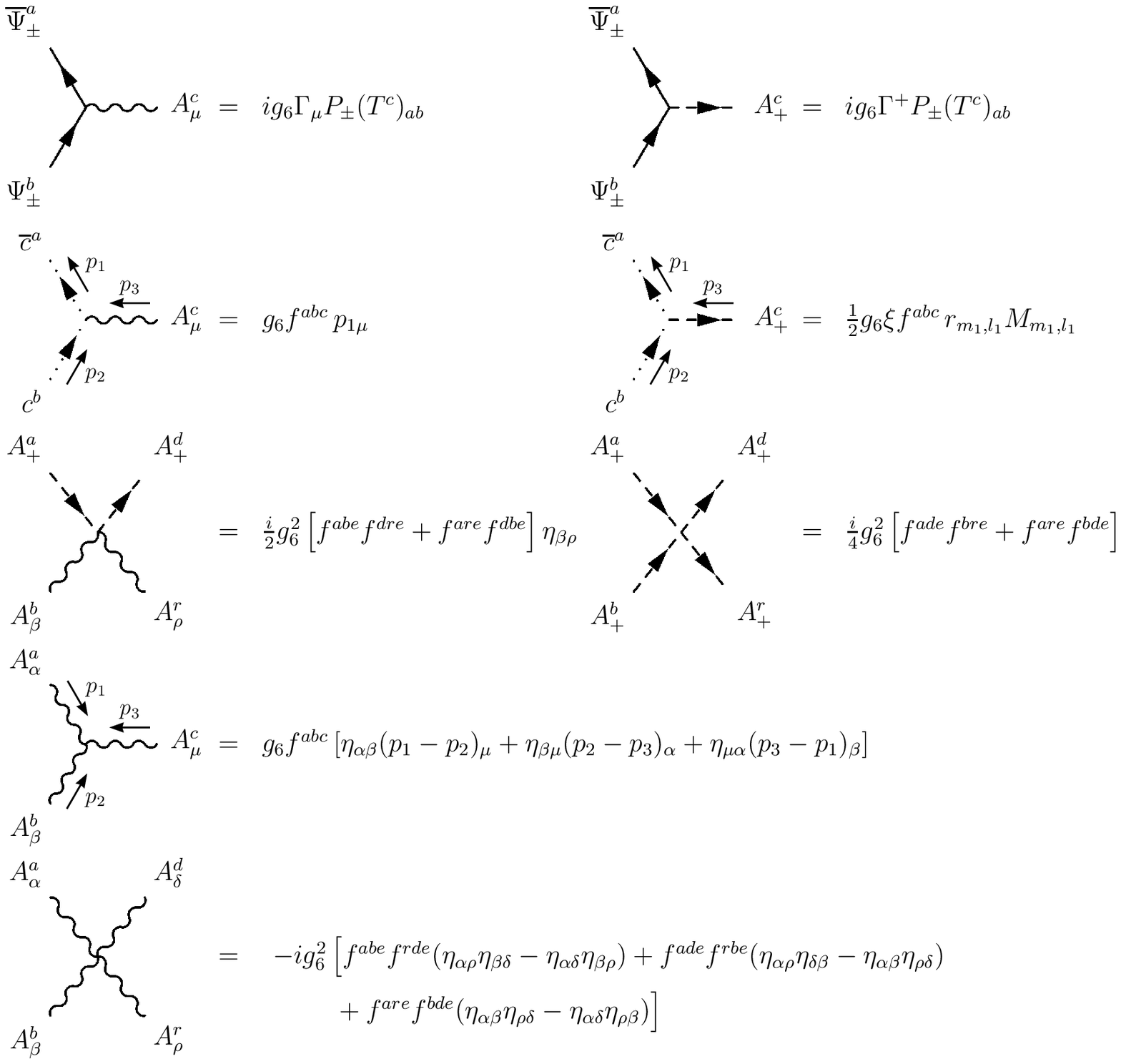}}
}
\caption{Feynman rules in the 6D gauge sector, including the
additional scalar degrees of freedom.  The dashed lines with the arrow
represent the propagation of $A_{+}$.  For the interaction of two
fermions or two ghosts with an outgoing $A_{+}$ there is a
corresponding vertex with an incoming $A_{+}$.  The corresponding
rules are obtained by $\Gamma^{+} \rightarrow \Gamma^{-}$, where
$\Gamma^{\pm} = \frac{1}{2} \left( \Gamma^{4} \pm i \Gamma^{5}
\right)$, and $r_{m,l} \rightarrow r^{*}_{m,l}$, respectively. The $r_{m,l}$ 
phases were defined in Eq.~(\ref{rjk}).}
\label{fig:Feynman1}
}
\FIGURE[t]{
\vspace*{-5mm}
\centerline{
   \resizebox{10.5cm}{!}{\includegraphics{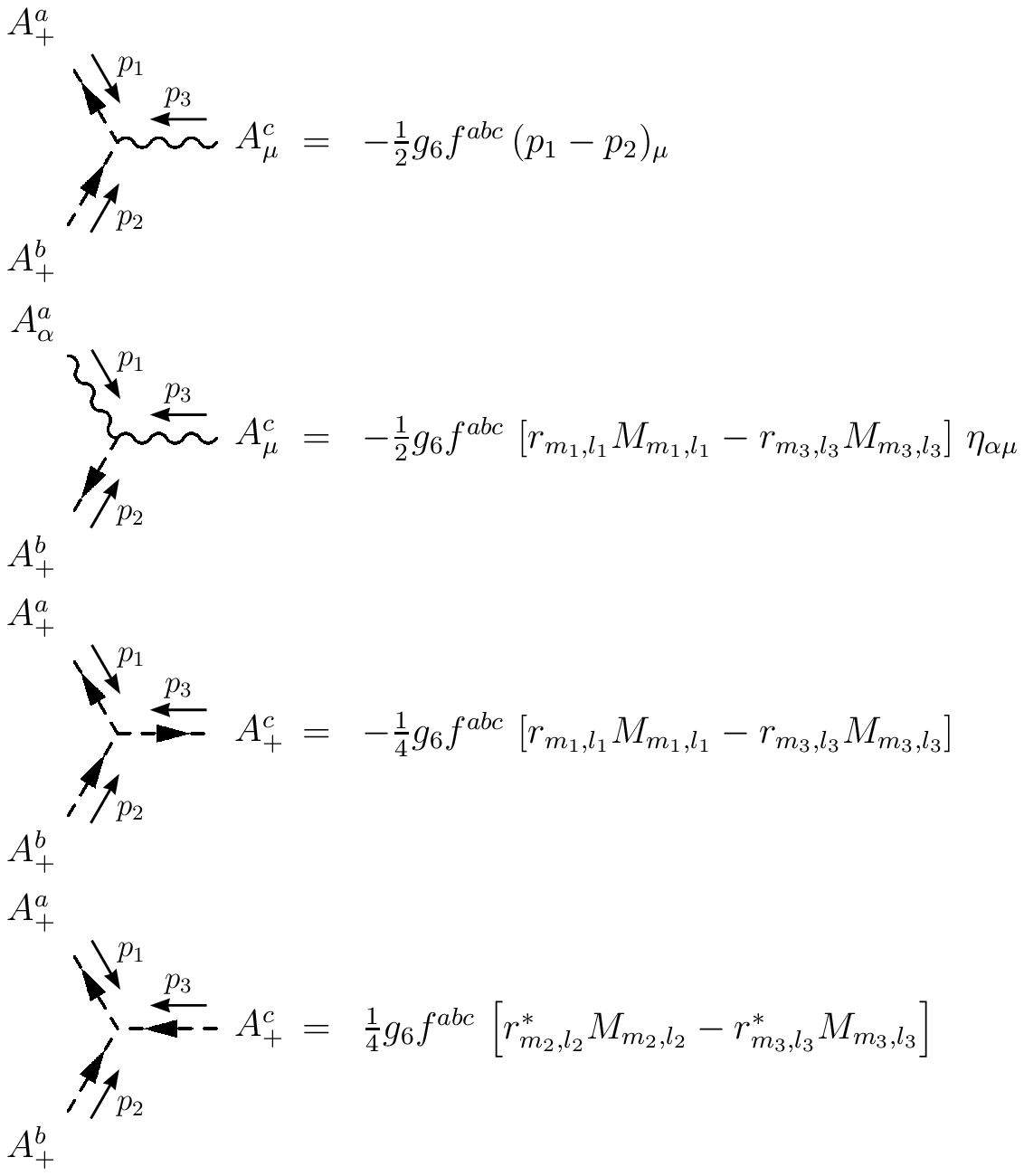}}
}
\caption{Feynman rules in the 6D gauge sector, including the
additional scalar degrees of freedom.  The dashed lines with the arrow
represent the propagation of $A_{+}$.  For the interaction of two
gauge bosons with a single outgoing $A_{+}$ there is a corresponding
vertex with an incoming $A_{+}$.  The corresponding rule is obtained
by the replacement $r_{m,l} \rightarrow r^{*}_{m,l}$. The $r_{m,l}$ 
phases were defined in Eq.~(\ref{rjk}).}
\label{fig:Feynman2}
}
%

%%%%%%%%%%%%%%%%%%%%%%%%%%%%%%%%%%%%%

\end{document}